\documentclass[journal=jctcce,manuscript=article]{achemso}

\usepackage[T1]{fontenc} 

\author{Hemani Chhabra}
\affiliation{Physical \& Theoretical Chemistry Laboratory, Department of Chemistry, University of Oxford, South Parks Road, Oxford OX1 3QZ, United Kingdom}
\author{Garima Mishra}
\affiliation{Department of Physics, Indian Institute of Technology Kanpur, Kanpur 208016, India}
\author{Yijing Cao}
\affiliation{Physical \& Theoretical Chemistry Laboratory, Department of Chemistry, University of Oxford, South Parks Road, Oxford OX1 3QZ, United Kingdom}
\author{Domen Pre\v{s}ern}
\affiliation{Physical \& Theoretical Chemistry Laboratory, Department of Chemistry, University of Oxford, South Parks Road, Oxford OX1 3QZ, United Kingdom}
\author{Enrico Skoruppa}
\affiliation{Laboratory for Soft Matter and Biophysics, KU Leuven, Celestijnenlaan 200D, 3001 Leuven, Belgium}
\author{Maxime M. C. Tortora}
\affiliation{Physical \& Theoretical Chemistry Laboratory, Department of Chemistry, University of Oxford, South Parks Road, Oxford OX1 3QZ, United Kingdom}
\altaffiliation{Laboratory of Biology and Modeling of the Cell,
\'{E}cole Normale Sup\'{e}rieure de Lyon, 46, all\'{e}e d'Italie, 69364 Lyon Cedex 07, France}
\author{Jonathan P. K. Doye}
\email{jonathan.doye@chem.ox.ac.uk}
\affiliation{Physical \& Theoretical Chemistry Laboratory, Department of Chemistry, University of Oxford, South Parks Road, Oxford OX1 3QZ, United Kingdom}

\title[Elastic properties of DNA nanostructures]
  {Computing the elastic mechanical properties of rod-like DNA nanostructures}


\begin{document}

\begin{tocentry}

\includegraphics[width=1\columnwidth]{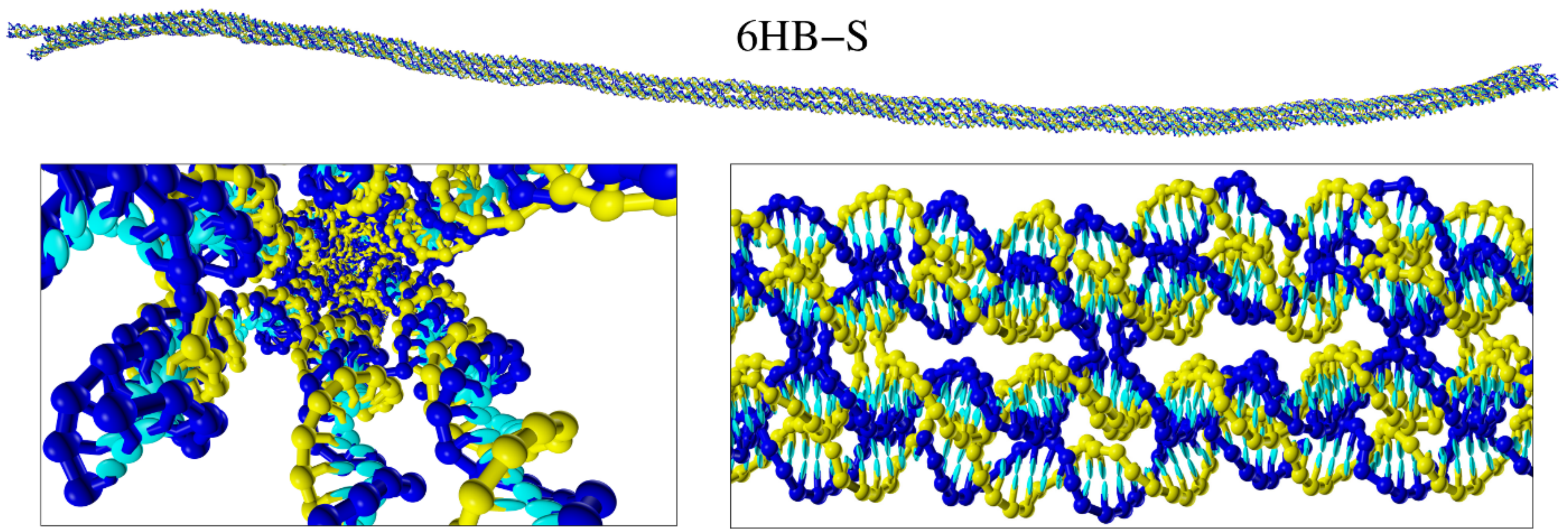}


%
%

\end{tocentry}

\begin{abstract}
To study the elastic properties of rod-like DNA nanostructures, we perform
long simulations of these structure using the oxDNA coarse-grained model.
By analysing the fluctuations in these trajectories we obtain 
estimates of the bend and twist persistence lengths, and the 
underlying bend and twist elastic moduli and couplings between them.
Only on length scales beyond those associated with the spacings between
the interhelix crossovers do the bending fluctuations behave like those of a worm-like chain.
The obtained bending persistence lengths are much larger than that for 
double-stranded DNA and increase non-linearly with the number of helices, 
whereas the twist moduli increase approximately linearly. 
To within the numerical error in our data, the twist-bend coupling constants are of order zero.
That the bending persistence lengths we obtain are generally somewhat 
higher than in experiment probably reflects both that the simulated origami 
have no assembly defects and that the oxDNA extensional modulus
for double-stranded DNA is too large.
\end{abstract}


\section{Introduction}

DNA nanotechnology exploits the programmability of DNA and its ability to self assemble 
to make nanoscale structures and devices. This is perhaps most famously achieved through 
the DNA origami approach, where a long single-stranded scaffold can be folded up into
almost any arbitrary shape made up of arrays of double helices by the addition of the
appropriate short staple strands.\cite{Rothemund06} 
Similar scaffold-free approaches where the size of the object is not constrained by 
the scaffold length are also available.\cite{Yin08,Wei12,Ke12b} 
Although a huge variety of structures have been produced by these 
techniques,\cite{Hong17,Ramezani20} one of the most common classes of structures
have been DNA rods or filaments;\cite{Pfeifer18b} these are the subject of the current paper. 

Understandably, much of the characterization of DNA origami and other large DNA
nano-assemblies initially focused on their structure. Cryo-electron microscopy,
for example, has revealed the detailed structure of DNA origami in exquisite
detail.\cite{Bai12} More recently there has been an increasing interest in
characterizing their mechanical properties.  One reason is the advent
of ``DNA mechanotechnology'', that is the use of DNA nanodevices to
generate, transmit and sense nanoscale forces.\cite{Blanchard19} These include
examples where DNA nanostructures are used as the ``hardware for single-molecule
investigations''.\cite{Gosse19} For example, in force spectroscopy, rod-like
origami have been used as stiff handles for force application with molecular
tweezers,\cite{Pfitzner13,Kilchherr16} and small origami have been used to
create multivalent force sensors.\cite{Dutta18}
Origami with inbuilt tension in components have been used to apply forces to unravel 
protein-DNA complexes\cite{Funke16b,Le16} and to study the force-dependent binding of proteins to 
DNA motifs.\cite{Nickels16b,Kramm20}
Protein-protein interactions have also been measured by the shape changes they induce 
in an origami to which they are attached.\cite{Funke16c}
Similarly, origami sensors have been designed that produce a change in FRET signal
when interactions are sufficiently strong to induce a shape change.\cite{Ke16}

For the above applications, it is important to have a good understanding of the
mechanical properties of the DNA origami; for example, knowing the stiffness of
the origami handles, knowing the compliance of the origamis with respect to the
internal forces that they are seeking to apply, being able to accurately
calibrate the magnitude of the forces that are being applied, knowing the
limits of the forces that can be applied whilst maintaining the structural
integrity of the origami.

Designing internal stresses into an origami also provides a means to 
further control its structure.
For example, twisting and bending of origami
can be achieved by changing the numbers of base pairs between junctions from their
ideal values.\cite{Dietz09} 
Another approach is to use the tension in extended single stranded sections; 
for example, to maintain the global shape in origami tensegrity structures,\cite{Liedl10} and 
to induce bending in the main body of an origami.\cite{Liedl10,Zhou14,Suzuki20} 
Bistable DNA origami have also been designed where transitions between the
two free-energy minima require bending of the different origami sections.\cite{Zhou15}
In the above examples, the elastic moduli for bending and twisting of the origami will 
control the resulting degree of bending or twisting, and the size of the free-energy
barrier in the bistable system. 

Understanding the mechanical properties of DNA origami is also of fundamental
interest. How does the coupling of the DNA double-helical elements lead to the
net mechanical behaviour of the origami? How can it be tuned or controlled?
The answers to these questions build on our fundamental understanding of
the mechanics of double-stranded and single-stranded DNA, which has been
revealed in detail through single-molecule experiments, for example using
optical and magnetic tweezers to pull and twist DNA,\cite{Bustamante03} 
and further analysed theoretically\cite{Brahmachari18b} and in
simulations.\cite{Romano13,MarinGonzalez17}

The most basic mechanical properties of DNA nanostructure that we might wish to
understand are their elastic moduli. The quantity that has been most commonly 
probed in experiments is the bending persistence length 
of elongated origami, particularly of DNA nanotubes. This has been done for
origami that are four-, six-, eight- and ten-helix 
bundles,\cite{Liedl10,Kauert11,Pfitzner13,Castro15,Lee19b} single-stranded-tile 
nanotubes with from five to ten helices\cite{Schiffels13} and nanotubes made from
double-crossover\cite{Rothemund04,ONeill06} or other\cite{Wang12b} larger
tiles, as well as more open wireframe origami.\cite{Benson18,Benson19}
The experimental approaches used include analyses of their molecular
contours when adsorbed on surfaces (either through their end-to-end 
distances\cite{Pfitzner13,ONeill06,Wang12b,Castro15,Lee19b} or from an analysis of their 
tangent-tangent correlation function\cite{Liedl10,Schiffels13}), 
their cyclization rates\cite{Rothemund04} 
and their force response in magnetic tweezers.\cite{Kauert11}  
For example, the values of the persistence length obtained are in the ranges 
740\,nm--1\,$\mu$m, 
1--3\,$\mu$m 
and 3--8.2\,$\mu$m for 
four-\cite{Kauert11,Lee19b}, 
six-\cite{Liedl10,Kauert11,Pfitzner13,Schiffels13,Wang12b,Castro15,Lee19b} 
and eight-helix\cite{Pfitzner13,Schiffels13} bundles, respectively.
These values are much larger than those for double-stranded DNA ($\sim$50\,nm), and show
how the coupling of double-helical elements in a DNA assembly can lead to new mechanical
behaviours.
Persistence lengths have also been measured for origami connected by linkers of varying 
stiffness.\cite{Pfeifer18} 
The magnetic tweezer study was
also able to measure the twist modulus of DNA origami, finding values of 390\,nm and 530\,nm for
four- and six-helix bundles, respectively.\cite{Kauert11}
The moduli associated with radial compression of rod-like origamis have also been probed by
atomic-force microscopy.\cite{Ma17}

Most explorations of the mechanical limits of DNA origami have focused on the effects of 
tension typically applied by optical tweezers or an atomic-force 
microscope.\cite{Bae14,Koirala14,Shrestha16,Engel18} These studies have revealed 
saw-tooth force-extension profiles, where the origamis yield through multiple unravelling
events. The buckling of DNA nanotubes under extreme twisting has also been briefly 
studied.\cite{Kauert11}

Modelling has the potential to play an important role in better understanding
the mechanics of DNA nanostructures. Firstly, the experimental measurement of mechanical 
properties can be challenging, so if modelling can provide a means to accurately estimate
such properties, this could be extremely useful---particularly as a means to pre-screen 
a design for the required mechanical properties, and to explore how those properties
can be modulated. Secondly, it can enhance our fundamental understanding, particularly
as it allows the mechanical and structural responses to stress to be directly related 
and the relative roles of different factors to be identified. Modelling
also allows theories that predict the mechanical properties of DNA nanostructures 
from duplex properties to be consistently tested.

Here, the focus is on computing the elastic mechanical properties of DNA
nanotubes.  To do this, one has to decide at what level of description it is
most appropriate to model the DNA. Of course, all-atom models will provide the
most detailed structural description, and all-atom calculations of full-size
DNA origami have now been performed,\cite{Yoo13,Gopfrich16} 
albeit at considerable computational cost, even
estimating the elastic moduli for cuboidal blocks.\cite{Yoo13,Sloane16}
However, the time scales associated with the long length-scale fluctuations of the
full-size DNA nanotubes studied here are likely to be prohibitive for all-atom
studies.  Very short sections of such DNA nanotubes have been studied by
all-atom molecular dynamics with estimates of the stretch moduli being obtained
from rapid pulling simulations and persistence lengths from bend angle
distributions.\cite{Joshi15,Joshi16,Naskar19} 
At the other limit are models that consider DNA helices as simple mechanical elements. 
Such models have been very successful at providing rapid structural predictions of 
DNA origami structure, including the effects of internal stresses on 
structure,\cite{Castro11,Kim12,Maffeo20} but their application to study mechanical properties
has been more limited.\cite{Kim16b}

Coarse-grained models at the nucleotide level have a number of attractive
features when considering origami mechanics. They are simple enough that
simulations can access sufficiently long time scales to characterize the
structural fluctuations of origamis, even in the case of high shape anisotropy. Also,
coupling of the stress response to internal degrees of freedom, such as
base-pair breaking and unstacking at nick sites, can emerge naturally
from the model.
However, it is important that the model can accurately reproduce the mechanical
properties of double-stranded (and in some instances, single-stranded) DNA, as
well as other motifs (e.g.\ nicks and junctions).  Here, we use the oxDNA model
as it possesses just such features.  It provides a good description of the
elastic properties of double-stranded DNA (dsDNA), including not only the persistence
length and torsional modulus\cite{Snodin15}, but also non-trivial 
features such as twist-bend coupling. \cite{Nomidis17,Skoruppa17}
It also captures well the yielding of dsDNA under 
tension \cite{Romano13} and twist.\cite{Matek15}

OxDNA has also been shown to provide a very good structural description of DNA 
origami\cite{Snodin15,Sharma17,Shi17,Zhou18,Snodin19,Berengut19,Suma20} and 
other large DNA nanostructures.\cite{Schreck16,Hong18,Matthies19} It has been used
to probe the mechanisms of origami failure under tension,\cite{Engel18}
and even to explain the origins of phase chirality of cholesteric phases of twisted DNA 
origami rods in terms of the net chirality of their thermal fluctuations.\cite{Tortora20}

The systems that we consider here focus on those for which there are experimental
measurements. As such rod-like origamis are often termed helix bundles we label each
system as $n$HB, where $n$ is the number of helices.
Firstly, we consider the 4HB and 6HB origami of Ref.\ \citenum{Kauert11} whose mechanical properties 
were probed by magnetic tweezers (MT). Secondly, we consider the set of four 6HB origamis
whose liquid-crystalline properties were experimentally studied in 
Ref.\ \citenum{Siavashpouri17} and analysed by simulation and theory in 
Ref.\ \citenum{Tortora20}. These are labelled by their designed global axial 
twist (2$\times$LH,1$\times$LH,1$\times$RH) or lack thereof (S: straight, i.e.\ untwisted).
The persistence length of 6HB-S was measured in Ref.\ \citenum{Pfitzner13}. We also consider
a 10HB that was also studied in Ref.\ \citenum{Siavashpouri17}.
Finally, we consider 6HB and 8HB systems made from single-stranded tiles (SST) 
whose mechanical properties were studied in Ref.\ 
\citenum{Schiffels13}, plus an additional equivalent 4HB example. 

\section{Methods}

\subsection{oxDNA}

\begin{figure}[t]
\centering
\includegraphics[width=5.3in]{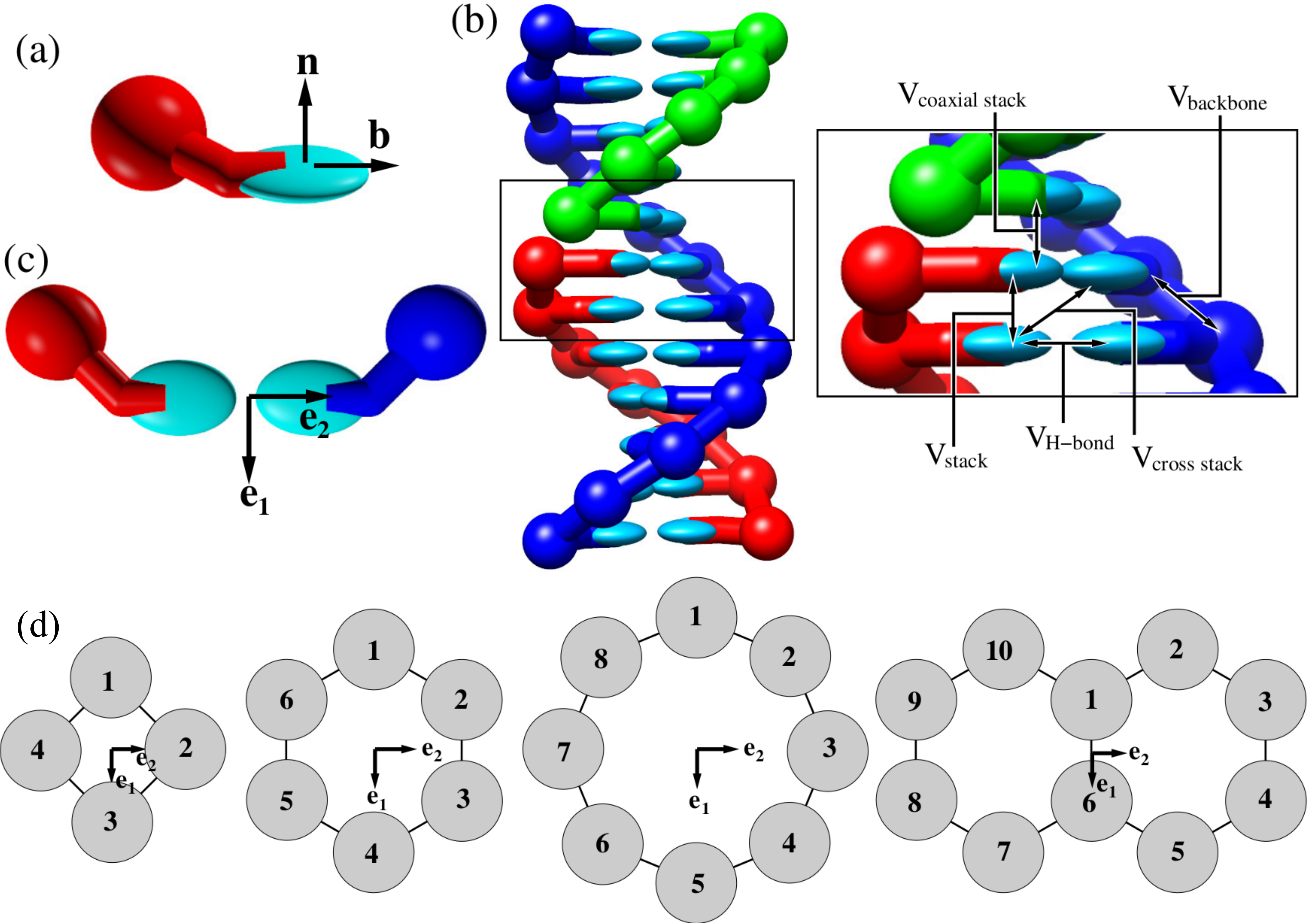}
\caption{(a) An oxDNA nucleotide along with the ``base'' and ``normal'' vectors used
to define its orientation. 
(b) A nicked double helix illustrating the different interactions in oxDNA. 
(c) An oxDNA base pair illustrating the directions of 
 $\widehat{\mathbf{e}}_1$ and $\widehat{\mathbf{e}}_2$.
(d) Schematics showing $\widehat{\mathbf{e}}_1$ and 
$\widehat{\mathbf{e}}_2$
for the different DNA nanostructures considered.
In (c) and (d) 
$\widehat{\mathbf{e}}_3$ is directed out of the plane of the paper.}
\label{fig:basics}
\end{figure}

OxDNA is a nucleotide-level coarse-grained model of DNA,\cite{Ouldridge11,Sulc12,Snodin15} 
where each nucleotide is a rigid body that has sites representing the backbone and base. 
The inter-nucleotide 
interactions include base stacking, hydrogen-bonding between complementary base pairs, a backbone
potential, excluded volume, electrostatic interactions between the charged backbones 
and cross stacking between diagonally opposite bases in dsDNA. These have been
parameterized to reproduce DNA structure, the thermodynamics of hybridization and the mechanical
properties of double- and single-stranded DNA. 
Here, we use the latest version of the model that has been fine-tuned to better reproduce
the properties of DNA origami.\cite{Snodin15}
A nucleotide and duplex as represented by oxDNA are depicted in Fig.\ 
\ref{fig:basics}.

The electrostatic potential is of a Debye-H\"{u}ckel form and has been fitted to reproduce the
[Na$^+$] dependence of hybridization thermodynamics.\cite{Snodin15} Here, we have run simulations 
at [Na$^+$]=0.5\,M,
which in oxDNA gives similar behaviour to the high-magnesium conditions typically used 
in origami assembly. 
As we are interested in the generic properties of DNA nanotubes, 
we use the sequence-averaged version of the model, where the interaction strengths of the stacking
and hydrogen-bond interactions do not depend on the identity of the base. For DNA origami, the
differences between the sequence-averaged model and using the specific M13 sequence are expected to 
be minimal.

In the oxDNA simulation code the positions and orientations of each nucleotide
are defined by $\mathbf{r}$ the notional centre of mass of the nucleotide
(this is collinear with the hydrogen-bonding and stacking sites),
$\widehat{\mathbf{b}}$ the ``base'' vector (this points from centre of mass towards the
base interaction sites), and $\widehat{\mathbf{n}}$ the base normal vector that is 
perpendicular to the notional plane of the base (Fig.\ \ref{fig:basics}(a)). 
These vectors will be used to define the local bend and twist angles for both a DNA
duplex and the DNA nanotubes.

\subsection{DNA nanotubes}

We study the mechanical properties of ten different elongated DNA
nanostructures that have been studied experimentally in three separate 
papers,\cite{Kauert11,Schiffels13,Siavashpouri17}
focussing mainly on systems where some mechanical properties have been
experimentally estimated. Seven of these are DNA origami. In a DNA origami,
there is a single long ``scaffold'' strand 
that runs through the entire structure, and is held together by many ``staple'' strands that
bind specifically to multiple domains of the scaffold. The scaffold strands in
origami typically derive from the genome of an M13 virus, which is just over 7000 
nucleotides long.
The remaining three DNA nanostructures are structurally very similar, 
but are made up of just a small number of short DNA strands 
(or single-stranded tiles\cite{Yin08}) with a repeating motif along the length
of the tube. 

\begin{figure}[t]
\centering
\includegraphics[width=5.3in]{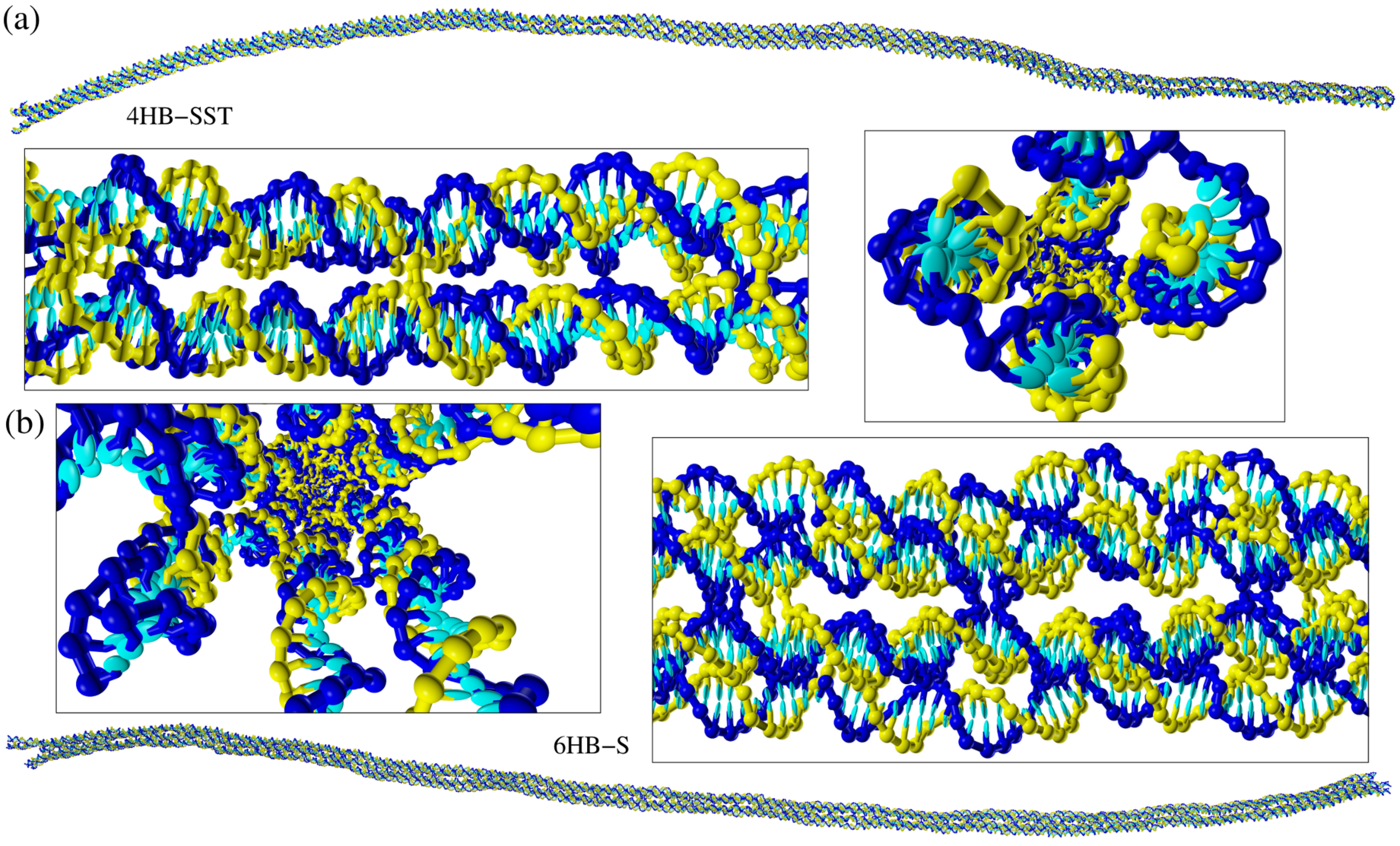}
\caption{Snapshots of two of the DNA nanostructures (a) 4HB-SST; (b) 6HB-S,
including close-ups from the ends and the side.}
\label{fig:snapshots}
\end{figure}

All the structures are helix bundles, i.e.\ they are made up of parallel arrays of
double helices. The points
at which strands pass between helices are termed crossovers or junctions. In the DNA origamis
these are typically ``double'' crossovers where two strands pass between the helices at the
junctions, and the ends of the staple strands occur between junctions. 
By contrast, in the SST nanotubes, the junctions are all ``single'' crossovers as the
ends of the strands are located at the junctions (Fig.\ \ref{fig:snapshots}(a)).

Non-planar DNA origami designs are typically based on either a hexagonal
or square lattice of DNA helices, where crossovers
occur between adjacent helices of the lattice, and the spacings between
junctions are chosen to best match the pitch of DNA (e.g.\ the junctions are
separated by seven base pairs in the hexagonal lattice giving an angle of exactly
$4\pi/3$ between consecutive junction if the 
pitch of DNA 
is 10.5 base pairs per turn).  The six-helix bundles origamis are based
on the hexagonal lattice, and the four-helix bundles on the square lattice. 

For the SST nanotubes, junctions are alternatively spaced 10 and then 11 base pairs apart. 
For a flat sheet with this design, one might expect that they would be flat and untwisted, as
the average crossover spacing matches the DNA pitch. However, as adjacent crossovers involve 
opposing strands, the groove structure of DNA origami also needs to be considered, and it
has been suggested that this leads to a natural curvature.\cite{Yin08} 
Even with this curvature, rolling up the sheets into tubes will likely introduce stress
into the tubes, as the interhelix angles will no longer match their relaxed angles.
We choose to study the tube isomer consistent with the presumed natural curvature, where 
the major groove is on the outside of the tube at the junctions. 

The 10HB origami is the only system considered that is not a tube. Instead, it has a double
hexagon cross-section (Fig.\ \ref{fig:basics}(d)). The anisotropy of the cross-section is expected
to lead to substantial differences in the two bending moduli.
The 4HB-MT and 6HB-MT origami are also slightly different from the rest in that they have
wider blocks at each end of the origami (Fig.\ S1) to facilitate attachment to the beads 
in the magnetic tweezer experiments that were performed on them.

Snapshots of example origami and SST nanotubes are given in Fig.\ \ref{fig:snapshots}, with 
snapshots of some of the remaining systems given in Figs.\ S1--S3.
These illustrate the typical scale of the bending fluctuations in these systems. 
The differences in the nature of the junctions for these two types of system should also 
be evident from the close ups.

\subsection{Simulations}
\label{sect:sims}

We use molecular dynamics to generate a large ensemble of thermalized
configurations of the DNA nanotubes that can then be analysed using the methods
described in the subsequent sections to extract persistence lengths and elastic
moduli. 
Our molecular dynamics approach uses an Andersen-like 
thermostat,\cite{Russo09} both to set the temperature and to generate 
diffusive motion as is appropriate for nucleotides in solution.
Simulations were run at a temperature of 20 or 23$^\circ$C. 
To generate an equilibrium ensemble that samples the large length-scale 
bending fluctuations well requires very long simulations (the total length of
the simulations for each system are between 3 and 18\,ms (Table S1)).
We monitor the sampling by following 
the distance between points near the opposite ends of 
the nanotubes as a function of time (see Fig.\ S4 for examples). 
The fastest equilibration occurs for the SST nanotubes. The reason for the 
somewhat slower dynamics for the origamis is not obvious. The main structural difference
in the origamis is that the junctions involve mainly double rather than single crossovers, 
and so there may be some coupling of bending to different internal geometric states
of these junctions that leads to the slower dynamics of the bending fluctuations. 
The slowest dynamics are for the 4HB-MT and 6HB-MT origamis where
the blocks at the end of the origami perhaps slow down the diffusive dynamics of the long 
length-scale fluctuations. Consequently, these two systems have the worst statistics, which
in turn impacts the precision of the computed elastic properties.
Further details of the simulations are given in the Supporting Information (Section S2).

\subsection{Elastic model}

The conformation of a stiff polymer 
can be described by an orthonormal reference frame
$\{\widehat{\mathbf{e}}_1(s),\widehat{\mathbf{e}}_2(s),\widehat{\mathbf{e}}_3(s)\}$
as a function of the arc length $s$ along the molecular contour, where
$\widehat{\mathbf{e}}_3(s)$ is tangential to the contour.  For a DNA duplex or
nanotube it is natural to discretize a configuration into steps along the contour. 
For the duplex a natural step is between consecutive base pairs. 
Similarly, the nanotubes considered here can be divided
up into well-defined ``slices'' transverse to the contour, where each helix of the bundle 
has a base pair in each slice, and a step corresponds to moving one base pair along each helix.
The junctions that interconnect the helices maintain the registry between the
base pairs in each slice. Note that, unlike for the duplex, the slices are not all strictly 
equivalent because of their different positions relative to the junctions.

A rotation vector $\mathbf{\Theta}^{(n)}$ can be defined that represents the rotation of the
triad
$\{\widehat{\mathbf{e}}_1(n),\widehat{\mathbf{e}}_2(n),\widehat{\mathbf{e}}_3(n)\}$
onto the adjacent triad
$\{\widehat{\mathbf{e}}_1(n+1),\widehat{\mathbf{e}}_2(n+1),\widehat{\mathbf{e}}_3(n+1)\}$.\cite{Skoruppa17}.
The components $\Theta_1$ and $\Theta_2$ represent local bending, whereas $\Theta_3$
is a twist angle about the system axis. If the system is on average straight
then one would expect $\langle \Theta_1 \rangle=\langle \Theta_2\rangle=0$. 
If we define $\Omega_i=(\Theta_i-\langle\Theta_i\rangle)/a$,
where $a$ is the average separation of successive steps, then
the harmonic form of the energy due to angular deformations is 
\begin{equation}
\beta E=\frac{a}{2} \sum^{N}_{n=1} \sum_{\mu,\nu=1}^3 \Omega_\mu^{(n)} M_{\mu\nu} \Omega_\nu^{(n)}.
\label{eq:harmonic_disc}
\end{equation}
These are the dominant terms at small angular deviations, but higher order terms may be relevant for 
larger deformations.

If there are no couplings between the different angular degrees of freedom, then 
\begin{equation}
\mathbf{M}=\left(
           \begin{array}{ccc}
             A_1 & 0 & 0 \\
             0 & A_2 & 0 \\
             0 & 0 & C
           \end{array}
           \right),
\end{equation}
as would be the case if $\Omega_1$ and $-\Omega_1$,
and $\Omega_2$ and $-\Omega_2$ were equivalent by symmetry.
However, in the general case
\begin{equation}
\mathbf{M}=\left(
           \begin{array}{ccc}
             A_1 & A_{12} & G_1 \\
             A_{12} & A_2 & G_2 \\
             G_1 & G_2 & C
           \end{array}
           \right).
\end{equation}
For duplex DNA, if $\Omega_2$ describes bending deformations into the grooves of the helix, there
is a symmetry with respect to the bending deformation $\Omega_1$ 
that implies $A_{12}=G_1=0$, 
and $G_2$, the twist-bend coupling constant, is the only non-zero off-diagonal element of $\mathbf{M}$.\cite{Marko94,Nomidis17}

For all the origamis we consider, the designs are based either on a hexagonal lattice (6HB) or square lattice (4HB). 
For a perfectly regular pattern of junctions and staple ends the highest possible $C_n$ axis for these two lattices would be $C_3$ or $C_2$, respectively. 
However, if the helix bundles were infinite in length, and ignoring strand ends (the scaffold runs antiparallel in adjacent helices), 
the tubes could also possess six-fold and four-fold screw axes, respectively.
Therefore, the 4HB and 6HB systems have approximate four-fold and six-fold symmetry.
The presence of 2-fold axial symmetry implies that $A_{12}=G_1=G_2=0$, and 
3-fold or higher rotational symmetry also implies that $A_1=A_2$. 
That the elastic constants do not fully obey the above is likely due to the 
breaking of the symmetry due to the irregular placement of junctions and strand ends. 

By contrast, adjacent helices in the SST nanotubes are not equivalent.  The
4HB-, 6HB- and 8HB-SST nanotubes thus have 2-fold, 3-fold and 4-fold symmetry,
respectively.  (Note that, unlike the origamis, these systems have a perfectly
regular pattern of junctions.) Furthermore, because the spacing between
junctions alternates between 10 and 11 base pairs, the inter-helix angles also
alternate,\cite{Yin08} and so the cross-sections of the tubes are not expected
to be on average regular polygons. Therefore, $A_1$ and $A_2$ are expected to
be different for the 4HB system, but identical for the larger tubes, whereas
one would expect $A_{12}=G_1=G_2=0$ for all SST systems.

\subsection{Persistence lengths and elastic moduli}
\label{sect:lb}

If a system behaves like an ideal worm-like chain, one expects
\begin{equation}
\langle \widehat{\mathbf{e}}_3(n) \cdot \widehat{\mathbf{e}}_3(n+m) \rangle = \langle \cos\theta(m) \rangle=e^{-m a/l_b},
\label{eq:tangent_corr}
\end{equation}
where $l_b$ is the bending persistence length 
and the averaging is performed over both different origins $n$ and different configurations.
Rearranging allows an $m$-dependent persistence length to be defined:
\begin{equation}
l_b(m)=-\frac{ma}{\log \langle \cos\theta(m) \rangle}.
\label{eq:lb}
\end{equation}
Note that, as the number of possible origins is fewer for larger $m$, and the longest length scale 
fluctuations will be least well sampled in the simulations, the statistical errors are expected 
to increase with $m$.
Similarly, one can define a twist persistence length in terms of the decay of the 
cumulative twist angle deviation
\begin{equation}
l_t(m)=-\frac{ma}{\log \langle \cos \left(\sum_{k=n}^{n+m-1} a \Omega_3^{(k)}\right) \rangle}.
\label{eq:lt}
\end{equation}

To compute the elastic constant matrix $\mathbf{M}$ we use the same approach as was used for
DNA duplexes in Ref.\ \citenum{Skoruppa17}.
Namely, we first calculate the correlation matrix $\mathbf{\Xi}$ which is defined below
\begin{equation}
\Xi_{\mu\nu}(m)=\left\langle
	\left[\sum_{k=n}^{n+m-1}\Omega_\mu^{(k)}\right]
	\left[\sum_{l=n}^{n+m-1}\Omega_\nu^{(l)}\right]\right\rangle.
\label{eq:Xi}
\end{equation}
Inversion of $\mathbf{\Xi}$ leads to the elastic constant matrix:
\begin{equation}
\mathbf{M}(m)=\frac{m}{a} \left[\mathbf{\Xi}(m)\right]^{-1}.
\label{eq:MasInv}
\end{equation}
The use of cumulative angular deformations in Eq.\ \ref{eq:Xi} is discussed further in 
Ref.\ \citenum{Skoruppa17b}.

Note that the local form of the elastic energy in Eq.\ \ref{eq:harmonic_disc}
would imply that the persistence lengths and elastic moduli are independent of $m$. In practice,
however, there are longer-range effective couplings that lead to a variation with $m$, 
where convergence to a constant value only emerges at larger $m$, 
as has previously been observed for the DNA duplex.\cite{Skoruppa17} 
Here, we are most interested in this asymptotic behaviour, as it describes the 
long length-scale bending and twisting fluctuations.

Another implication of Eq.\ \ref{eq:harmonic_disc} 
is that the probability distributions for $\Omega_i$ would be expected to
be Gaussian. These distributions are illustrated in Figs.\ S5(a)-S8(a) for the
4HB-MT, 6HB-MT and 6HB-SST systems. 
For the 6HB-SST nanotube, 
the $\Omega_i$ distributions show some deviations from Gaussianity 
with large angular deformations being somewhat more likely than would be expected.
By contrast, the two origami have much more significant deviations 
in the tails of these distributions, with the 6HB-MT origami showing clear sub-peaks
in the distributions for the bending deformations. It might be that these more extreme local 
fluctuations are related to the slower dynamics of the origamis which were previously noted in 
Section \ref{sect:sims}. Figs.\ S5-S8 also show the distributions for the cumalative angular
deformations used in Eq.\ \ref{eq:Xi} for different values of $m$. As expected,\cite{Wiggins06} 
these distributions converge to a Gaussian form as $m$ increases, with this convergence 
occuring more rapidly for the 6HB-SST nanotube. 
Thus, although there are deviations from worm-like chain behaviour at the level of a single
step, on longer renormalised length scales worm-like chain statistics apply.

When applying the above formulae to a DNA duplex or the DNA nanotubes, we choose to
not include a certain number of base pairs or tube slices at either end, as the
ends will typically be somewhat more flexible. In the nanotubes, this general effect is 
exacerbated by the significant structural differences at the ends. The helices
splay out much more at the ends due to the lack of constraining junctions, as is
evident from the snapshots in Fig.\ \ref{fig:snapshots}. 
The number of tube slices at each end that are ignored is given in Table S1 for each
system.

\subsection{Triad definitions}
In Ref.\ \citenum{Skoruppa17} a number of different triads were considered when analysing the
elastic properties of the DNA duplex. Here, we recap the definitions for their ``Triad III'', 
as this approach will be most naturally adapted to the DNA nanotubes.
In this triad, the positions of the centres of mass of the base pairs are used to define
the vector $\widehat{\mathbf{e}}_{3}$. Namely, 
\begin{equation}
\widehat{\mathbf{e}}_{3}(i)= \frac{\mathbf{R}_{\mathrm{duplex}}(i+1)-\mathbf{R}_{\mathrm{duplex}}(i-1)}
{\left|\left|\mathbf{R}_{\mathrm{duplex}}(i+1)-\mathbf{R}_{\mathrm{duplex}}(i-1)\right|\right|}
\label{eq:e3_TIII}
\end{equation}
where $\mathbf{R}_{\mathrm{duplex}}$ is given by: 
\begin{equation}
\mathbf{R}_{\mathrm{duplex}}= \frac{\mathbf{r}_{\mathrm{nuc1}}+\mathbf{r}_{\mathrm{nuc2}}}{2}.
\label{eq:rduplex}
\end{equation}
The vector between the centres of mass of the nucleotides in the base pair is used to define
$\widehat{\mathbf{y}}$ 
\begin{equation}
\widehat{\mathbf{y}}= \frac{\mathbf{r}_{\mathrm{nuc1}}-\mathbf{r}_{\mathrm{nuc2}}}
            {\left|\left| \mathbf{r}_{\mathrm{nuc1}}-\mathbf{r}_{\mathrm{nuc2}} \right|\right|}
\label{eq:y_TIII}
\end{equation}
whose orthonormalization with respect to $\widehat{\mathbf{e}}_{3}$ leads to 
$\widehat{\mathbf{e}}_{2}$: 
\begin{equation}
\widehat{\mathbf{e}}_2=\frac{\widehat{\mathbf{y}}-\left(\widehat{\mathbf{y}} \cdot \widehat{\mathbf{e}}_3\right) \widehat{\mathbf{e}}_3}
                       {\left|\left| \widehat{\mathbf{y}}-\left(\widehat{\mathbf{y}} \cdot \widehat{\mathbf{e}}_3\right) \widehat{\mathbf{e}}_3 \right|\right|}
\label{eq:e2}
\end{equation}
Finally, $\widehat{\mathbf{e}}_{1}$ follows from the cross product: 
\begin{equation}
\widehat{\mathbf{e}}_1=\widehat{\mathbf{e}}_2 \times \widehat{\mathbf{e}}_3.
\label{eq:e1}
\end{equation}

\begin{figure}[t]
\centering
\includegraphics[width=5.5in]{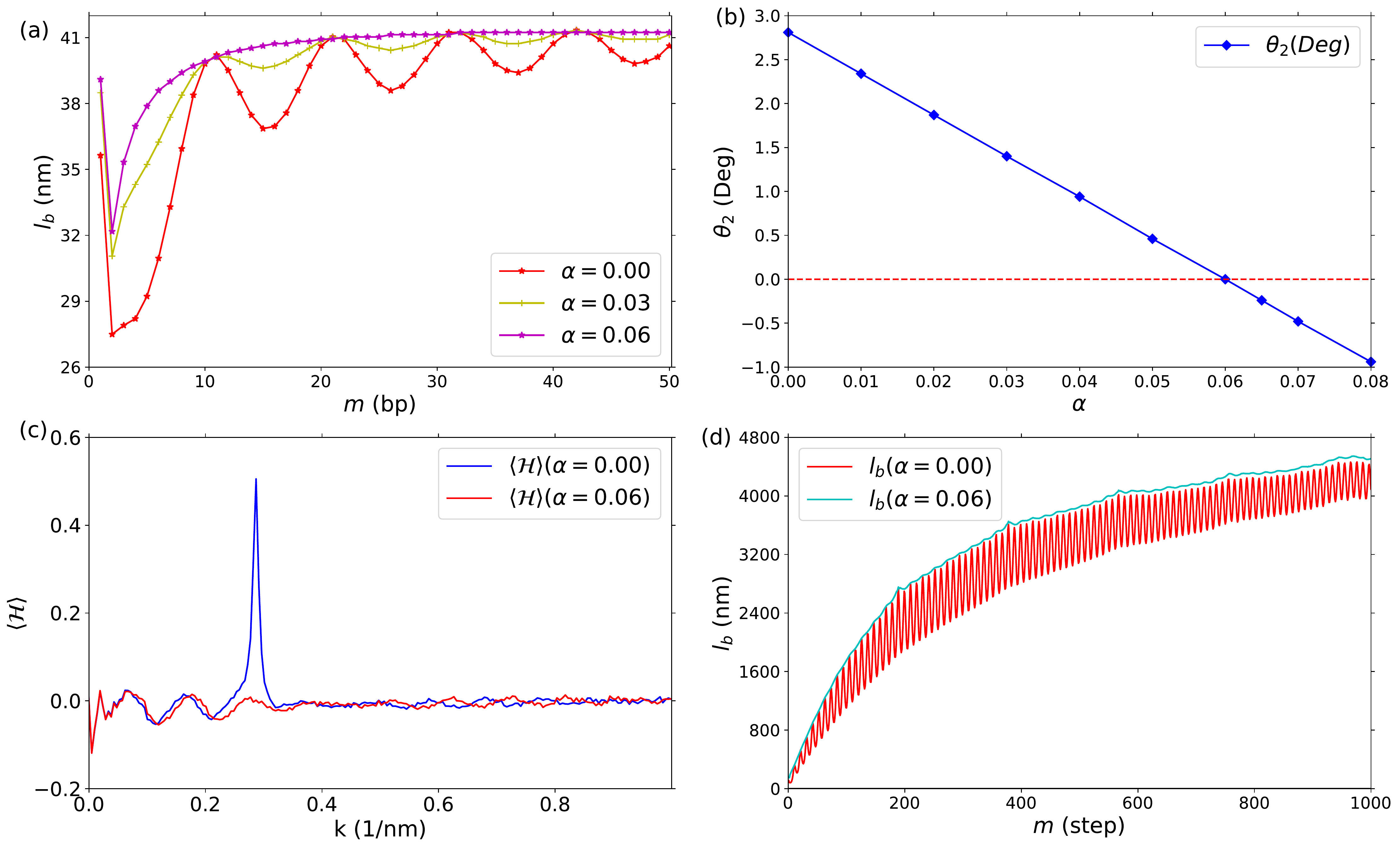}
\caption{$l_b(m)$ for (a) dsDNA and (d) the 6HB-S origami at different 
values of $\alpha$, as labelled.
(b) $\langle \Theta_2\rangle$ for dsDNA as a function of $\alpha$.
(c) Helicity measure for dsDNA as a function of inverse wavelength at $\alpha=0$ and 0.06.
}
\label{fig:alpha}
\end{figure}

The results for $l_b(m)$ using this triad show an oscillation on the length
scale of the pitch length of DNA that dies away at large $m$ (Fig.\
\ref{fig:alpha}(a)).  Also, $\langle \Theta_2 \rangle$ was found to have a
non-zero value (Fig.\ \ref{fig:alpha}(b)).\cite{Skoruppa17} The reason is that 
$\mathbf{R}_{\mathrm{duplex}}$
does not exactly lie at the centre of the DNA double helix, but is
slightly displaced towards the minor groove. Thus, the line of base-pair
centres will be helical, and the correlations in $\widehat{\mathbf{e}}_{3}$
will be strongest for an integer multiple of the pitch length. Similarly, the
helicity measure introduced in Ref.\ \citenum{Tortora20} when applied to the
contour defined by $\mathbf{R}_{\mathrm{duplex}}$ shows a sharp peak at the
reciprocal of the pitch length (Fig.\ \ref{fig:alpha}(c)). 
(Note, the helicity measure has magnitude one for a perfect helix, and positive/negative values correspond to right-handed/left-handed
helicity, respectively.)

Although the above oscillation is not problematic for computing $l_b$ for duplex DNA as
the oscillation dies away and the bulk limit is reached on reasonable length
scales, when performing a similar analysis for the DNA nanotubes, the
issue is much more troublesome. 
If we define, the centre of the DNA nanotube as
the average over the centres of mass of the duplexes in a given slice:
\begin{equation}
\mathbf{R}_\mathrm{tube}=\frac{1}{N_\mathrm{helix}} \sum_{j=1}^{N_\mathrm{helix}} \mathbf{R}
_{\mathrm{duplex},j}
\end{equation}
and $\widehat{\mathbf{e}}_{3}(i)$ in an analogous manner to the duplex  
\begin{equation}
\widehat{\mathbf{e}}_{3}(i)= \widehat{\mathbf{z}}(i)= \frac{\mathbf{R}_{\mathrm{tube}}(i+1)-\mathbf{R}_{\mathrm{tube}}(i-1)}
{\left|\left|\mathbf{R}_{\mathrm{tube}}(i+1)-\mathbf{R}_{\mathrm{tube}}(i-1)\right|\right|}
\label{eq:e3_tube}
\end{equation}
the persistence length can be calculated using Eq.\ \ref{eq:lb}, as illustrated
for one of the origami six-helix bundles in Fig.\ \ref{fig:alpha}(d). The
magnitude of the oscillation on the pitch length is much stronger and persists
over the whole range of $m$ that can be sampled.  This is for two reasons.
Firstly, the helices in the nanotubes are held in registry by the four-way
junctions present in the origamis, and so the helical paths of the 
centrelines of the individual helices in the nanotubes add up coherently.
Secondly, the helix bundles are much stiffer than a DNA duplex and so the
variations in the tangent-tangent correlation function due to the helicity of the centerline
both make a much more significant contribution and do not significantly die
away because the persistence length of the tubes is much longer than the
contour length of the nanotubes. Unlike for duplex DNA, a similar oscillation
is also seen in the bending elastic constants calculated using Eq.\
\ref{eq:MasInv} (Fig.\ S8).  Interestingly, for the SST nanotubes no
pitch-length oscillation is observed, presumably because 
the single-crossover junctions place fewer constraints on the relative orientations of 
adjacent helices, leading to a loss of coherence between the twist of the 
individual helices (Fig.\ S8).  

In Ref.\ \citenum{Nomidis19}, an approach to correct for the helicity of the
duplex contour was introduced that involved transforming from the helical frame
of reference to a non-helical one; this transformation led to a small correction to
the elastic moduli.
Here, we explore an alternative approach that will be more convenient for the DNA nanotubes 
and instead uses a new definition of the duplex centre 
$\mathbf{R}_{\mathrm{duplex}}$ 
that attempts to remove the helicity by adding a small offset in the direction of the major groove.
Namely, 
\begin{equation}
\mathbf{R}_{\mathrm{duplex}}(\alpha)= \frac{\mathbf{r}_{\mathrm{nuc1}}+\mathbf{r}_{\mathrm{nuc2}}}{2}+
		    \frac{\alpha}{2}\left({\widehat{\mathbf{b}}_{\mathrm{nuc1}} \times \widehat{\mathbf{n}}_{\mathrm{nuc1}} + \widehat{\mathbf{b}}_{\mathrm{nuc2}} \times \widehat{\mathbf{n}}_{\mathrm{nuc2}}}\right)
\label{eq:alpha}
\end{equation}
where for $\alpha=0$ the original definition of Eq.\ \ref{eq:rduplex} is
recovered.

Using this new definition of 
$\mathbf{R}_{\mathrm{duplex}}(\alpha)$ in Eq.\
\ref{eq:e3_TIII}, the oscillation in $l_b(m)$ is reduced as $\alpha$ is
increased until the oscillation disappears at $\alpha=0.06$ (Fig.\ \ref{fig:alpha}(a)). 
Similarly, if we examine $\langle \Theta_2(\alpha)\rangle$,
it passes through zero at $\alpha=0.06$ (Fig.\ \ref{fig:alpha}(b)). Furthermore,
the peak in the helicity measure at the reciprocal of the pitch length disappears for
$\alpha=0.06$ (Fig.\ \ref{fig:alpha}(c)).
Although the value of $\alpha$ does not affect the limiting value of the bending persistence 
length for the duplex, it does lead to small changes in the elastic constants, as was also
found when switching to a non-helical frame of reference.\cite{Nomidis19}
The variation of the duplex elastic constants with $\alpha$ is shown in 
Fig.\ S9.

Applying the new definition of the duplex centre with $\alpha=0.06$ 
to the origami nanotubes leads to the removal of the
oscillations in $l_b(m)$ (Fig.\ \ref{fig:alpha}(d)). 
Note that the $\alpha=0.06$ line for $l_b(m)$ is an upper envelope function to the curves
for other values of $\alpha$, because the helicity of the centerline leads to a periodic 
loss in correlations that is not representative of the true stiffness of the system.
For all further results we will use $\alpha=0.06$.

Now having a non-helical definition of the tube centre, we consider
our definitions of $\widehat{\mathbf{e}}_{1}$ and $\widehat{\mathbf{e}}_{2}$.
Following the recipe of Triad III for the duplex, 
we could define $\widehat{\mathbf{y}}$ for a six-helix bundle as
\begin{equation}
\widehat{\mathbf{y}}= \frac{\mathbf{R}_{\mathrm{duplex}}(2) + \mathbf{R}_{\mathrm{duplex}}(3) -\mathbf{R}_{\mathrm{duplex}}(5) - \mathbf{R}_{\mathrm{duplex}}(6)}
            {\left|\left| \mathbf{R}_{\mathrm{duplex}}(2) + \mathbf{R}_{\mathrm{duplex}}(3) -\mathbf{R}_{\mathrm{duplex}}(5) - \mathbf{R}_{\mathrm{duplex}}(6) \right|\right|}
\label{eq:y_tube}
\end{equation}
and then use Equations \ref{eq:e2} and \ref{eq:e1} to obtain 
$\widehat{\mathbf{e}}_{1}$ and 
$\widehat{\mathbf{e}}_{2}$.
However, when we did so for the 
6HB-SST system
we found that $A_1>A_2$ (Fig.\ S10(a)) in contrast to our expectations based on the symmetry of the 
system. 
We suspected this may be because of the ``inequitable'' treatment of the three directions in the 
orthogonalization scheme.

As a test we therefore defined
\begin{equation}
\widehat{\mathbf{x}}= \frac{\mathbf{R}_{\mathrm{duplex}}(4)-\mathbf{R}_{\mathrm{duplex}}(1)}
            {\left|\left| \mathbf{R}_{\mathrm{duplex}}(4)-\mathbf{R}_{\mathrm{duplex}}(1) \right|\right|},
\label{eq:x_tube}
\end{equation}
obtained $\widehat{\mathbf{e}}_{1}$ by orthogonalization of
$\widehat{\mathbf{x}}$ to $\widehat{\mathbf{e}}_{3}$, and
$\widehat{\mathbf{e}}_{2}$ by the vector product of $\widehat{\mathbf{e}}_{3}$
and $\widehat{\mathbf{e}}_{1}$.  With this scheme the order of the bending
moduli was reversed with $A_2>A_1$ (Fig.\ S10(b)), showing this not
to be a property of the tube but a reflection of the order in which we
performed the orthogonalization. 

\begin{figure}[t]
\centering
\includegraphics[width=3.3in]{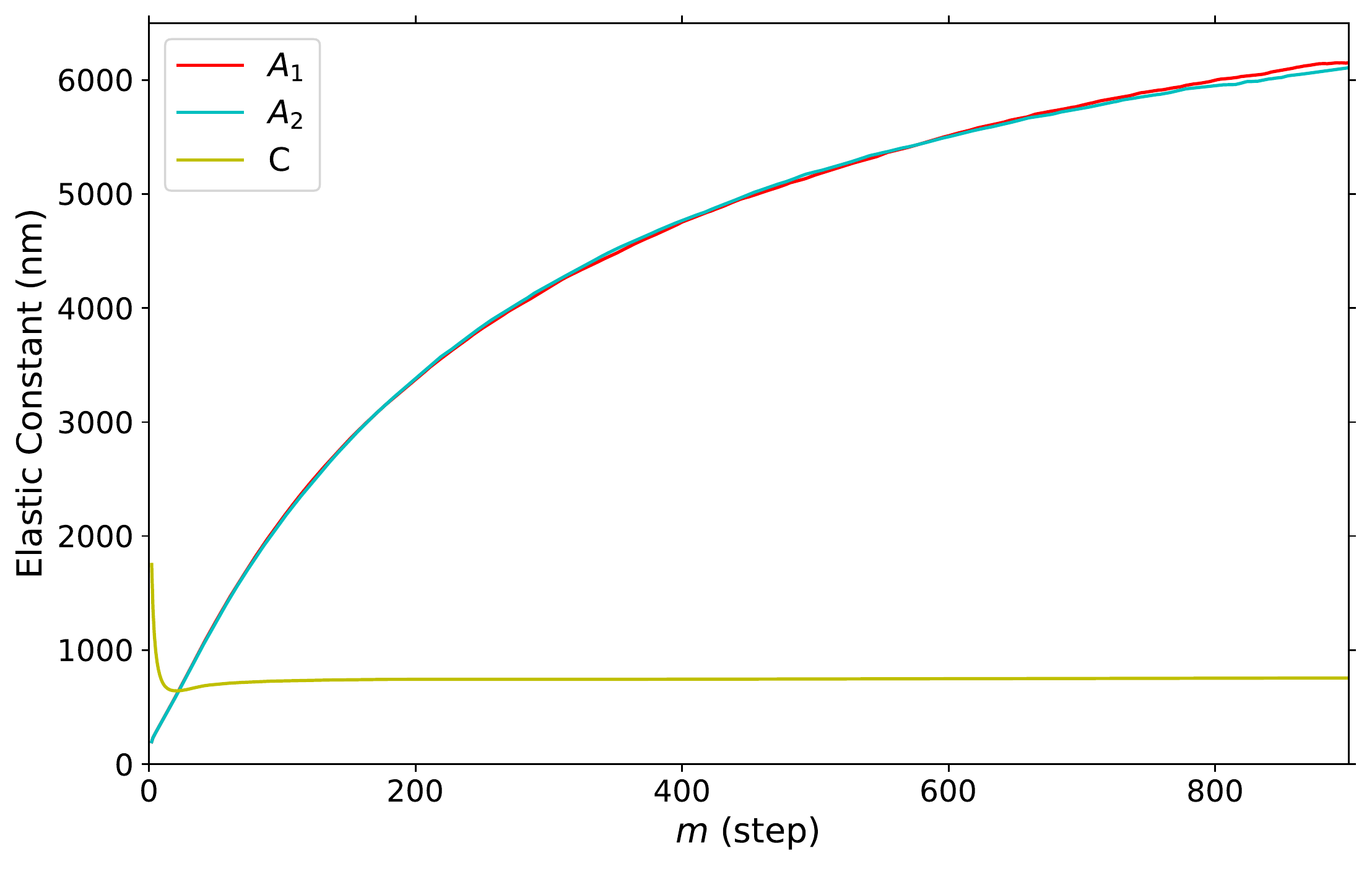}
\caption{Elastic moduli for 6HB-SST using
singular value decomposition.
$\alpha=0.06$.  }
\label{fig:svd_main}
\end{figure}

Therefore, we instead explored using singular value decomposition 
to find the orthogonal set 
$\{\widehat{\mathbf{e}}_{1},\widehat{\mathbf{e}}_{2},\widehat{\mathbf{e}}_{3}\}$
that are closest to the set of vectors 
$\{\widehat{\mathbf{x}},\widehat{\mathbf{y}},\widehat{\mathbf{z}}\}$.\cite{Strang06}
As one can see from Fig.\ \ref{fig:svd_main}, for the 6HB-SST system, 
within the statistical errors, there is now essentially no
difference between $A_1$ and $A_2$, in line with our expectations based on the
symmetry of these nanotubes. For the 8HB-SST, this is also true (Fig.\ S24(c)). 
Note that for the origamis, the non-regular pattern of junctions is one factor that can somewhat 
break this potential symmetry.
For all further results we use the singular value decomposition approach. 

The definitions of $\widehat{\mathbf{x}}$ and $\widehat{\mathbf{y}}$ for the other
systems are given in the Supporting Information (Section S3.2) and 
illustrated on the tube cross-sections in Fig.\ \ref{fig:basics}(d).
All the definitions that we have introduced are designed for DNA that adopts a
base-paired double-helical configuration. However, for structures as large as DNA origamis,
there will always be some designed base-pairs that are not intact due to
thermal fluctuations.  These broken base pairs are most likely to be at nicks
and junctions. When such ``fraying'' occurs, it is likely to lead to anomalous
values for the angular deviations $\Omega_i$. However, 
we do not attempt to treat such cases differently or exclude them from our averaging,
because both fraying is a natural 
part of the dynamics of the systems and the number of broken base pairs is likely to 
be a very small fraction of the total number of base pairs. 

\subsection{Extracting limiting values}
\label{sect:bulk}

The $m$-dependent definitions of the persistence lengths (Eqs.\ \ref{eq:lb} and
\ref{eq:lt}) and elastic moduli (Eq.\ \ref{eq:MasInv}) both for the duplexes
and the nanotubes show a similar behaviour rising up towards a limiting value
at large $m$ (Fig.\ \ref{fig:alpha}(d) and \ref{fig:svd_main}).  
However, unlike for the duplex, we cannot simply obtain the bulk
values of the bending persistence length and the bend elastic moduli from their
values at the largest $m$ value that we sample, because the bending persistence
lengths of the nanotubes are much larger than their lengths, and so have not yet 
converged to their limiting values. 

\begin{figure}[t]
\centering
\includegraphics[width=3.3in]{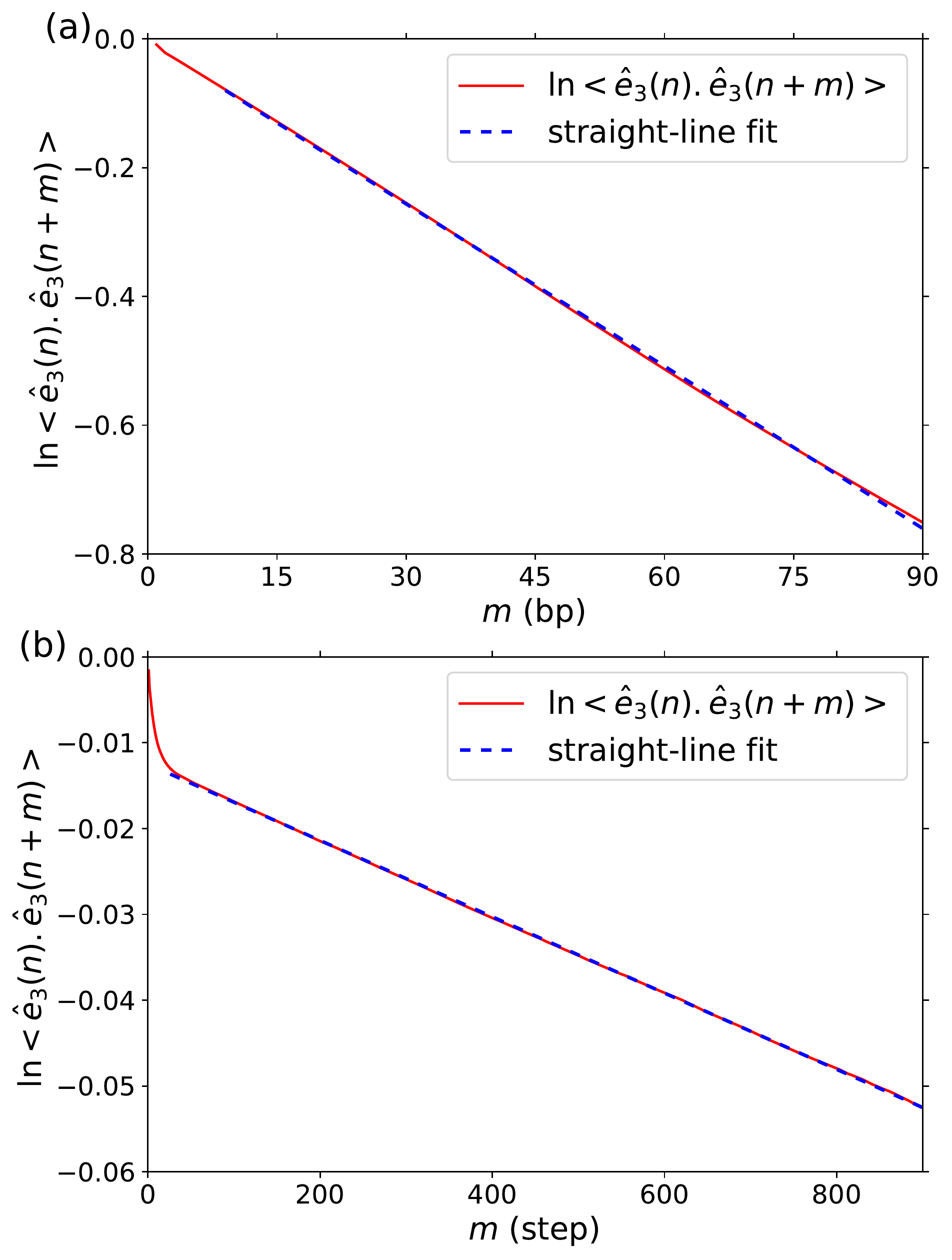}
\caption{$\log\langle \cos\theta(m)\rangle$ versus $m$ for (a) duplex
and (b) 6HB-SST along with fit lines to the exponential tail. 
}
\label{fig:lb_defs}
\end{figure}

The reason for this strong $m$-dependence is that the local fluctuations in the angles
$\Theta_1$ and $\Theta_2$ are greater than would be expected from the 
bending of the nanotube contour on long length scales. This effect can be seen in Fig.\
\ref{fig:lb_defs}, which depicts the tangent-tangent correlation function. The
behaviour is not perfectly exponential, as would be the case for an ideal
worm-like chain (Eq.\ \ref{eq:tangent_corr}), but shows an initial more rapid decay, which
for the duplex is only a few base pairs but for the example
nanotube is on the order of twenty-five slices. 
While for the duplex all base pairs are equivalent, all slices of a nanotube 
are not equivalent because of their different positions with respect to the junctions. 
This provides an additional source of heterogeneity that we average over in our analysis. 
For example, it is well known that the adjacent helices in an origami splay apart slightly 
as one moves away from the junctions\cite{Rothemund06,Baker18,Snodin19} (Fig.\ \ref{fig:snapshots}) 
leading to a local bending of the helices that is not representative of their 
longer-length-scale behaviour. 

Fig.\ \ref{fig:lb_defs} suggests a simple way to extract the bulk persistence length. 
As beyond the transient behaviour at small $m$ the correlation function shows a near
perfectly exponential tail, $l_b$ can be extracted from the limiting slope of the graph.
The same approach is used to calculate $l_t$ (Fig.\ S11). 

Extracting bulk bending elastic moduli is more difficult.
That the fluctuations of $\Theta_1$ and $\Theta_2$ are not fully representative of the 
bending behaviour at long length scales 
implies that there are correlations beyond a single step. 
These can be quantified by the following correlation coefficients:
\begin{equation}
\Lambda_{\mu\nu}^{(n)}=\left\langle \Omega_\mu^{(i)}\Omega_\nu^{(i+n)}\right\rangle.
\label{eq:lambda}
\end{equation}
In Ref.\ \citenum{Skoruppa17b}, it has been shown that the matrix $\mathbf{\Xi}(m)$ can 
be constructed from these correlation coefficients, as follows. 
\begin{equation}
\Xi_{\mu\nu}(m)=m \Lambda_{\mu\nu}^{(0)} + 
	 \sum_{n=1}^{m-1} (m-n)\left(\Lambda_{\mu\nu}^{(n)}+\Lambda_{\nu\mu}^{(n)}\right). 
	\label{eq:Xi_f(lambda)}
\end{equation}
In Fig.\ S12 we show that the elastic constants obtained from the above formula
reproduce well the elastic constants obtained from Eq.\ \ref{eq:Xi} for 6HB-SST.
However, when we use this approach to extrapolate to larger $m$, the behaviour
becomes less sensible, and so does not provide a method to estimate the bulk
elastic constants. The issue is that the weighting of the longer-range (i.e.\ larger $n$) 
correlation coefficients in Eq. \ref{eq:Xi_f(lambda)} becomes more
significant at large $m$, and although these coefficients might be expected to converge
to zero as $n$ increases, in practice the statistical noise in their values leads
to somewhat random behaviour in the predicted elastic constants when extrapolating beyond
the length of the nanotubes.

\begin{figure}[t]
\centering
\includegraphics[width=3.3in]{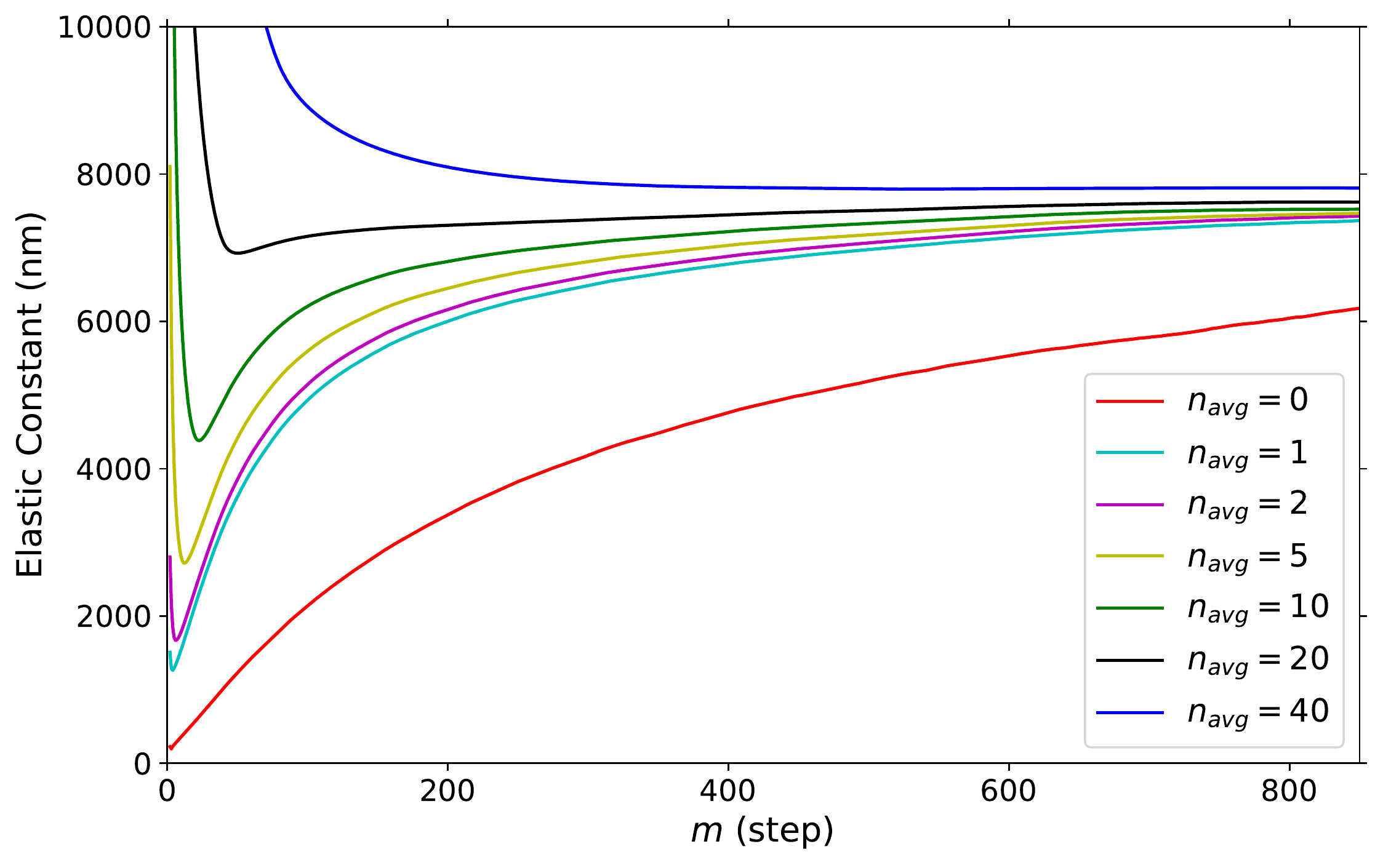}
\caption{The effect of averaging on $A_1$ for different $n_\mathrm{avg}$ for the 6HB-SST nanotube.}
\label{fig:navg}
\end{figure}

Therefore, we have explored an alternative approach in which we perform
averaging over longer length scales to reduce the enhanced local fluctuations that are at 
the root of the issue, and hence to expedite convergence of the elastic moduli to their bulk
limit. This idea is similar to using a renormalized length scale; as we noted in 
Section \ref{sect:lb} the cumulative angular deformations exhibit behaviour increasingly 
consistent with worm-like chain statistics as $m$ increases.
We do this averaging by defining our local axes at each slice
to be an average of the current slice and $n_\mathrm{avg}$ slices on either
side, i.e.
\begin{equation}
	\widehat{\mathbf{x}}^\prime(i)=
	\frac{\sum_{j=i-n_\mathrm{avg}}^{j=i+n_\mathrm{avg}}\widehat{\mathbf{x}}(j)}
	{\left|\left|{\sum_{j=i-n_\mathrm{avg}}^{j=i+n_\mathrm{avg}}\widehat{\mathbf{x}}(j)}\right|\right|}
\end{equation}
and equivalently for $\widehat{\mathbf{y}}^\prime$ and
$\widehat{\mathbf{z}}^\prime$. As before we then use singular value decomposition
to find the $\widehat{\mathbf{e}}_i$ vectors from the triad
$\{\widehat{\mathbf{x}}^\prime,\widehat{\mathbf{y}}^\prime,\widehat{\mathbf{z}}^\prime\}$.
Fig.\ \ref{fig:navg} illustrates the behaviour of the elastic constant $A_1$ for the
6HB-SST system for different values of $n_\mathrm{avg}$. As $n_\mathrm{avg}$
increases, the stiffness at small $m$ increases quickly, as intended. For sufficiently
large $n_\mathrm{avg}$ the $m$-dependent elastic moduli even become larger than their
bulk values and convergence to the limiting value as $m$ increases occurs from
above. It is also clear from the figure that the curves for different
$n_\mathrm{avg}$ are converging to the same limiting value. Therefore, 
we equate the bulk elastic moduli with their value at large $m$ (typically at an $m$ value 
somewhat lower than the largest we sample because the latter are most affected by statistical
noise; the values used are given in Table S1) when using the value of $n_\mathrm{avg}$ for 
which convergence is most rapid. Figures equivalent to Fig.\ \ref{fig:navg} but for $A_2$ and 
$C$ are given in the Supporting Information (Fig.\ S13), along with the effects of this 
averaging on the persistence length measurements.
For example, for the 6HB-SST nanotube we use $n_\mathrm{avg}=20$ to obtain bulk $A_1$ and $A_2$
values, and $n_\mathrm{avg}=5$ to obtain $C$ (Table S1). 

\section{Results}
A range of structural and mechanical properties of the ten systems that we studied are given in 
Table \ref{table:all}.
Firstly, $\langle \Theta_1 \rangle$ and $\langle \Theta_2 \rangle$ for all
systems are zero within statistical error; i.e.\ all the tubes are
on average straight, as expected. The behaviour of $\langle \Theta_3 \rangle$ is more
interesting.  
We should first note that the current
oxDNA potential 
has been fine-tuned to reproduce the
experiments of Ref.\ \citenum{Dietz09} which suggested that origami based on
the hexagonal lattice and without any insertions or deletions should be
approximately untwisted. Consistent with this, the 6HB-S, 6HB-MT and 10 HB origami
are all fairly near to being untwisted, albeit with all showing a slight
right-handed twist; this twist is lowest for the 10HB reflecting its
significantly larger twist modulus.  

\begin{table*}[t]
\renewcommand{\arraystretch}{1.2}
\tiny
\begin{tabular}{ccccccccccccccccccc}
\hline
& Ref.\  & $n_\mathrm{helix}$ & $N_\mathrm{slice}$ & $a$ & $\langle \Theta_1 \rangle$ & $\langle \Theta_2 \rangle$ & $\langle \Theta_3 \rangle$  & $A_1$ & $A_2$ & $C$ & 
	 $l_b$  &  $l_t/2$ & $l_b^\mathrm{expt}$ & $C_\mathrm{expt}$ \\
\hline
4HB-MT & \citenum{Kauert11} & 4 & 1354 & 0.3373 & 0.006 & 0.010 & 0.035 & 2150 & 2020 & 330 & 1040 & 330 & 740 & 390 \\
4HB-SST & & 4 & 1029 & 0.3400 & -0.004 &  0.002 & 0.136 & 3120 & 1750 & 370 & 2030 & 280 & & \\
6HB-2$\times$LH & \citenum{Siavashpouri17} & 6 & 1259 & 0.3371 & 0.002 & -0.005 & -0.884 & 4480 & 5300 & 610  & 4500 & 620  \\
6HB-1$\times$LH & \citenum{Siavashpouri17} & 6 & 1261 & 0.3376 & 0.002 & -0.003 & -0.419 & 5520 & 5980 & 580 & 5740 & 680 \\
6HB-S & \citenum{Siavashpouri17} & 6 & 1260 & 0.3383 & -0.001 & -0.001 &  0.046 &  6400 & 5780 & 670 & 5430 & 660 & 2400 \\
6HB-1$\times$RH & \citenum{Siavashpouri17} & 6 & & 0.3457 & 0.004 & 0.000 & 0.492 & 5590 & 4860 & 650 & 5950 & 690 \\
6HB-MT & \citenum{Kauert11} & 6 & 1085 & 0.3385 & -0.013 & -0.011 &  0.071 & 6960 & 5560 & 800 & 4140 & 810 & 1880 & 530 \\
6HB-SST & \citenum{Schiffels13} & 6 & 1070 & 0.3389 & 0.000 &  0.000 & 0.115 & 7560 & 7630 & 750 & 7620 & 740 & 3300 & \\
8HB-SST & \citenum{Schiffels13} & 8 & 1050 & 0.3383 & 0.000 & 0.001 & 0.111  & 16\,980 & 16\,880 & 1290 & 16620 & 1270 & 8200 \\
10HB & \citenum{Siavashpouri17} & 10 & 705 & 0.3379 & -0.008 & 0.002 & 0.012 & 37\,580 & 13\,440 & 2230 & 19440 & 1780 \\
\hline
\end{tabular}
\caption{Structural and mechanical properties of all the systems studied. All angles are 
	given in degrees and elastic moduli and persistence lengths in nanometres. 
	The $n_\mathrm{avg}$ and $m_\mathrm{limit}$ values used to extract the bulk values of
	the elastic constants are given in Table S1. 
}
\label{table:all}
\end{table*}

The series of twisted six-helix bundles of Ref.\
\citenum{Siavashpouri17} show the expected linear variation of the twist (Fig.\ S14) 
across the series, albeit 
with the nanotube that is designed to be untwisted being slightly right-handed, as noted above. 
The designed twist in these structures is based on modelling with CanDo.\cite{Castro11,Kim12} 
Here, we find that the insertions and deletions (i.e.\ increasing or decreasing the number of
base pairs between these junctions to induce internal twist stress\cite{Dietz09}) in these 
structures have a larger angular effect with 30 sets of insertions or deletions causing a 
change in twist of just over one and a half turns rather than a single turn.

Interestingly, the single-stranded tiles systems all show significant
right-handed twist 
(equivalent to about 120$^\circ$ degrees per 1000 base pairs).  Like for the
hexagonal lattice origamis, the average junction spacing in the planar sheets
on which they are based matches the presumed pitch length of DNA (10.5 base
pairs). Therefore, they might be expected to show a similar twist as those
origamis; instead it is over twice as large. This is mostly because the twist
angle at a junction with a single-crossover is larger than for one with a double-crossover 
for the oxDNA model.

\begin{figure}[t]
\centering
\includegraphics[width=3.3in]{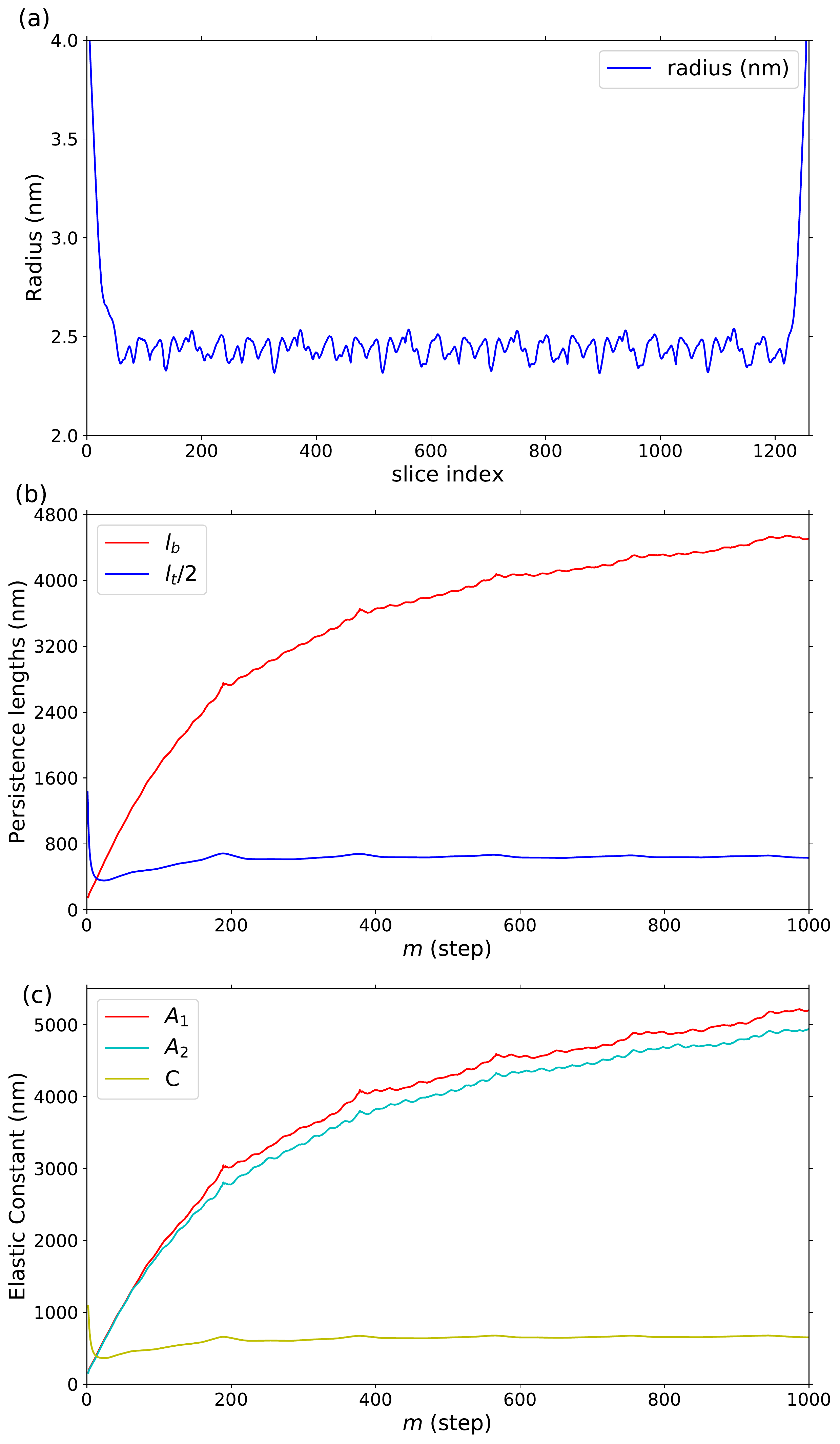}
\caption{(a) Radius as a function of position, 
(b) $m$-dependent persistence lengths and (c) elastic moduli for the 6HB-S origami.}
\label{fig:s6hb_all}
\end{figure}

Another basic structural property that we considered was the radius of the tubes defined
as the average distance of the centres of the DNA helices from the tube centre:
\begin{equation}
r=\frac{1}{N_\mathrm{helix}} \sum_{j=1}^{N_\mathrm{helix}}  \left| \mathbf{R}_\mathrm{duplex}(j)-\mathbf{R}_\mathrm{tube} \right|.
\end{equation}
Note, to get a measure of the limit of the excluded volume of the tube one
should add the radius of the DNA duplex. The dependence of the
radius on position along the tube is shown in Fig.\ \ref{fig:s6hb_all}(a) for
6HB-S; equivalent figures for all the other systems are depicted in Figs.\
S15--S17 of the Supporting Information. Typical of all systems, the tube has a
significantly larger radius at either end. This is simply a reflection of the
significant splaying out of the helices of the origami at the ends that is evident from the
snapshots in Fig.\ \ref{fig:snapshots} and is a result of these sections 
being less constrained by junctions. 
The radius measurements allows us to decide how many slices from each end to ignore when performing
the persistence length and elastic moduli calculations. 

The radii are also larger than one would expect from close-packing of the helices.
This is a result of the well-documented slight splaying out of helices between
junctions\cite{Rothemund06,Baker18,Snodin19}
that is also visible from the close-ups in Fig.\ \ref{fig:snapshots}.
Baker {\it et al.}\ estimated the inter-helix separation in a 10-helix bundle nanotube to 
be 2.75\,nm from SAXS measurements.\cite{Baker18}
This is fairly similar to the 2.4--2.5\,nm radius that we find for the six-helix bundles 
(Figs.\ \ref{fig:s6hb_all}(a), S15,S17).
The small variations in this radius along the length of the tube are also due to the variations
in the interhelix distance caused by this splaying.
In the example in Fig.\ \ref{fig:s6hb_all}(a), the relatively irregular
placement of these junctions means there is little clear pattern in this variation,
however for the origamis of Ref.\ \citenum{Kauert11} there is a much clearer repeating
pattern (Fig.\ S15). 

\begin{figure}[t]
\centering
\includegraphics[width=3.3in]{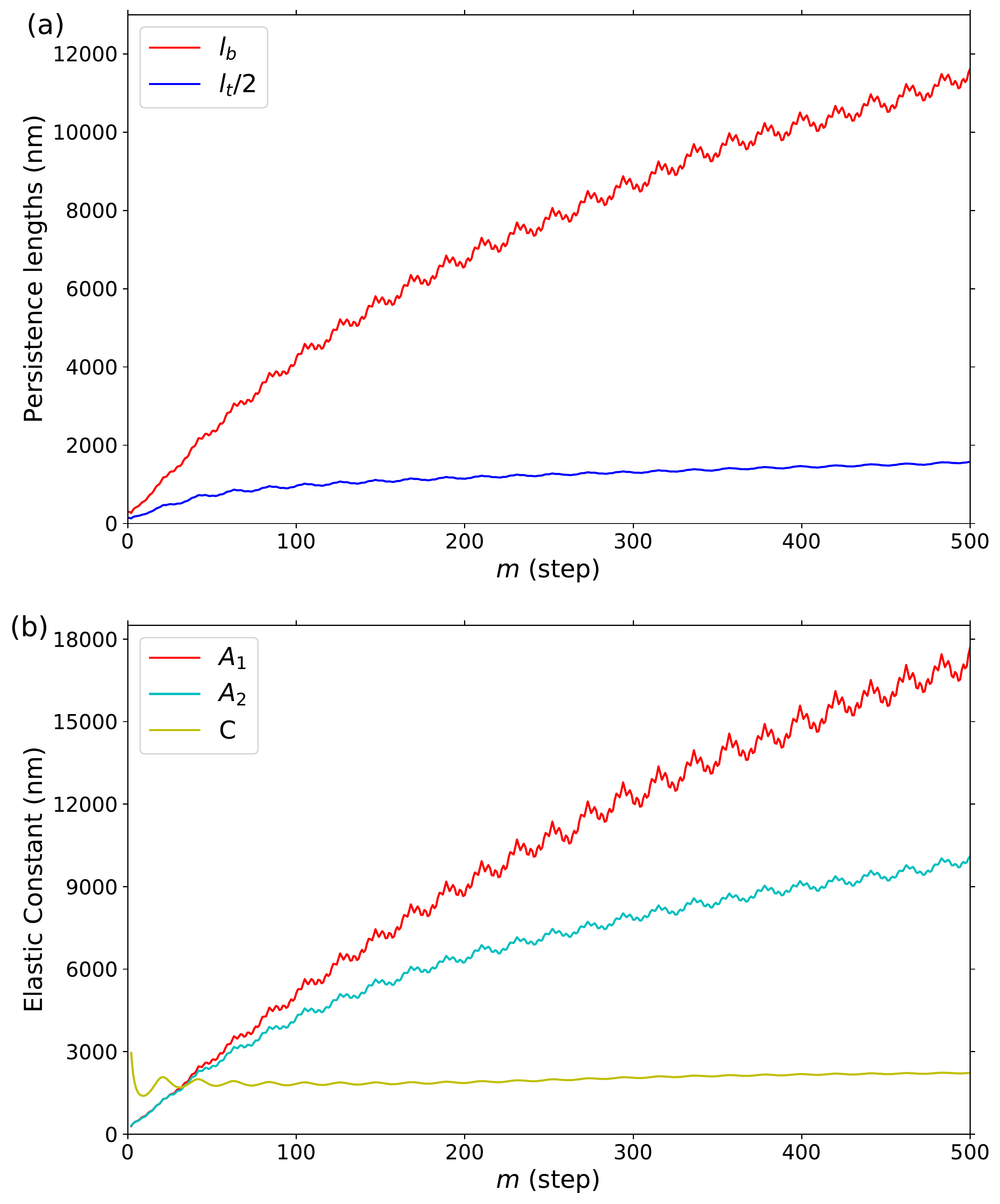}
\caption{(a) $m$-dependent persistence lengths and (b) elastic moduli for the 10HB origami.}
\label{fig:10hb}
\end{figure}

\begin{figure}[t]
\centering
\includegraphics[width=3.3in]{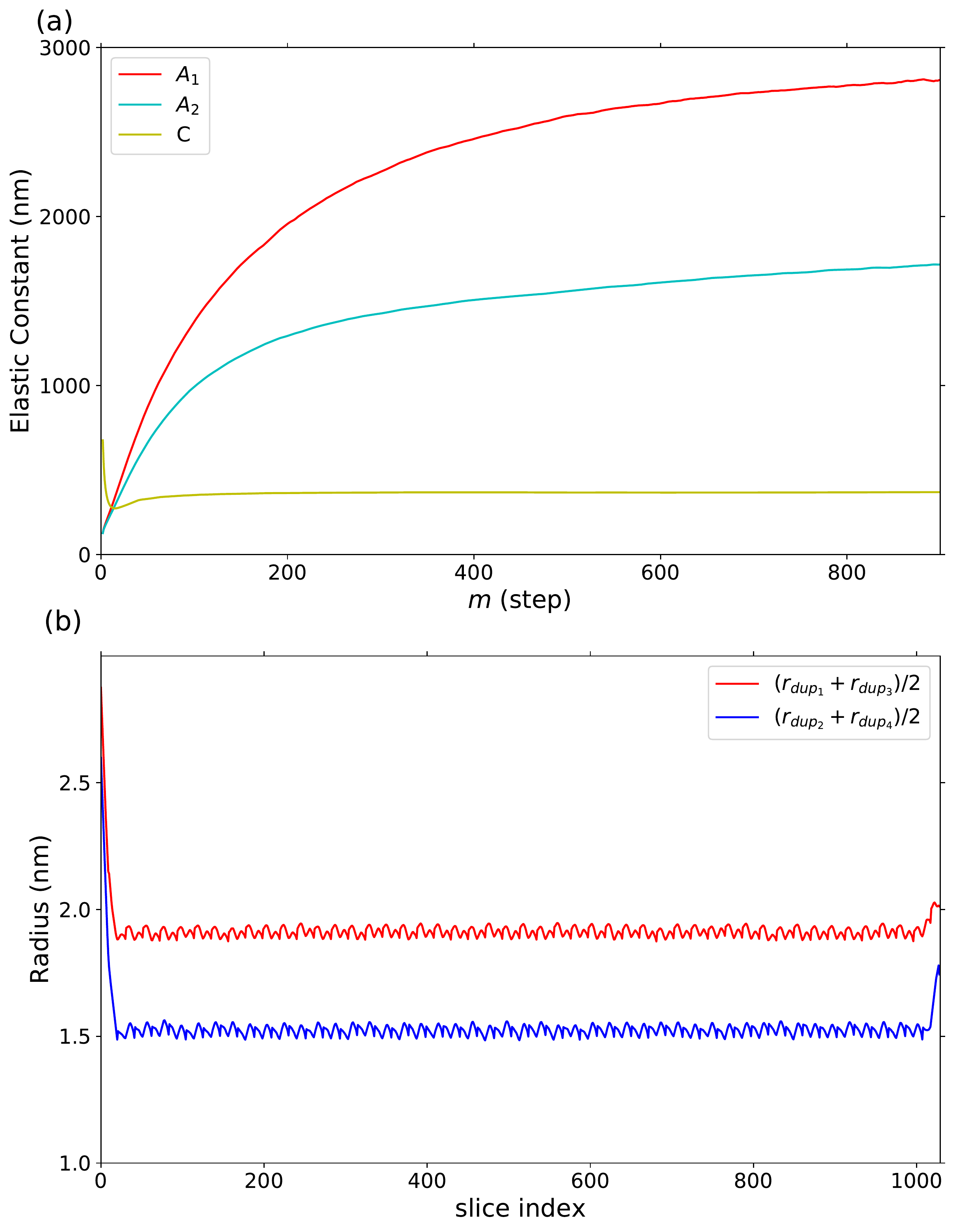}
\caption{(a) $m$-dependent elastic moduli and 
(b) radii for equivalent helices as a function of position for the 4HB-SST nanotube.}
\label{fig:4HB_deform}
\end{figure}

The $m$-dependent persistence lengths and elastic moduli are shown for 6HB-S,
10HB and 4HB-SST in Figs.\ \ref{fig:s6hb_all}, \ref{fig:10hb} and
\ref{fig:4HB_deform}, respectively; those for the other systems are shown in
Figs.\ S18--S21. The limiting values of these quantities
extrapolated by the approaches outlined in Section \ref{sect:bulk} are given in
Table \ref{table:all} for all systems. 
As noted already, the $m$-dependent bending persistence lengths show a slow
convergence to their limiting values due to the enhanced flexibility on short
length scales.  The twist persistence length, partly reflecting its smaller
value, converges much more rapidly.
All the origamis show some degree of repetitive fine structure in these curves.
For example, there is a small peak every about 190 slices for 6HB-S (Fig.\
\ref{fig:s6hb_all}(b)) and a stronger oscillation every 21 slices for 10HB
(Fig.\ \ref{fig:10hb}(a)).  The enhanced correlations at the peaks are
associated with length scales that match a periodicity or approximate periodicity
in the pattern of junctions; these periodicities are also evident in the repeating 
patterns seen in the tube radii.
By contrast, the virtual absence of any fine structure, and consequent
smoothness of all the data, for the SST nanotubes (e.g.\ Fig.\ \ref{fig:4HB_deform}(a))
is noteworthy, and reflects the much
smaller variations in the tube radii due to the lesser constraints associated with their 
single-crossover junctions (Fig.\ \ref{fig:4HB_deform}(b)). 

\begin{figure}[t]
\centering
\includegraphics[width=3.3in]{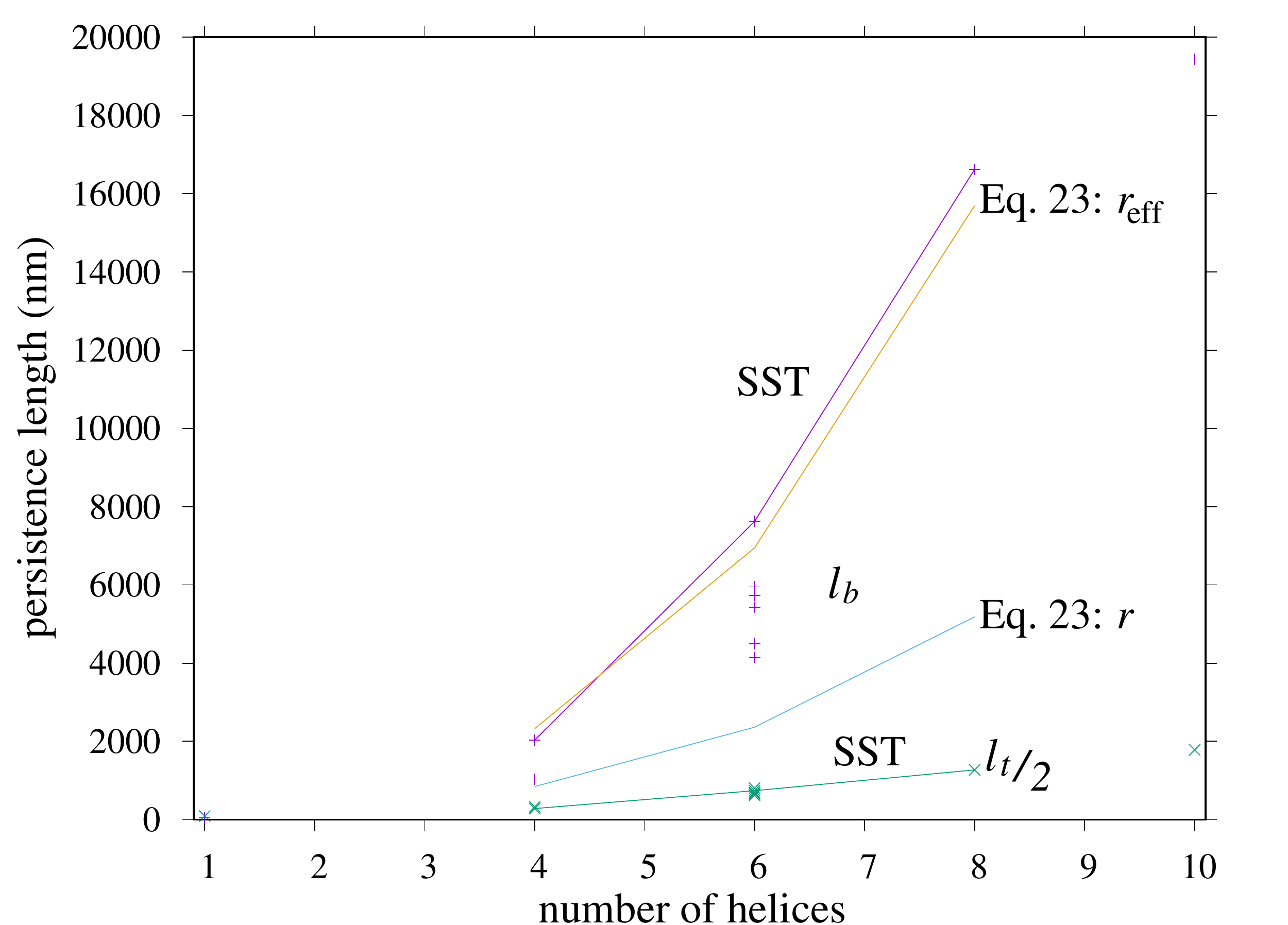}
\caption{The variation of the bend and twist persistence lengths
with $n_\mathrm{helix}$ for all systems including
dsDNA. Lines have been added for the SST series of nanotubes. There are also 
two additional lines (with no data points) that represent the predictions of Eq.\ 23 for
the SST nanotubes using both the actual radius of dsDNA or an 
effective radius that is chosen to fit the observed data.}
\label{fig:nhelix}
\end{figure}

In Fig.\ \ref{fig:nhelix} we plot the bulk persistence lengths for bending and twisting
versus $n_\mathrm{helix}$ for all systems, including dsDNA. 
There is a clear non-linear increase in the bending persistence length with 
the number of helices, whereas the twist persistence length increases roughly linearly.
For example, the bending persistence lengths of the six-helix bundles are 110--190 times
that of dsDNA, whereas the twist persistence length increases only by
a factor of 6--9. This basic behaviour was previously noted in Refs.\ 
\citenum{Kauert11} and \citenum{Schiffels13}. 
The variation of the bending persistence length can be described theoretically if one
assumes that dsDNA behaves as a uniform elastic cylindrical rod, 
leading to the expression:\cite{Wang12b}
\begin{equation}
l_b^\mathrm{tube}=n_\mathrm{helix} l_b^\mathrm{duplex}\left(1+2\left(\frac{R}{r}\right)^2\right).
\label{eq:lb_scaling}
\end{equation}
where $r$ is the radius of dsDNA and $R$ the radius of the
nanotube (see 
Section S4.2).
From this equation it can be seen that the non-linear variation is due to the
additional effect of the helices being a distance $R$ from the
centreline of the nanotube.  One difficulty when applying this to experimental systems is that
of measuring $R$.  However, all quantities in this expression can be measured
in simulations, allowing the consistency of this expression to be tested. The
predicted persistence lengths when applied to the series of three SST nanotubes
significantly underestimates the nanotube persistence lengths (Fig.\ \ref{fig:nhelix}).  
A significant factor in the over-prediction likely lies in the assumption that 
dsDNA can be considered as a uniform elastic rod. Given the structure of 
DNA---it has a tightly-packed core of stacked and paired bases and, due to the major and minor
grooves, a shape more akin to an asymmetric double-threaded screw than a
cylindrical rod---it is perhaps not surprising this may not be the best approximation.
If we consider $r$ to be a parameter, we can find the value that best fits the
SST nanotube data; this fitting gives an effective radius of 0.65\,nm, with the
resulting form capturing well the dependence on $n_\mathrm{helix}$. This
compares to 1.15\,nm used in the original prediction, which is a measure of the
limits of the excluded volume of an oxDNA double helix.

There is some variation of the persistence lengths for those helix bundles with
the same numbers of helices. For example, for the six-helix bundles $l_b$
varies between 4160 and 7650\,nm and $l_t/2$ between 620 and 810\,nm.
Although one in principle expects the persistence lengths to also depend somewhat on the
types of junction (single or double crossover), the patterns of nicks and
junctions, tube radius (Eq.\ \ref{eq:lb_scaling}) 
and the degree of internal stress, a significant part of this variation is probably
down to errors due to imperfect sampling; as mentioned earlier, the 4HB-MT and 6HB-MT 
examples suffer particularly in this regard with the tangent-tangent correlations
lacking the clear exponential tail that is expected for a worm-like chain. 
One systematic feature, which seems robust, is that the SST nanotubes have a 
larger bending persistence. This is perhaps surprising, as one might have
expected them to be more flexible given that they have single rather than
double crossovers at the junctions. The most likely reason is that the extra
freedom associated with the single crossover allows the nucleotides at the
junctions to better optimize their stacking interactions, causing the 
junctions to actually be stiffer. 

We can also compare our results to the available experimental data for these
systems, which are included where available in Table \ref{table:all}. Although
the persistence length of the 10HB origami design considered here has not been 
determined experimentally, a value of 5430\,nm was obtained for a very similar
10HB design.\cite{Lee19b}
In all cases the oxDNA values for the bending persistence length are larger. 
One clear factor is
that the simulations have been performed for origamis with the ideal
structures, whereas there is likely to be imperfections in the experimental
origamis due both to assembly errors, be it a result of missing staples or the
incorrect routing of the strands, and to errors in the synthesis of the
staples. Quantitative evidence has been obtained for the presence of such
assembly errors,\cite{Wagenbauer14,Myhrvold17,Strauss18} but estimates of their
prevalence have varied significantly. For example, Ref.\ \citenum{Strauss18} 
estimated that on average only 84\% of staples were incorporated into a 2D
origami, whereas Ref.\ \citenum{Wagenbauer14} obtained folding qualities 
of 96\% or greater for 3D origamis.
The resulting defects could well lead to both local static structural
irregularities and enhanced fluctuations, both of which would contribute to
lowering the measured persistence lengths.
From this viewpoint, the experimentally measured values should be seen as lower
bounds to the persistence lengths of ideal structures.

The substantial effect of defects on the mechanical properties has been confirmed
in experiments that deliberately introduce defects into origami structures. For
example, gaps between staple ends reduce bend persistence lengths\cite{Lee19b} 
and gaps and nicks reduce the effects of torsional stresses\cite{Kim19,Lee19} 
(i.e.\ they reduce the local twist modulus of the origami).
Similarly, missing staples have been used to achieve designs with higher local
mechanical compliance.\cite{Lee17,Turek18}
Radiation damage has also been found to make a 6HB origami more flexible.\cite{Fang20}

Another factor that we should mention is that in our calculations we choose to ignore the 
ends of the origami due to their enhanced fluctuations, which would otherwise introduce 
some degree of  non-worm-like chain behaviour and a lowering of the persistence lengths. 
However, these 
effects are relatively small. Also, our $m$-dependent definition of the bending persistence
length shows a significant length dependence. However, this is due to enhanced local fluctuations
at the shortest length scales, whereas the experimental measurements are unlikely to be 
sensitive to these effects, both due to the lower resolution of the molecular contour and 
the analysis methods used. Similarly, these length-dependent effects are dramatically
reduced when we use the averaging approach of Section \ref{sect:bulk} (Fig.\ S13(c))

The most accurate persistence length measurements are probably those of Ref.\
\citenum{Schiffels13} because the lengths of the SST nanotubes can be significantly
larger than an origami.  By visualizing the molecular contours of individual
tubes as a function of time they were able to measure the tangent-tangent
correlation of individual nanotubes over length scales of up to about 10\,$\mu$m.
One interesting feature of the results was that the variation of the
persistence lengths between different examples of the same design was greater
than the estimated error in the persistence length for an individual tube. So
for the 6HB, the persistence lengths varied between 1.6 and 4.7\,$\mu$m, and
between 5.7 and 10.1\,$\mu$m for the 8HB. This indicates that the population
of nanotubes has a heterogeneous set of mechanical properties.  In this case, there
is a possible isomerism (associated with whether the strands are in registry or
offset as one goes around the circumference of the tube) that may be partly
responsible for this heterogeneity, but it is also likely to reflect
differences in the assembly quality of individual nanotubes. For this reason, the
highest values may provide the best estimates of the persistence lengths of the
ideal nanotubes (the average values are reported in Table \ref{table:all}).

The difference between the experimental and oxDNA values could also reflect
deficiencies in the oxDNA model. One known shortcoming is that although oxDNA
reproduces well the bend and twist persistence lengths of dsDNA,
it considerably overestimates the stretch modulus. For the original oxDNA
model the extensional modulus is about 2100\,pN,\cite{Ouldridge11} but it has not previously been
calculated for the second version of the model used here (sometimes called
``oxDNA2''). Fitting the force-extension curve of dsDNA to the
extensible worm-like chain form (Fig.\ S22) gives an estimate of about 2700\,pN for oxDNA2.
This compares to experimental estimates of about 1200\,pN at high 
salt conditions.\cite{Wang97} The
bending of a helix bundle not only involves the bending of the individual
helices, but also leads the helices on the inside of the bend to be compressed,
and those on the outside to be in tension. Thus, that the oxDNA extensional modulus
is too large is likely to contribute to an overestimation of the bending persistence
lengths. Note, though, that the tube can also respond to the tension and compression by
changing its local structure, e.g.\ by changing the splay angle of the helices at junctions 
and hence the interhelix separation.

In Table 1, we report the limiting values of the elastic moduli for all systems. Note that we do
not give values for the coupling terms $A_{12}$, $G_1$ and $G_2$ as there is no strong evidence 
that non-zero values are more than statistical noise. For completeness, we plot their $m$-dependent
values in Figs.\ S23--S26, but it is noteworthy that rather than rising smoothly to their limiting 
values---as is the case for $A_1$, $A_2$ and $C$, as well as the twist-bend coupling constant for dsDNA\cite{Skoruppa17}---they are generally close to zero for small 
values of $m$, where the statistics are best, with any significant deviations from zero 
occurring at larger $m$. However, as we estimate the limiting values of $A_1$, $A_2$ and 
$C$ from the large-$m$ behaviour, the non-zero values of the off-diagonal terms are likely to 
somewhat affect the accuracy of these elastic moduli.

It was expected based on approximate symmetry that for all systems, except for 4HB-SST and 
the 10HB origami, $A_1$ and $A_2$ would be approximately equal. This is generally the case with
differences being at most 25\%. For the 6HB and 8HB SST nanotubes for which there is an exact symmetry due
to the regularity of their junctions, $A_1$ and $A_2$ are virtually exactly equal. 

The large difference in $A_1$ and $A_2$ for the 10HB is expected given its very anisotropic 
cross-section. 
The origin of this difference for the 4HB-SST system is more subtle. As a consequence of the 
alternating pattern of 10 and 11 base-pair spacings between junctions in the SST nanotubes, 
the interhelix angles alternate on going around the nanotube. This leads to the diamond, rather
than square, tube cross-section apparent in Fig.\ \ref{fig:snapshots}(a).
This effect can be quantified by measuring the distance of equivalent helices in the
SST nanotubes from the tube centre. These two distances are shown in Fig.\ 
\ref{fig:4HB_deform}(b) for the
4HB-SST and in Fig.\ S28 for the other SST tubes. The difference in the two distances is roughly
0.4--0.5\,nm for the 4HB-SST tube.

If $A_{12}=G_1=G_2=0$ then the bending persistence length should simply be the
harmonic mean of the bend elastic constants $A_1$ and $A_2$.  This generally
holds to within 10\%, the exceptions being the MT systems for which the
sampling issues affecting the values of $l_b$ have already been noted.
Similarly, if the coupling terms are zero then $l_t/2=C$; this generally holds
to within less than 10\%, the exceptions being 4HB-SST and 10HB, for which
the large differences between $A_1$ and $A_2$ make the obtained limiting value of 
$C$ more sensitive to the behaviour of the coupling terms. 
For the two experimental systems for which the twist modulus has been measured, the oxDNA
values are in reasonable agreement, being a bit smaller for the 4HB-MT origami and somewhat
larger for the 6HB-MT origami, which happens to have the largest $C$ of the 6HB systems that we
have studied.

\section{Conclusions}

In this paper we have introduced approaches to compute persistence lengths and elastic moduli associated
with bending and twisting for rod-like DNA nanostructures.
One feature is that only on length scales beyond those
associated with the spacings between the junctions 
do the bending fluctuations behave like those of a worm-like chain. As our discretization was 
performed at the level of single base-pair steps along the molecular contour, averaging methods 
had to be introduced to extract bulk values of the elastic moduli.
The approaches, although computationally intensive due to the long simulations required to 
adequately sample the long length-scale fluctuations, are straightforwardly 
applicable to other DNA nanostructures. Thus, they could be used as part of a rational design 
process to produce DNA nanostructures with tailored mechanical properties.

In line with previous experiments,\cite{Kauert11,Schiffels13,Lee19b} we found that 
the persistence lengths and elastic moduli associated with bending exhibit a 
strong non-linear dependence on the number of helices, 
whereas the twist persistence lengths increase approximately linearly. Consequently, the 
DNA nanotubes are in a very different mechanical regime with respect to angular 
deformations compared to duplex DNA. For example, for the six-helix bundles the persistence 
length is on average about eight times larger than the twist modulus, whereas for dsDNA the 
twist modulus is over twice as large.

In all cases, the values for the bending persistence lengths that we obtain are 
larger than those measured experimentally. The main reasons for these differences 
are firstly that our
calculations are performed for ideal structures, whereas the structures in experiments
will inevitably have some degree of defects due to both assembly and strand
synthesis. Secondly, one of the known deficiencies of the oxDNA model is that
it overestimates the stretch modulus of dsDNA, which in turn is likely to lead
to the cost of nanotube bending being overestimated as the individual helices
will be somewhat compressed on the inside of a bend and stretched on the
outside. Note that there are also uncertainties in the accuracy of oxDNA's description of the
mechanical properties of junctions and nicks, as there is relatively little 
experimental information to which to compare.
The relative contributions of these two main factors is not fully clear. 
However, it is interesting to note that oxDNA is able to reproduce well the 
bending and twisting of the example origamis in Ref.\ \citenum{Dietz09} 
that result from internal stresses induced by insertions and deletions, and 
that partly reflect the elastic moduli of the origami.\cite{Snodin15}

Here, we have characterized the elastic fluctuations of a set of DNA nanostructures. It would 
also be of interest to go beyond the elastic regime to explore their response to external
stresses, in particular how they buckle under extreme twist and bending. Their generic 
behaviour could well be quite different from dsDNA, because of the very different ratios
of the bend to twist moduli. For example, would they form plectonemes in the same way that 
dsDNA does under extreme twists, or would the costs of bending the nanotube back on itself
at the tip of the plectoneme be prohibitive?

\begin{acknowledgement}
We thank Magdalen College, Oxford for the award of a Perkin Research Scholarship (HC), and 
are grateful for support from the EPSRC Centre for Doctoral Training, Theory and Modelling 
in Chemical Sciences, under grant EP/L015722/1 (HC and DP), 
from the Department of Science and Technology India under 
grant DST/INSPIRE/04/2014/002085 (GM), from the European 
Union's Horizon 2020 research and innovation programme under 
Marie Sk\l{}odowska-Curie Grant Agreement No.\ 641839 (MMCT), and 
from ``Fonds voor Wetenschappelijk Onderzoek'' under project number 1SB4219N (ES).
We acknowledge the use of computational facilities provided by 
the University of Oxford Advanced Research Computing [doi:10.5281/zenodo.22558]
and Calcul Quebec. 

\end{acknowledgement}

\begin{suppinfo}

The Supporting Information is available free of charge on the
ACS Publications website at DOI: 10.1021/acs.jctc.XXXX

\begin{itemize}
  \item Additional details on the simulations and their equilibration; 
        triad vector definitions; further snapshots of the nanostructures; 
        further results both to support the methodological developments
        and to provide a fuller characterization of the structure and
        mechanical properties of all systems.
\end{itemize}

\end{suppinfo}


\providecommand{\latin}[1]{#1}
\makeatletter
\providecommand{\doi}
  {\begingroup\let\do\@makeother\dospecials
  \catcode`\{=1 \catcode`\}=2 \doi@aux}
\providecommand{\doi@aux}[1]{\endgroup\texttt{#1}}
\makeatother
\providecommand*\mcitethebibliography{\thebibliography}
\csname @ifundefined\endcsname{endmcitethebibliography}
  {\let\endmcitethebibliography\endthebibliography}{}

\setcounter{figure}{0}
 \makeatletter
 \renewcommand{\thefigure}{S\@arabic\c@figure}
 \setcounter{equation}{0}
 \renewcommand{\theequation}{S\@arabic\c@equation}
 \setcounter{table}{0}
 \renewcommand{\thetable}{S\@arabic\c@table}
 \setcounter{section}{0}
 \renewcommand{\thesection}{S\@arabic\c@section}

\section*{Supporting Information}

\section{Further simulation snapshots}

\begin{figure}
\centering
\includegraphics[width=5.3in]{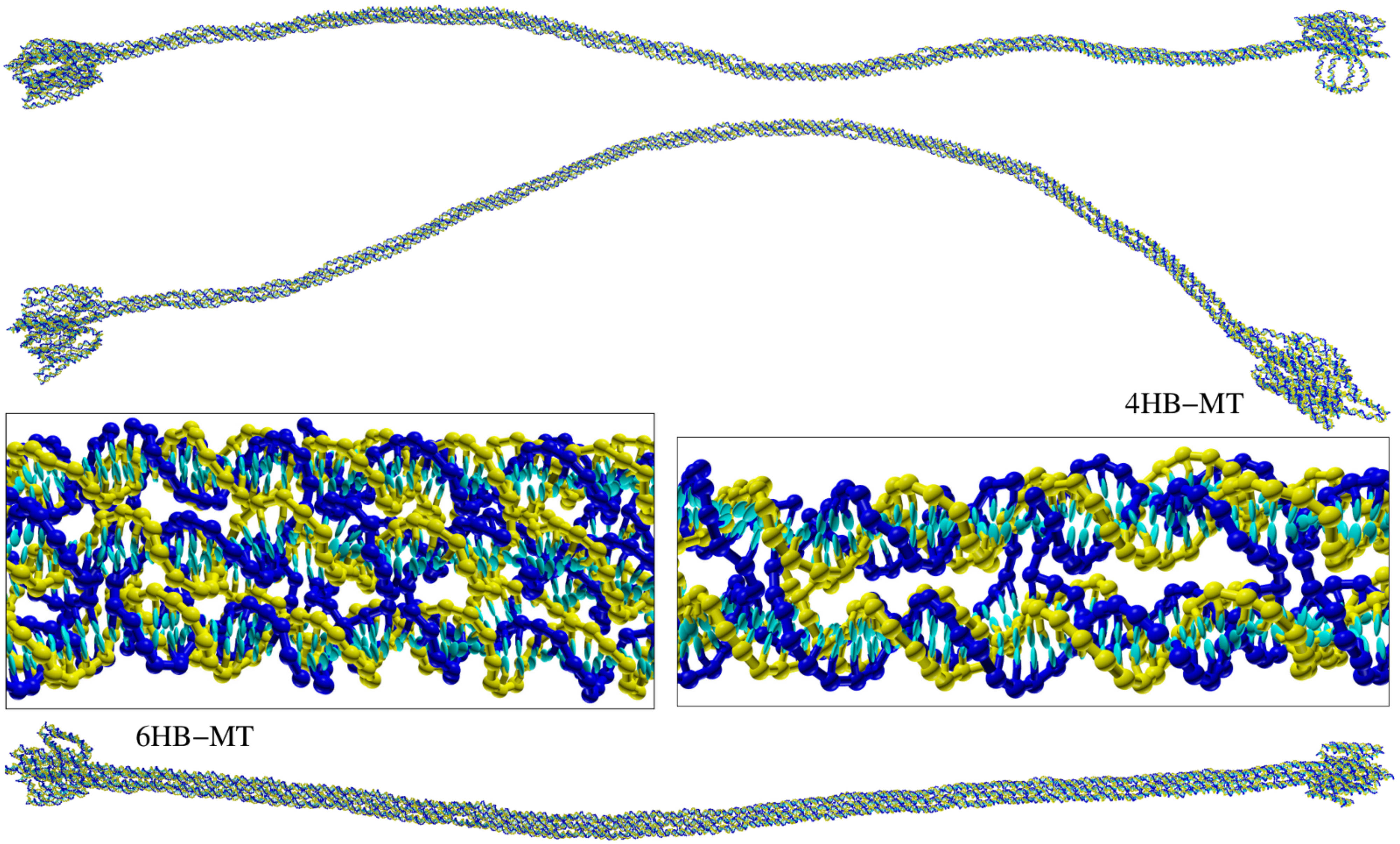}
\caption{Snapshots of the 4HB-MT and 6HB-MT origami.
The second image of the 4HB-MT origami shows an exceptionally bent configuration.
The blocks at each end of the origamis are to facilitate attachment to the beads in
magnetic tweezer experiments.}
\label{sfig:MT_pics}
\end{figure}

\begin{figure}
\centering
\includegraphics[width=5.3in]{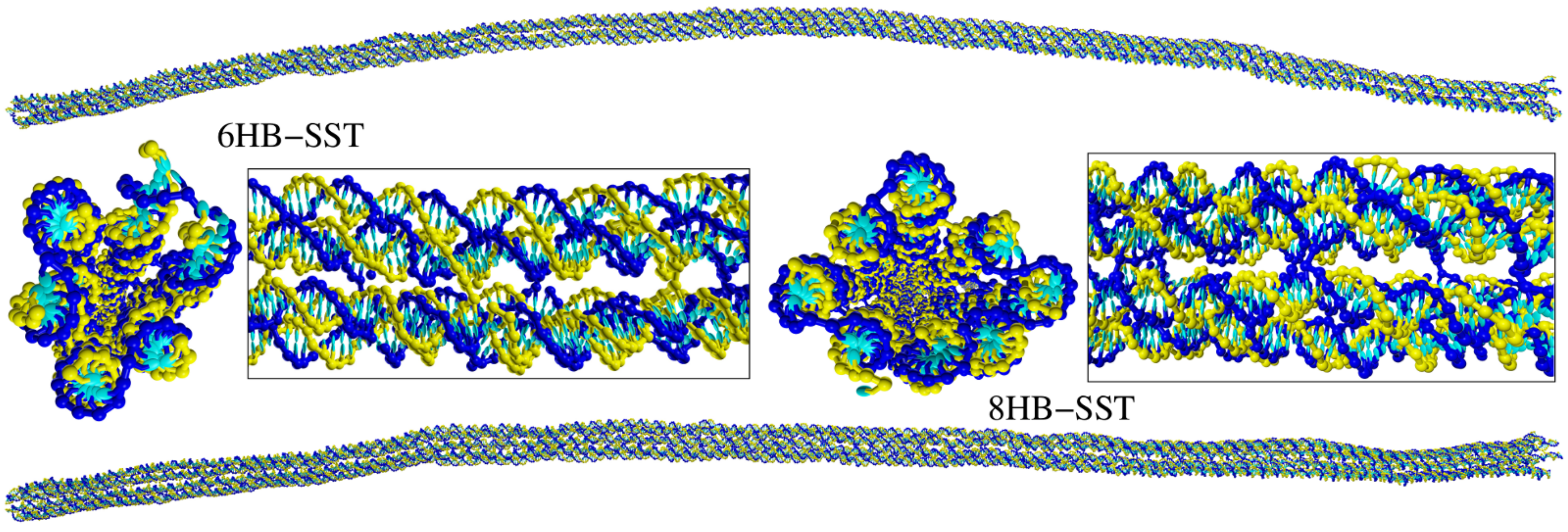}
\caption{Snapshots of the 6HB-SST and 8HB-SST nanotubes.
The end views clearly show the alternating distances of the helices 
from the centres of the tubes.}
\label{sfig:SST_pics}
\end{figure}

\begin{figure}
\centering
\includegraphics[width=5.3in]{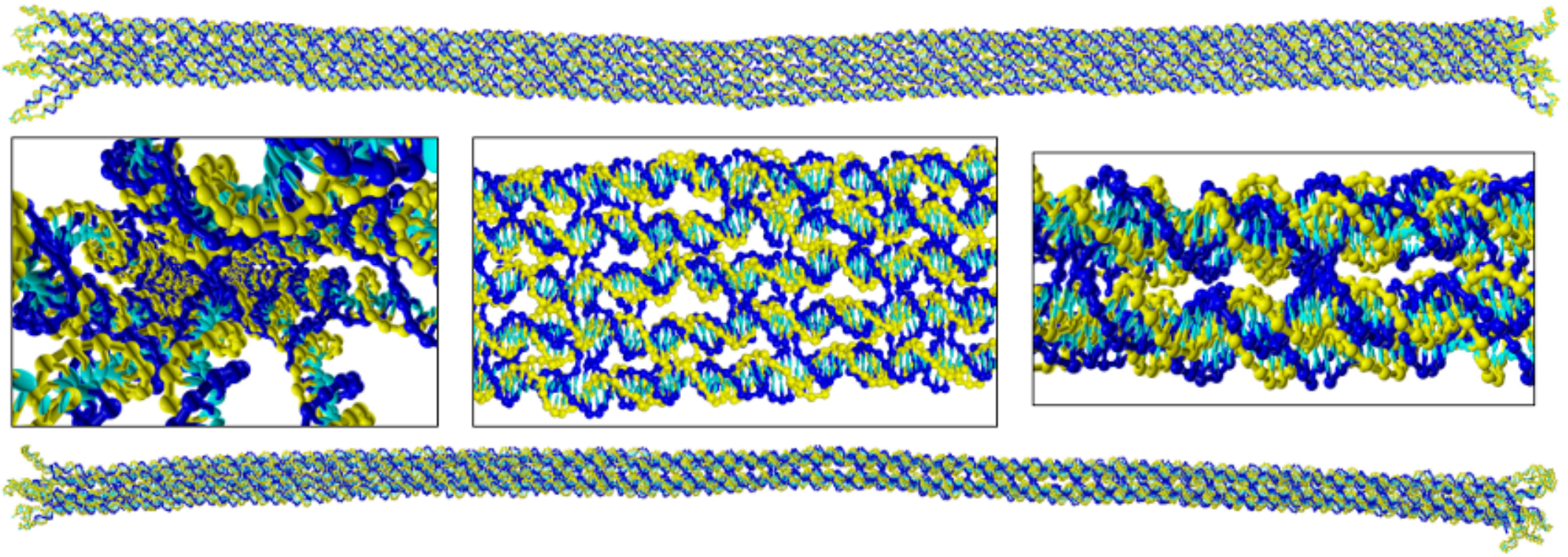}
\caption{Snapshots of the 10HB origami. The two side views are roughly perpendicular to
their thick and thin dimensions. Note this is not perfectly maintained because of the slight
right-handed twist of the origami.}
\label{sfig:10HB_pics}
\end{figure}

Simulation snapshots of some of the systems not visualized in the main text are
given in Figures \ref{sfig:MT_pics}, \ref{sfig:SST_pics} and
\ref{sfig:10HB_pics}.  (Some snapshots of the twisted origami nanotubes can be
found in Ref.\ \citenum{Tortora20}.) The snapshots give an impression of the
scale of typical bending fluctuations.  Apparent from the end views of 6HB-SST
and 8HB-SST is the departure of their cross-sections from a regular hexagonal
and octagonal form, respectively, due to the alternating inter-helix angles in
these systems, an effect noted for 4HB-SST in the main text.

\section{Further Equilibration and Simulation details}

\begin{figure}
\centering
\includegraphics[width=6.6in]{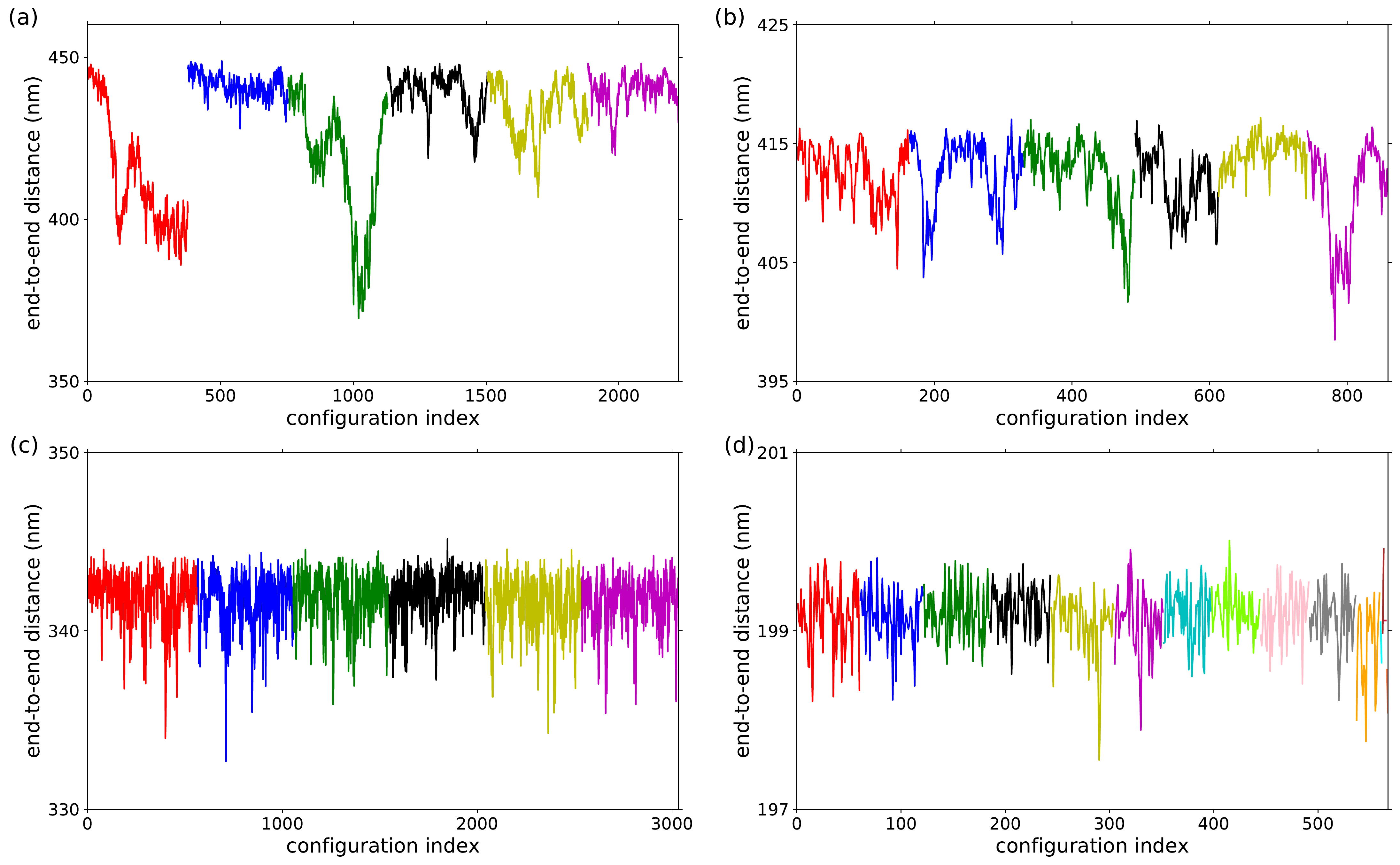}
\caption{The fluctuation of the ``end-to-end'' distance during the molecular dynamics
trajectories for (a) 4HB-MT, (b) 6HB-S, (c) 6HB-SST and (d) 10HB. The different colours represent
separate trajectories. Each configuration is separated by $6.06\,\mu$s.
In order for the measure to capture the fluctuations of the molecular contour in the bulk of 
the tube, the distance is measured between slices that are 100 slices in from each end.
}
\label{sfig:end-to-end}
\end{figure}

Figure \ref{sfig:end-to-end} shows the end-to-end distance in the molecular
dynamics trajectories for four example systems. The 6HB-SST nanotube shows
relatively fast fluctuations over the range of end-to-end distances it samples.
An origami system with the same number of helices (6HB-S) exhibits both a wider
range of end-to-end distances (reflecting its lower bend persistence length)
and a slower sampling of those distances.  In the main text we have suggested
the latter may be due to the more constrained nature of the junctions in
origamis and a coupling of bending to changes in junction geometry.  
It is noticeable that distributions of $\Omega_1$ and $\Omega_2$ for the origamis have a 
greater degree of non-Gaussianity than for the SST nanotubes 
(Figs.\ \ref{sfig:pOmega_4HBMT}-\ref{sfig:pOmega_6HBSST}). 
Another possible source of slow dynamics is if there is a coupling of more extreme
bending with base-pair breaking (note that all the 6HB origamis have some very short
2 base-pair binding domains). 

The plot for the 4HB-MT shows even slower dynamics and parts of the trajectory
show fluctuation to anomalously bent configurations. The former may be partly
due to the effects of the blocks on the ends of these origamis, but the reason
for the extremely bent configurations (e.g.\ the second configuration in Fig.\
\ref{sfig:SST_pics}) is less clear (one unique feature of the 4HB-MT design is
that it has some junctions where one helix has a nick on the strand not
involved in the crossover).  As a result of this, the statistics for this
system are probably the worst.  Furthermore, the extreme bending leads to a
breakdown of worm-like chain behaviour. For example, the tangent-tangent
correlation function lacks a perfectly exponential tail, and that the
$m$-dependent persistence length exhibits a broad maximum (Fig.\
\ref{sfig:lp_MT}(a)), rather than smoothly increasing to its limiting value, is
a sign that the fluctuations on the longest length scales are greater than one
would expect from the (better sampled) shorter length scale fluctuations.
That the 6HB-MT origami also exhibits a similar broad maximum (Fig.\
\ref{sfig:lp_MT}(a)), in contrast to all the other systems, perhaps suggests
that this may not simply be a case of insufficient sampling, but that the
``MT'' origami are behaving somewhat differently. The obvious difference from
the other origamis is the blocks on the end, but it is hard to see why these
might facilitate anomalously large bending fluctuations on long length scales.

\begin{table}[t]
\renewcommand{\arraystretch}{1.2}
\begin{tabular}{cccccccc}
\hline
 & $t_\mathrm{sim}$/ms & $N_\mathrm{config}$ & $N_\mathrm{ends}$ & $n_\mathrm{avg}^{A_1}$ & $n_\mathrm{avg}^{A_2}$ & $n_\mathrm{avg}^C$ & $m_\mathrm{limit}$ \\
\hline
4HB-MT          & 13.48 & 2225 & 60 & 20 & 20 & 5 & 600 \\
4HB-SST         &  6.81 & 1124 & 60 & 15 & 15 & 5 & 600 \\
6HB-2$\times$LH &  8.52 & 1406 & 60 & 20 & 20 & 5 & 600 \\
6HB-1$\times$LH &  8.05 & 1328 & 60 & 20 & 20 & 5 & 600 \\
6HB-S           &  5.21 &  859 & 60 & 20 & 20 & 5 & 600 \\
6HB-1$\times$RH &  8.34 & 1376 & 60 & 20 & 20 & 5 & 600 \\
6HB-MT          & 13.33 & 2199 & 60 & 20 & 20 & 5 & 600 \\
6HB-SST         & 18.38 & 3034 & 60 & 20 & 20 & 5 & 600 \\
8HB-SST         & 13.91 & 2295 & 60 & 40 & 40 & 5 & 600 \\
10HB            &  3.44 &  567 & 60 & 40 & 20 & 5 & 400 \\
\hline
\end{tabular}
\caption{Details of the simulations and the parameters used when calculating elastic properties.
$t_\mathrm{sim}$ is the total simulation time of the trajectories. 
$N_\mathrm{config}$ is the number of configurations in the ensemble used to calculate the
elastic properties of each system. $N_\mathrm{ends}$ is the number of slices at each end 
that are ignored in these calculations. $2 n_\mathrm{avg}^{X}+1$ is the number of slices over 
which averaging is performed in Eq.\ (21) for the elastic constant $X$.
$m_\mathrm{limit}$ is the value of $m$ used when taking the limiting values of the
elastic constants.
}
\label{table:details}
\end{table}

We also include data for the 10HB origami in Fig.\ \ref{sfig:end-to-end}. This illustrates
that as the origami become stiffer the fluctuations become faster and easier to sample.

From the above it is clear that the fluctuations of the SST nanotubes are the
most well sampled, and hence the ensembles of configurations are least correlated. 
It is also noticeable that these systems also have values
for the coupling terms that are closest to zero, and in the cases where
expected (6HB-SST and 8HB-SST) $A_1$ and $A_2$ values that are most similar
(Fig.\ \ref{sfig:elastic_SST}). The greater regularity of their pattern of 
junctions also likely plays a role. 

The total simulation time used to generate the configurations and the consequent
size of the configurational ensembles used for each system are given in 
Table \ref{table:details}. It also contains details of the parameters used in our
calculations of the mechanical properties.

\section{Further details of mechanical properties calculations}

\subsection{Angular deformations distributions}
\begin{figure}
\centering
\includegraphics[width=6.6in]{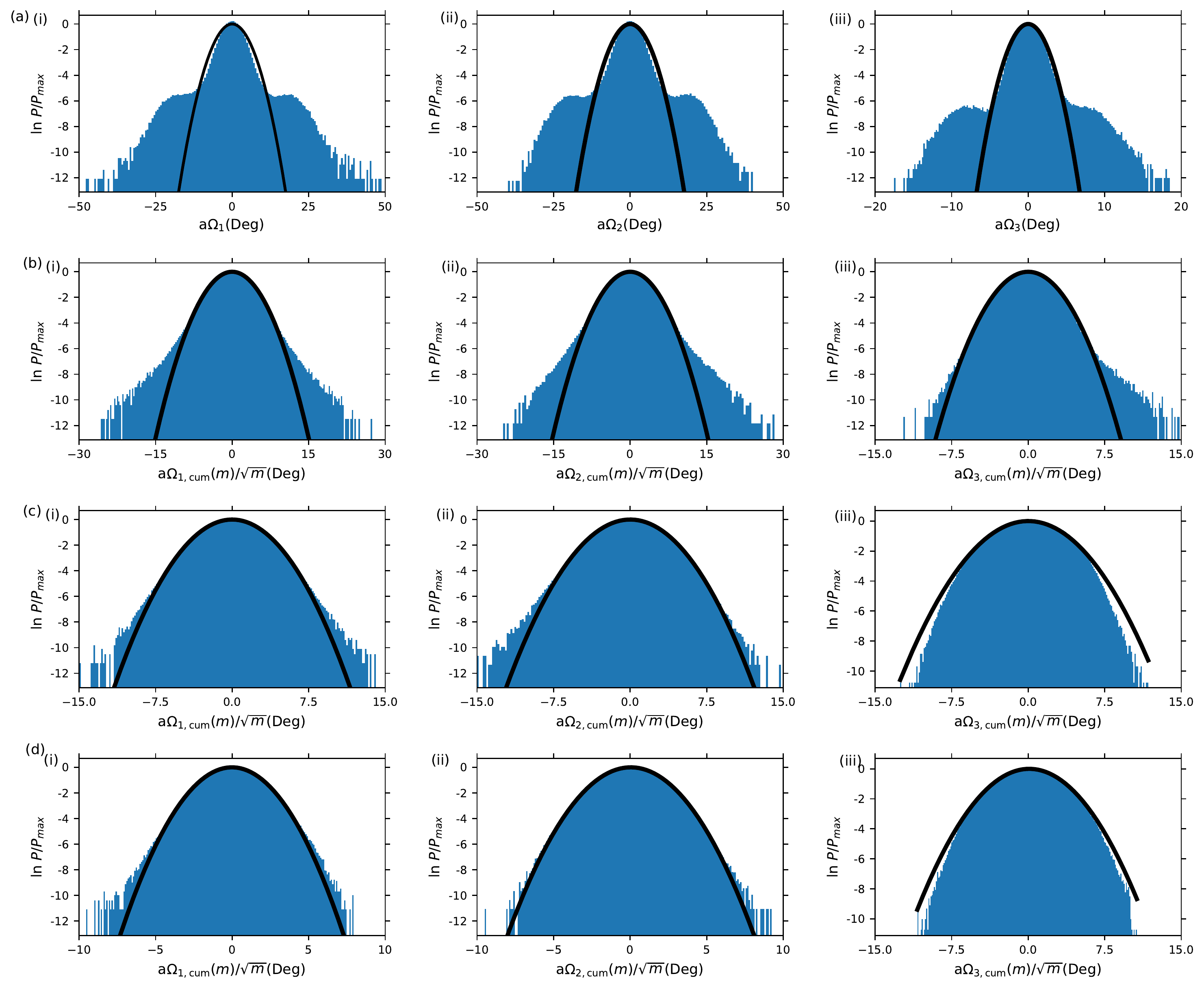}
\caption{Probability distributions for (i) $a\Omega_{1,\mathrm{cum}}(m)/\sqrt{m}$, 
(ii) $a\Omega_{2,\mathrm{cum}}(m)/\sqrt{m}$, 
and (iii) $a\Omega_{3,\mathrm{cum}}(m)/\sqrt{m}$
for the 4HB-MT origami at different values of $m$:
(a) $m=1$,
(b) $m=3$,
(c) $m=11$ and
(d) $m=41$.
The solid lines are Gaussian fits to the distributions.
}
\label{sfig:pOmega_4HBMT}
\end{figure}

\begin{figure}
\centering
\includegraphics[width=6.6in]{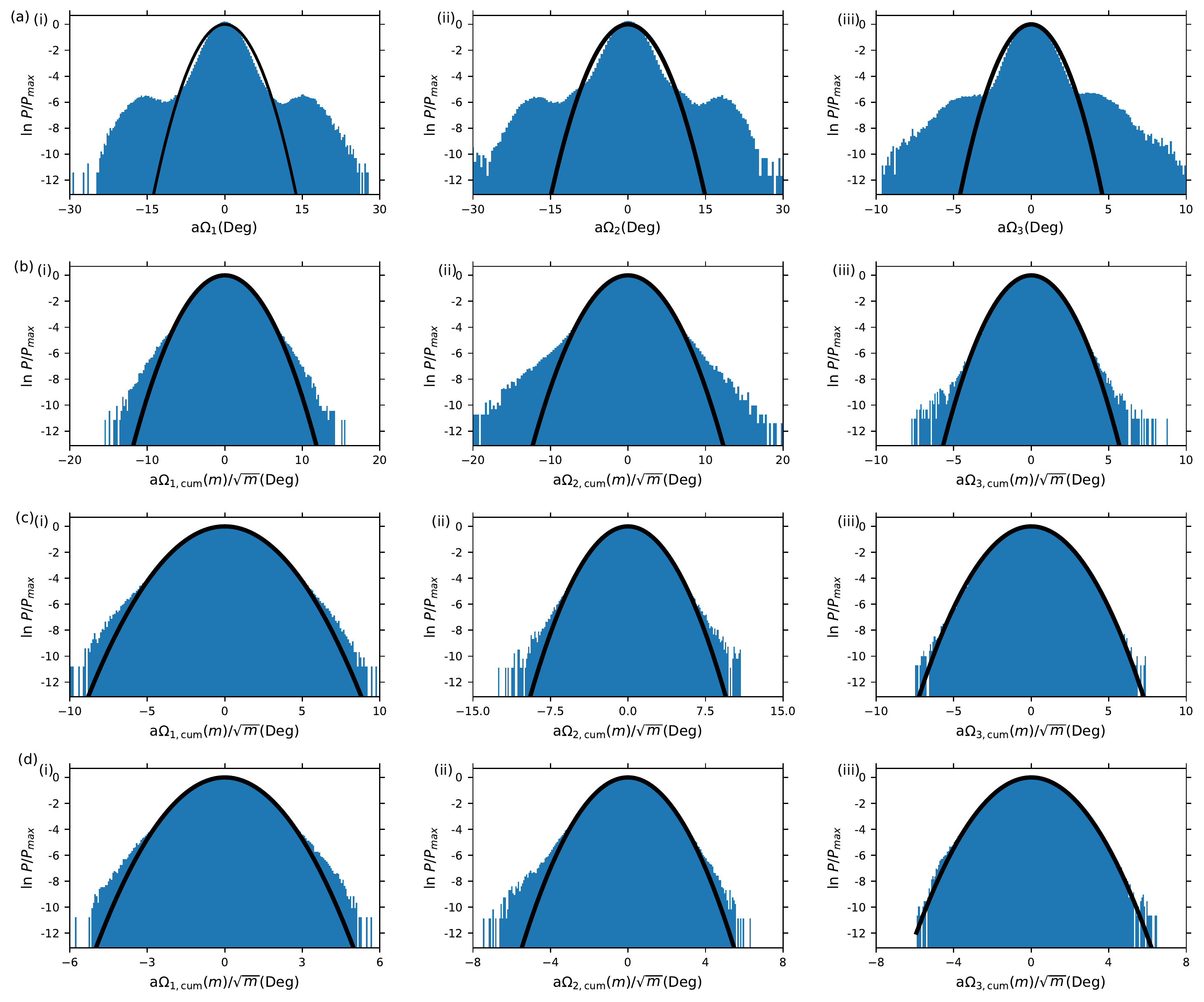}
\caption{Probability distributions for (i) $a\Omega_{1,\mathrm{cum}}(m)/\sqrt{m}$, 
(ii) $a\Omega_{2,\mathrm{cum}}(m)/\sqrt{m}$, 
and (iii) $a\Omega_{3,\mathrm{cum}}(m)/\sqrt{m}$
for the 6HB-MT origami 
at different values of $m$:
(a) $m=1$,
(b) $m=3$,
(c) $m=11$ and
(d) $m=41$.
The solid lines are Gaussian fits to the distributions.
}
\label{sfig:pOmega_6HBMT}
\end{figure}

\begin{figure}
\centering
\includegraphics[width=6.6in]{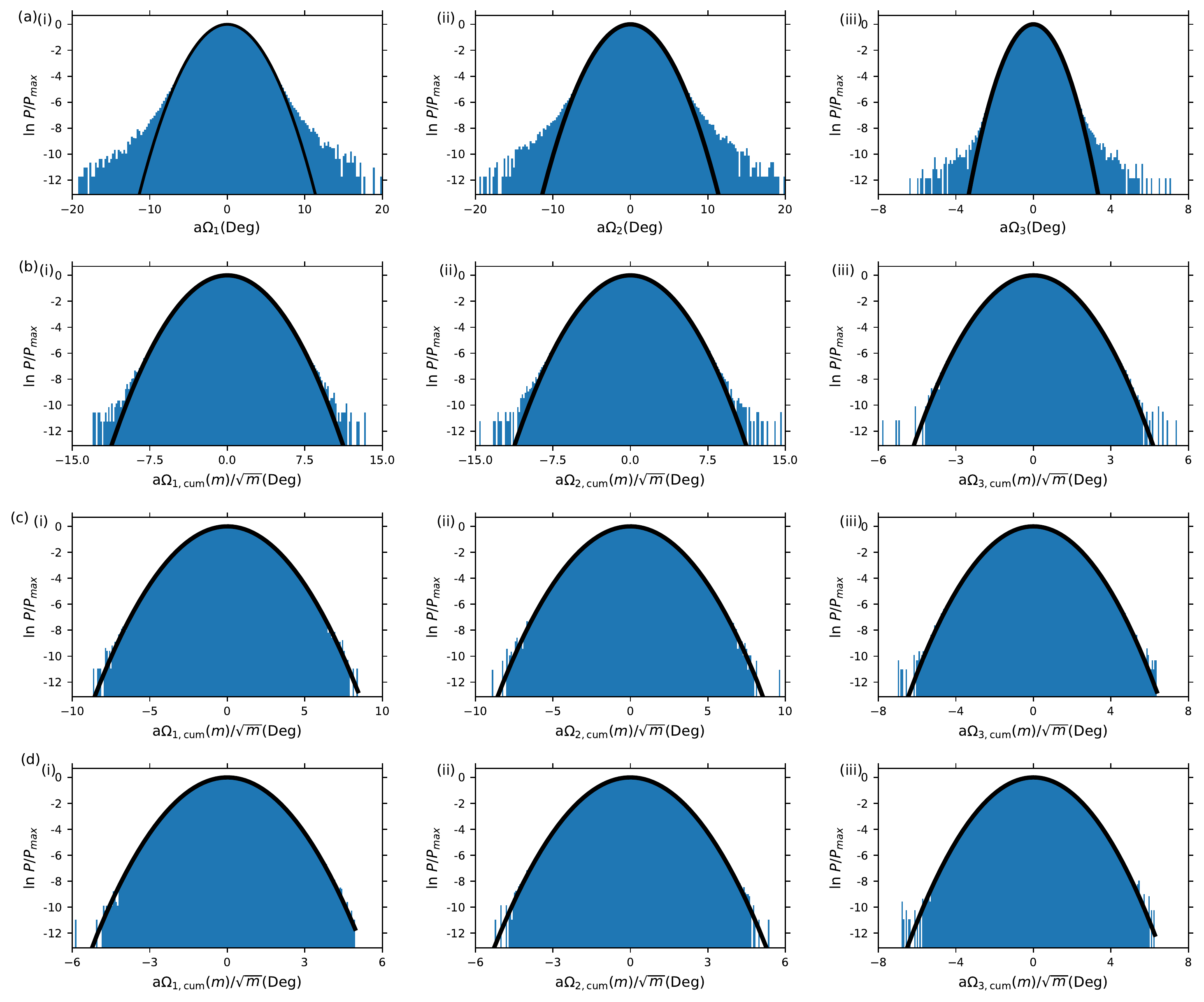}
\caption{Probability distributions for (i) $a\Omega_{1,\mathrm{cum}}(m)/\sqrt{m}$, 
(ii) $a\Omega_{2,\mathrm{cum}}(m)/\sqrt{m}$, 
and (iii) $a\Omega_{3,\mathrm{cum}}(m)/\sqrt{m}$
for the 6HB-SST nanotube 
at different values of $m$:
(a) $m=1$,
(b) $m=3$,
(c) $m=11$ and
(d) $m=41$.
The solid lines are Gaussian fits to the distributions.
}
\label{sfig:pOmega_6HBSST}
\end{figure}

The cumulative angular deformations used in Eq.\ \ref{eq:Xi} are defined by 
\begin{equation}
\Omega_{i,\mathrm{cum}}^{(k)}(m)=\sum_{k=n}^{n+m-1}\Omega_i^{(k)}
\end{equation}
The probability distributions of $a\Omega_{i,\mathrm{cum}}(m)/\sqrt{m}$ for the
4HB-MT and 6HB-MT origamis and the 6HB-SST nanotube for are plotted in Figs.\
\ref{sfig:pOmega_4HBMT}--\ref{sfig:pOmega_6HBSST}, respectively. 
As expected from the central-limit theorem, the distributions become more Gaussian-like 
as $n_\mathrm{avg}$ increases, confirming that worm-like chain behaviour holds at longer
renormalized length scales.
The distributions for the bending deformations become narrower as 
$m$ increase, consistent with the increase in the $m$-dependent bending elastic 
constants with $m$.

\subsection{Further effects of $\alpha$}

\begin{figure}
\centering
\includegraphics[width=6.6in]{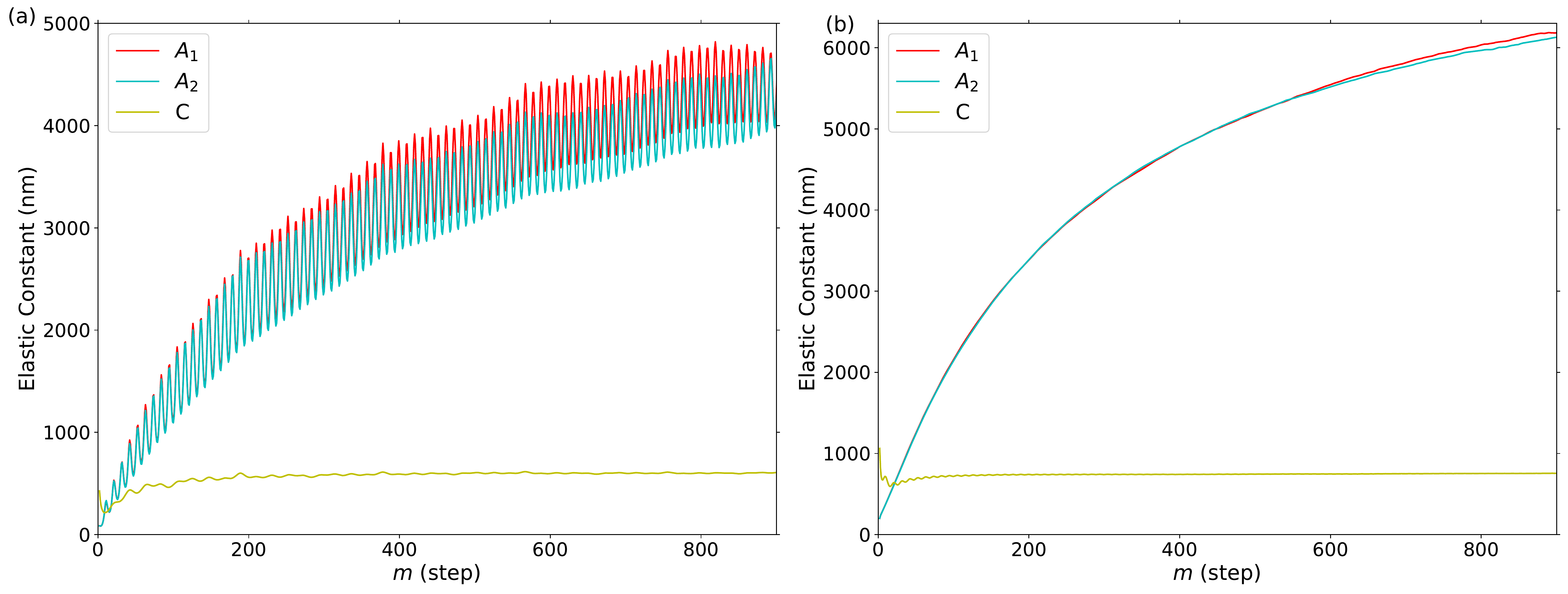}
\caption{Elastic moduli for (a) 6HB-S and (b) 6HB-SST at $\alpha=0.0$.
The origami exhibits an oscillation on the pitch length of double-stranded DNA, whereas the
elastic moduli increase smoothly for the SST nanotube.
}
\label{sfig:elastic_oscillation}
\end{figure}

In Fig.\ \ref{sfig:elastic_oscillation} we show plots of the $m$-dependent elastic
moduli for example origami and SST systems at $\alpha=0$, i.e.\ without the correction to
remove the helical component of the centreline of double-stranded DNA in oxDNA.
Like the plot of the $m$-dependent persistence length for 6HB-S in the main text (Fig.\ 3(d))
the bending elastic moduli for this system show strong oscillations on the length scale of the 
DNA pitch. By contrast, the SST systems show a smooth monotonic rise even at $\alpha=0$.
The reason for this is most likely because the single crossovers in these systems place fewer
constraints on the relative twist of different helices. The strong correlations between
the orientations of the individual helices in the origamis give an additive effect
causing the helix-bundle centerlines to have a significant helical character. By contrast,
the reduced local orientation correlations in the SST systems cause the helical character
of the centerlines of the individual DNA helices to cancel out giving a smooth nanotube centreline.

\begin{figure}
\centering
\includegraphics[width=3.3in]{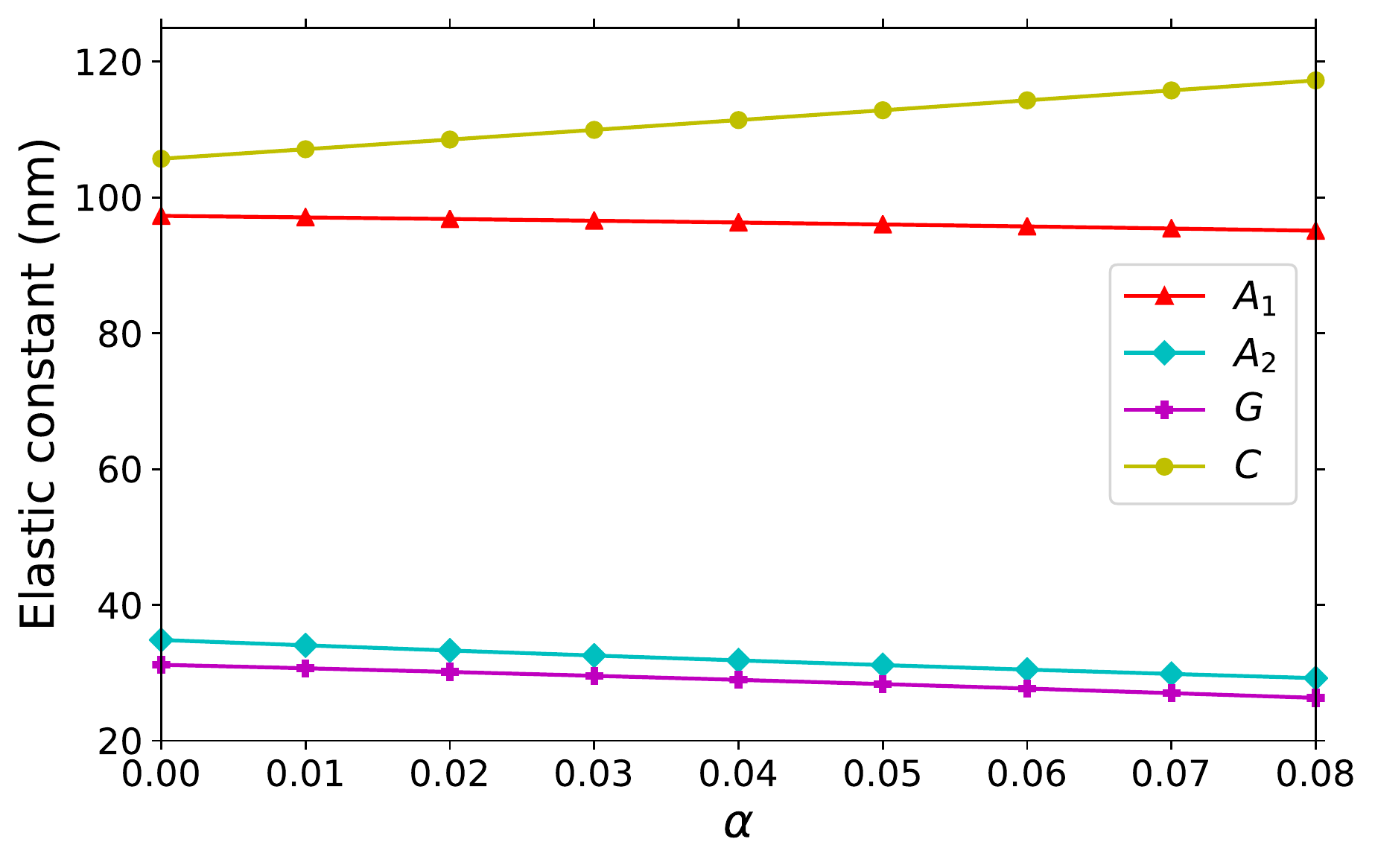}
\caption{Double-stranded DNA elastic constants $A_1$, $A_2$, $C$, $G$ as a function of $\alpha$.}
\label{sfig:duplexmech_alpha}
\end{figure}

The dependence of the elastic constants of double-stranded DNA on $\alpha$
using Triad III of Ref.\ \citenum{Skoruppa17} are depicted in Fig.\
\ref{sfig:duplexmech_alpha}. The structural and mechanical properties of
double-stranded DNA at $\alpha=0$, i.e.\ without any correction for the
helicity of the centreline, and at $\alpha=0.06$ (the value that removes the
net helicity) are given in Table \ref{table:duplex}. Also included in the table
are the predicted elastic constants obtained by applying a transformation from a helical
to a straight coordinate system, the expressions for which were derived in Ref.\
\citenum{Nomidis19}. Comparing the two methods for removing the helicity of
the centreline, the changes in $A_2$, $C$ and $G$ are in the same direction,
but with some differences in the magnitudes of the changes, and although the
coordinate transformation approach predicts no change in $A_1$, we observe a small
decrease on going from $\alpha=0$ to $\alpha=0.06$.

The removal of the helicity centreline makes little difference to the bending persistence length, but
leads to a significant increase in the measured twist persistence length.
Note also that compensating for the helicity reduces the measured pitch 
($360/\langle\Theta_3\rangle$) from 10.54 to 10.51\,bp.

\begin{table*}[t]
\renewcommand{\arraystretch}{1.2}
\begin{tabular}{ccccccccccccccc}
\hline
  & $\langle \Theta_1 \rangle$ & $\langle \Theta_2 \rangle$ & $\langle \Theta_3 \rangle$  & $A_1$ & $A_2$ & $C$ & $G$ & $l_b$ & $l_t/2$ \\
\hline
$\alpha=0.00$ & 0.00 & 2.81 & 34.16 & 97.3 & 34.8 & 105.7 & 31.2 & 40.1 & 81.1 \\
$\alpha=0.06$ & 0.00 & 0.00 & 34.26 & 95.7 & 30.5 & 114.3 & 27.7 & 41.2 & 93.2 \\
helically corrected & &     &       & 97.3 & 30.2 & 110.3 & 25.0 \\
\hline
\end{tabular}
\caption{Structural and mechanical properties of double-stranded DNA before and after
applying the correction to remove the helicity of the centreline.
The elastic constants obtained from those at $\alpha=0$ by applying the transformation
from a helical to a straight coordinate system outlined in Ref.\ \citenum{Nomidis19} are also
given. The results are for [Na$^+$]=0.5\,M and a temperature of either 23$^\circ$C.
$a=0.3462$\,nm.}
\label{table:duplex}
\end{table*}

\subsection{Triad definitions}

The definitions of $\widehat{\mathbf{x}}$ and $\widehat{\mathbf{y}}$ for the six helix bundles
are given in the main text, and schematically indicated in Fig.\ 1(d) for the other systems. 
Here, we give their mathematical definitions, where the numbering of the helices is as in 
Fig.\ 1(d).

For the 4HB systems:
\begin{equation}
\widehat{\mathbf{x}}= \frac{\mathbf{R}_{\mathrm{duplex}}(3) -\mathbf{R}_{\mathrm{duplex}}(1) }
            {\left|\left| \mathbf{R}_{\mathrm{duplex}}(3) -\mathbf{R}_{\mathrm{duplex}}(1) \right|\right|}
\label{eq:x_4HB}
\end{equation}
and
\begin{equation}
\widehat{\mathbf{y}}= \frac{\mathbf{R}_{\mathrm{duplex}}(2)-\mathbf{R}_{\mathrm{duplex}}(4)}
            {\left|\left| \mathbf{R}_{\mathrm{duplex}}(2)-\mathbf{R}_{\mathrm{duplex}}(4) \right|\right|}.
\label{eq:y_4HB}
\end{equation}
For the 8HB systems:
\begin{equation}
\widehat{\mathbf{x}}= \frac{\mathbf{R}_{\mathrm{duplex}}(5) -\mathbf{R}_{\mathrm{duplex}}(1) }
            {\left|\left| \mathbf{R}_{\mathrm{duplex}}(5) -\mathbf{R}_{\mathrm{duplex}}(1) \right|\right|}
\label{eq:x_8HB}
\end{equation}
and
\begin{equation}
\widehat{\mathbf{y}}= \frac{\mathbf{R}_{\mathrm{duplex}}(3)-\mathbf{R}_{\mathrm{duplex}}(7)}
            {\left|\left| \mathbf{R}_{\mathrm{duplex}}(3)-\mathbf{R}_{\mathrm{duplex}}(7) \right|\right|}.
\label{eq:y_8HB}
\end{equation}
For the 10HB system 
\begin{equation}
\widehat{\mathbf{x}}= \frac{\mathbf{R}_{\mathrm{duplex}}(5) + \mathbf{R}_{\mathrm{duplex}}(7) -\mathbf{R}_{\mathrm{duplex}}(2) - \mathbf{R}_{\mathrm{duplex}}(10)}
            {\left|\left| \mathbf{R}_{\mathrm{duplex}}(5) + \mathbf{R}_{\mathrm{duplex}}(7) -\mathbf{R}_{\mathrm{duplex}}(2) - \mathbf{R}_{\mathrm{duplex}}(10) \right|\right|}
\label{eq:x_10HB}
\end{equation}
and
\begin{equation}
\widehat{\mathbf{y}}= \frac{\mathbf{R}_{\mathrm{duplex}}(3) + \mathbf{R}_{\mathrm{duplex}}(4) -\mathbf{R}_{\mathrm{duplex}}(8) - \mathbf{R}_{\mathrm{duplex}}(9)}
            {\left|\left| \mathbf{R}_{\mathrm{duplex}}(3) + \mathbf{R}_{\mathrm{duplex}}(4) -\mathbf{R}_{\mathrm{duplex}}(8) - \mathbf{R}_{\mathrm{duplex}}(9) \right|\right|}.
\label{eq:y_10HB}
\end{equation}

\begin{figure}
\centering
\includegraphics[width=6.6in]{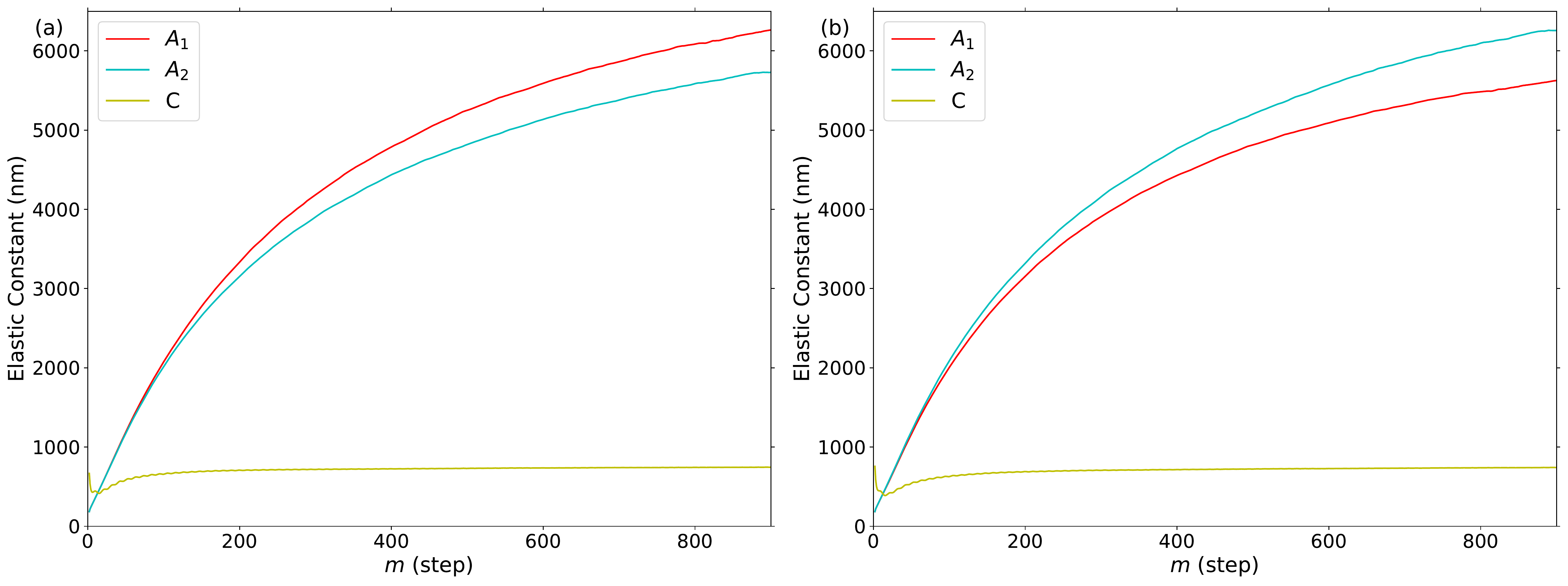}
\caption{
The effect of the orthogonalization scheme to obtain the triad 
$\{\widehat{\mathbf{e}}_{1},\widehat{\mathbf{e}}_{2},\widehat{\mathbf{e}}_{3}\}$ on the
elastic moduli for the 6HB-SST nanotube. In (a) 
$\widehat{\mathbf{y}}$ is orthogonalized to $\widehat{\mathbf{e}}_{3}$, 
whereas in (b) $\widehat{\mathbf{x}}$ is orthogonalized to $\widehat{\mathbf{e}}_{3}$.
}
\label{sfig:no_svd}
\end{figure}

Figure \ref{sfig:no_svd} shows the effect of the orthogonalization scheme on the 
elastic moduli. 
If $\widehat{\mathbf{e}}_{2}$ is obtained by 
the orthogonalization of $\widehat{\mathbf{y}}$ to $\widehat{\mathbf{e}}_{3}$, 
and $\widehat{\mathbf{e}}_{1}$ from the vector product 
of $\widehat{\mathbf{e}}_{2}$ and $\widehat{\mathbf{e}}_{3}$ 
(as in Triad III of Ref.\ \citenum{Skoruppa17}), then $A_1>A_2$ for the 6HB-SST system.
By contrast, if $\widehat{\mathbf{e}}_{1}$ is obtained by 
the orthogonalization of $\widehat{\mathbf{x}}$ to $\widehat{\mathbf{e}}_{3}$, 
and $\widehat{\mathbf{e}}_{2}$ from the vector product 
of $\widehat{\mathbf{e}}_{3}$ and $\widehat{\mathbf{e}}_{1}$, then 
$A_2>A_1$. These results shows the biases introduced by these orthogonalization schemes. 
Applying the singular value decomposition technique to find the triad avoids these 
problems, and results in virtually equal values of $A_1$ and $A_2$ (Fig.\ 4), as is expected from
the symmetry of this system.

\subsection{Limiting values}

\begin{figure}
\centering
\includegraphics[width=3.3in]{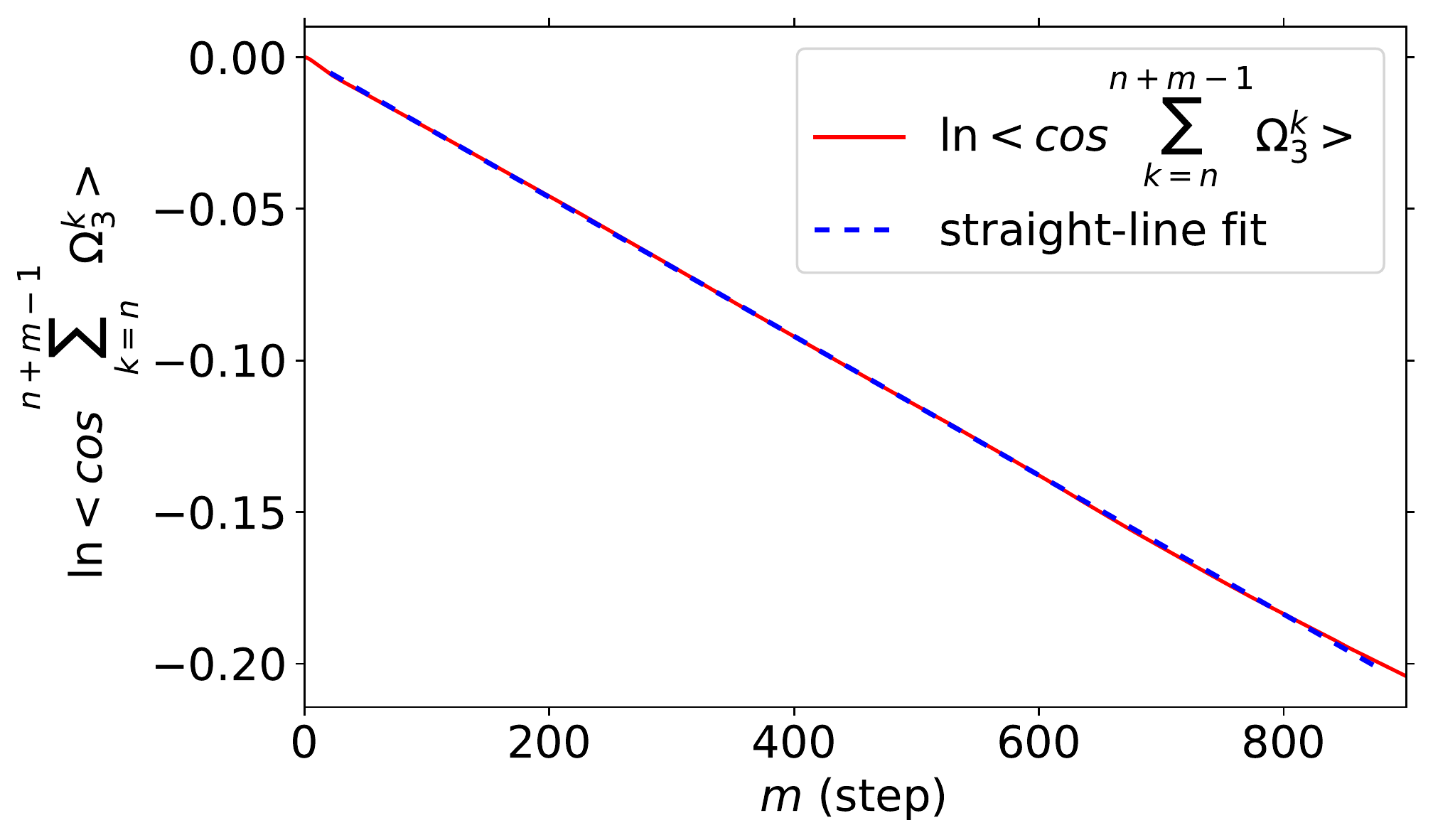}
\caption{The exponential decay of the correlation function used to 
calculate the twist persistence length for the 6HB-SST nanotube.}
\label{sfig:twist_tail}
\end{figure}

Fig.\ \ref{sfig:twist_tail} illustrates the behaviour of the correlation function
used to obtain the twist persistence length (Eq.\ 6) for the 6HB-SST system.
It has a clear exponential tail making it straightforward to extract a limiting value
of the twist persistence length. Moreover, unlike for the tangent-tangent correlation
function, the deviations from exponential behaviour at small $m$ are barely noticeable. 
Consequently, the $m$-dependent twist persistence lengths quickly converge
to their limiting values.

\begin{figure}
\centering
\includegraphics[width=3.3in]{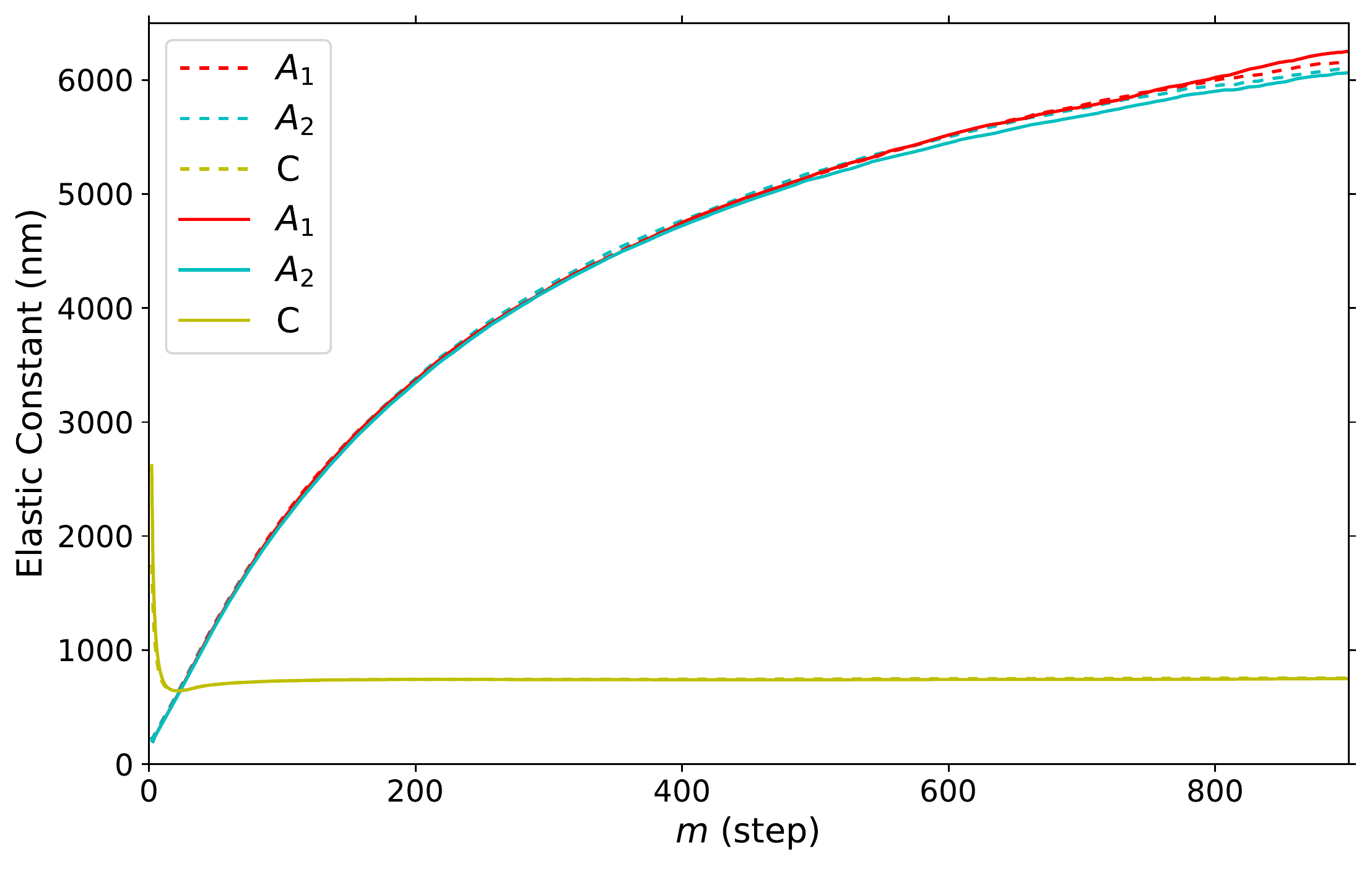}
\caption{A comparison of elastic constants calculated using Equations 
7
(solid lines) and 
20
(dashed lines)
for the 6HB-SST nanotube.}
\label{sfig:elastic_correlation}
\end{figure}

The elastic moduli can be calculated from the correlation coefficients describing
the correlations between angular deformations separated by $n$ steps (Eq.\ 19) using
the expression in Eq.\ 20. The resulting values provide a good description of the elastic
moduli over the range of $m$ sampled in simulations (Fig.\ \ref{sfig:elastic_correlation}). 
However, when one tries to extrapolate the predicted elastic constants
to their limiting values, the results become sensitive to the statistical noise in the 
values of the correlation coefficients. 

\begin{figure}
\centering
\includegraphics[width=6.6in]{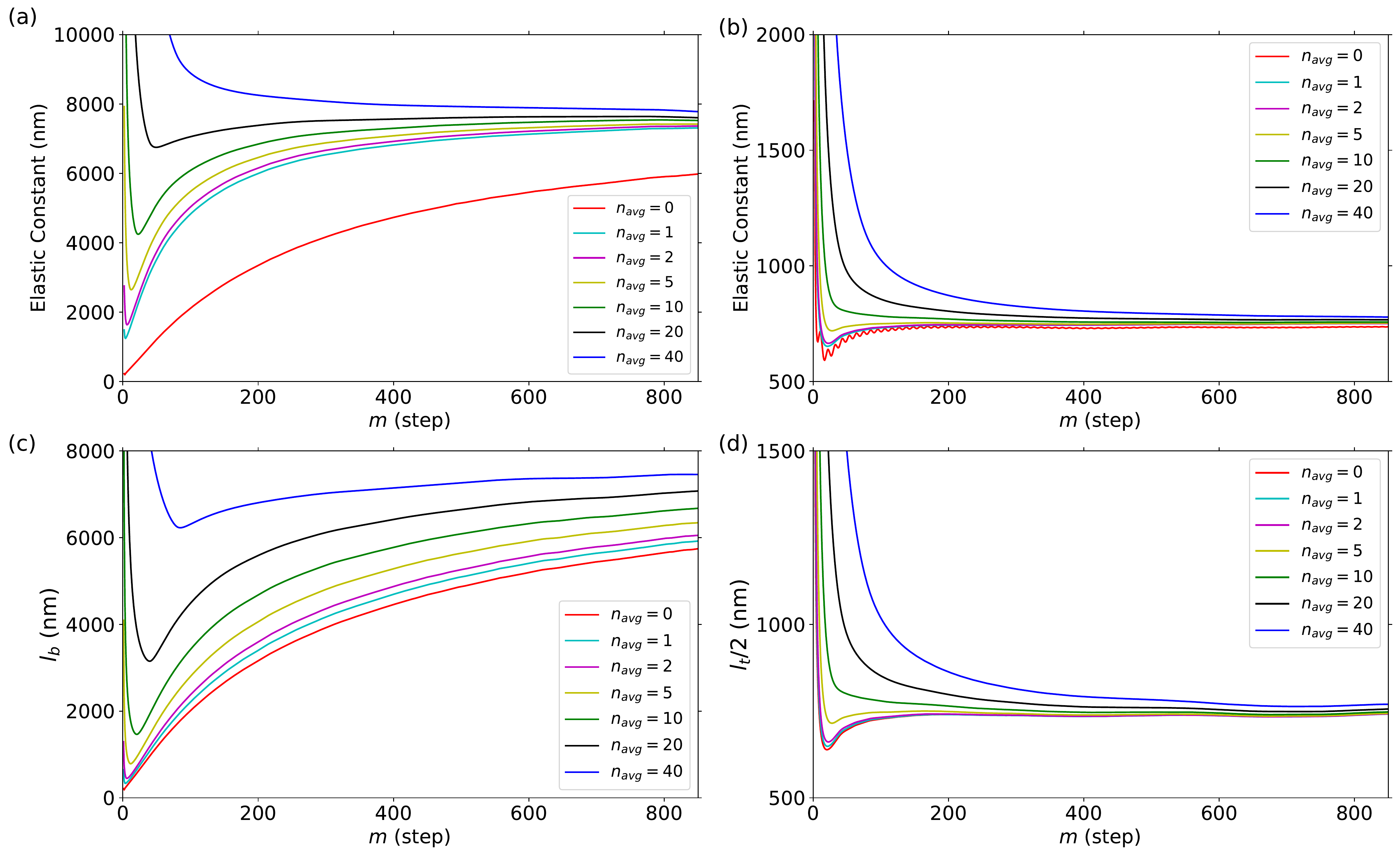}
\caption{The effect of averaging on (a) $A_2$, (b) $C$, (c) $l_b$ and (d) $l_t$ 
for different $n_\mathrm{avg}$ for the 6HB-SST nanotube.}
\label{sfig:navg_sup}
\end{figure}

In Fig.\ 5 of the main text, we illustrate the effect of the ``averaging'' approach we use to obtain
the limiting values of the elastic constants for $A_1$. Equivalent figures for $A_2$ and $C$ 
are given in Fig.\ \ref{sfig:navg_sup}. The dependence of the behaviour on the 
parameter $n_\mathrm{avg}$ for $A_2$ is, unsurprisingly, very similar to that for $A_1$, and 
except for the 10HB origami, which has the largest $A_1/A_2$ ratio, we always use the same value
of $n_\mathrm{avg}$ when computing the limiting values of these two elastic constants 
(Table \ref{table:details}). Performing this averaging is not so important for the twist elastic
constant as it converges reasonably rapidly to its limiting value. This can be accelerated somewhat
by using a small value of $n_\mathrm{avg}$; we always use $n_\mathrm{avg}=5$. Although we do not use
this averaging approach to obtain limiting values of the persistence lengths, 
Fig.\ \ref{sfig:navg_sup} also illustrates its effects on the $m$-dependent persistence lengths.
Notably, the effect on the convergence of the bend persistence length is less pronounced than 
it is for the bending moduli.

\section{Further results}

\begin{figure}
\centering
\includegraphics[width=3.3in]{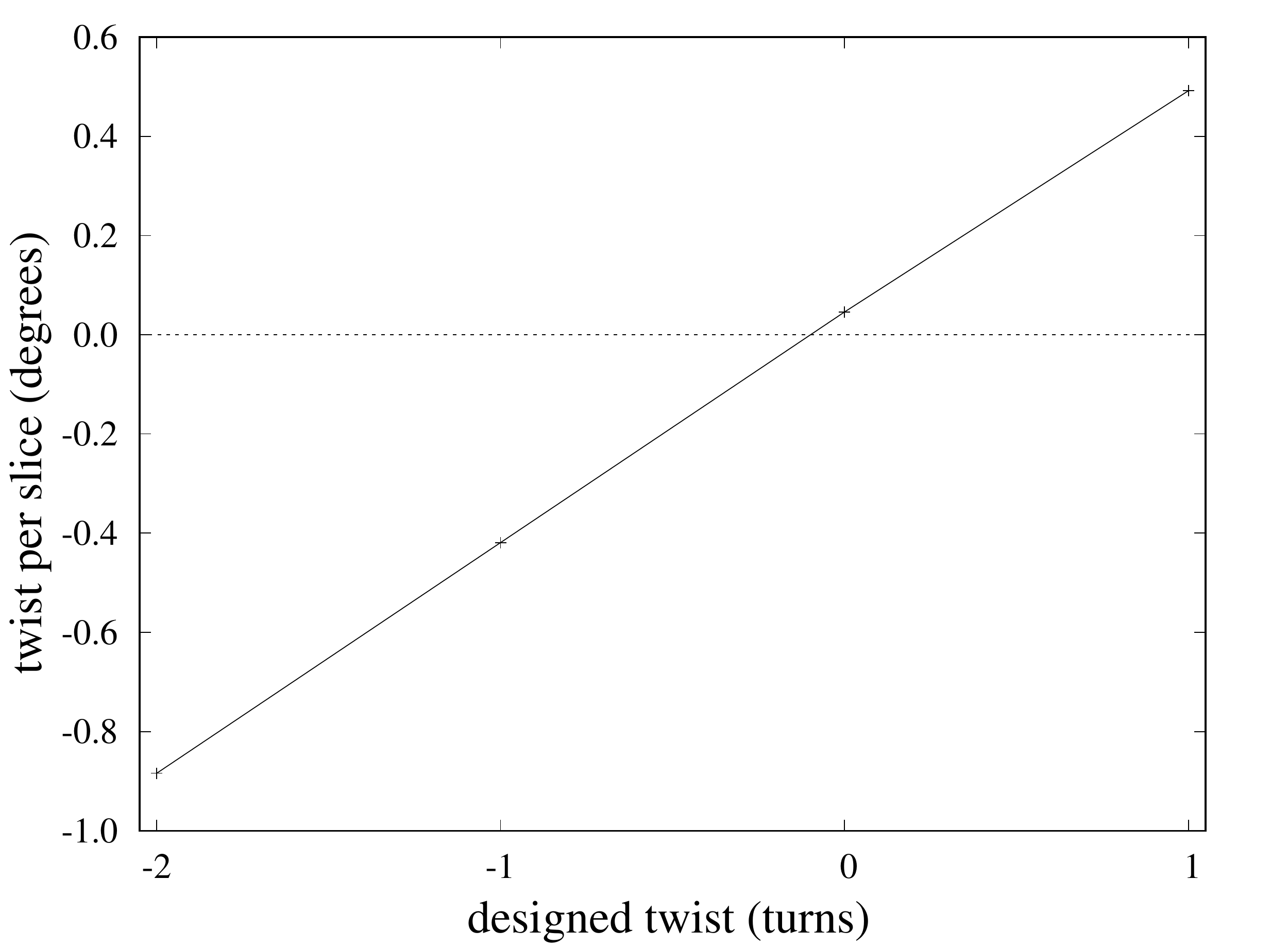}
\caption{Twist per slice  as a function of the designed twist in 
turns (i.e.\ 2$\times$LH is -2, 1$\times$RH is +1) for the series of twisted 6HB origamis.}
\label{sfig:twist_sup}
\end{figure}

Fig.\ \ref{sfig:twist_sup} depicts the equilibrium twist of the series of twisted origamis versus
their designed twist. There is an almost perfect linear relationship, albeit with the 
origami that is designed to be untwisted having a small right-handed twist. Note that the oxDNA 
model has been fitted to reproduce the reported zero twist of an origami in Ref.\ \citenum{Dietz09}.
Interestingly, our slope is larger than would be expected from the designed
twist, which is derived from the predictions of the CanDo model.\cite{Castro11,Kim12}

\subsection{Nanotube radii}

\begin{figure}
\centering
\includegraphics[width=6.6in]{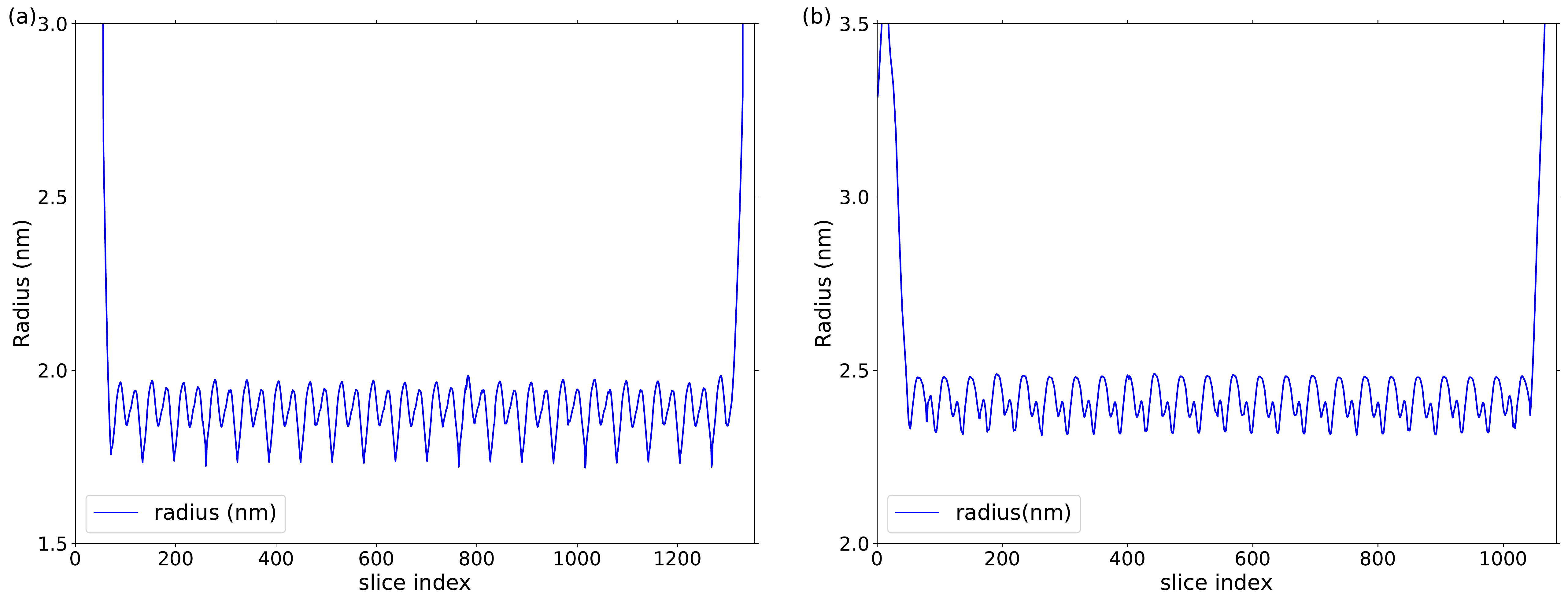}
\caption{Radii as a function of position in the nanotube for 
(a) 4HB-MT and (b) 6HB-MT.
}
\label{sfig:radius_MT}
\end{figure}

\begin{figure}
\centering
\includegraphics[width=7.0in]{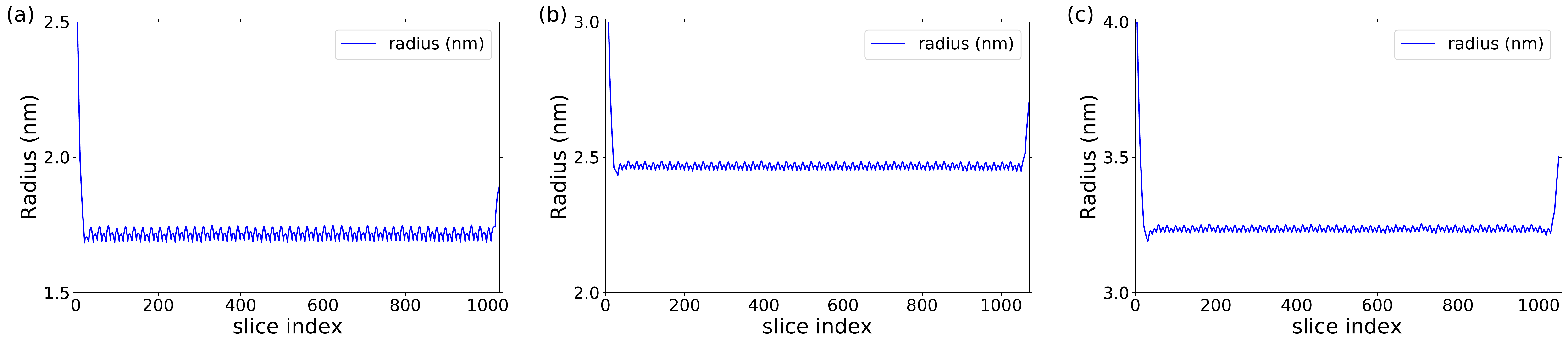}
\caption{Radiii as a function of position in the nanotube for 
(a) 4HB-SST, (b) 6HB-SST and (c) 8HB-SST.
}
\label{sfig:radius_SST}
\end{figure}

\begin{figure*}
\centering
\includegraphics[width=6.6in]{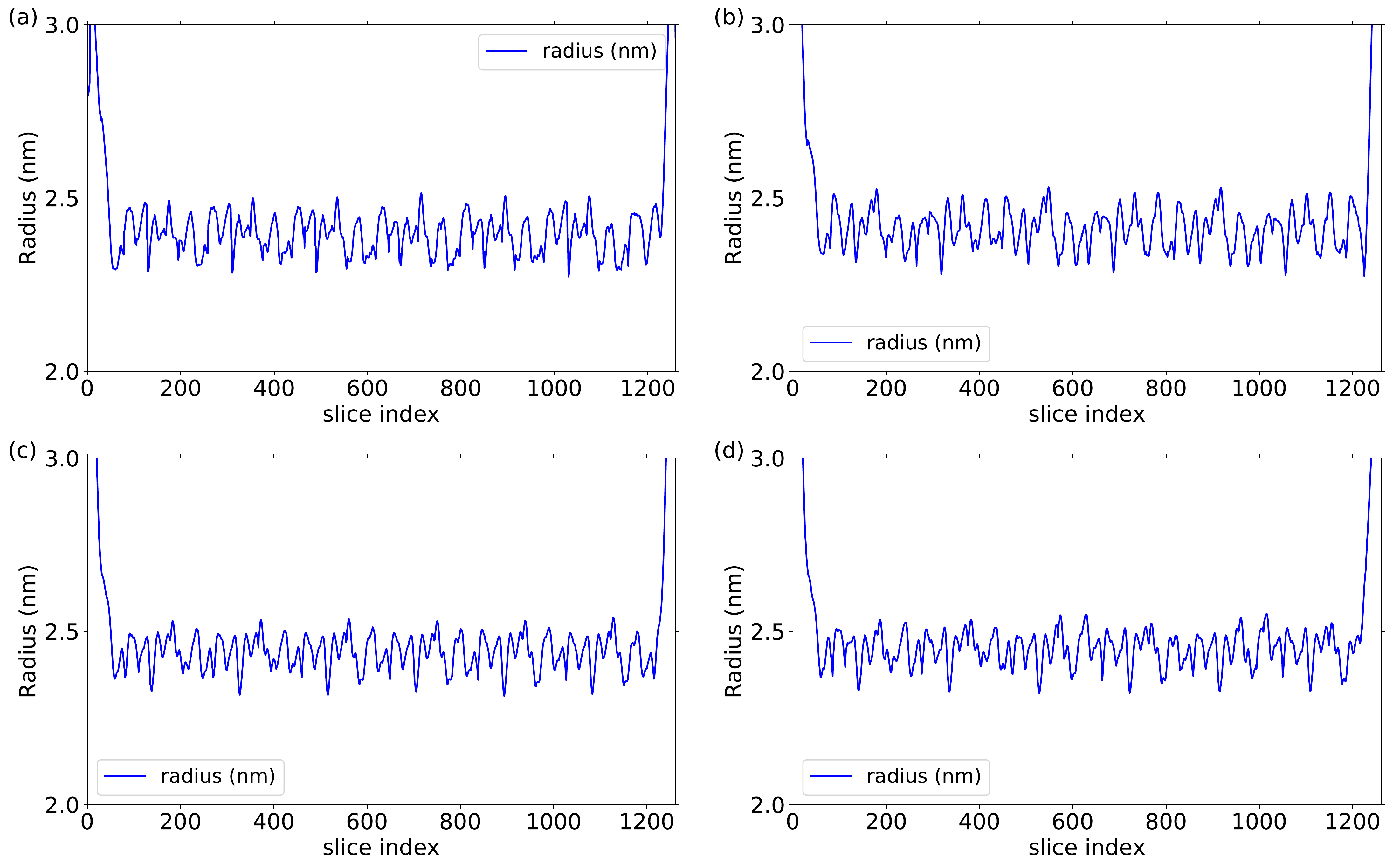}
\caption{Radii as a function of position in the nanotube for 
(a) 6HB-2$\times$LH, (b) 6HB-1$\times$LH, (c) 6HB-S and (d) 6HB-1$\times$RH.
}
\label{sfig:radius_twist}
\end{figure*}

The radii of the nanotubes are depicted in Figs.\ \ref{sfig:radius_MT}-\ref{sfig:radius_twist} 
as a function of position along the tube. Note that we do not provide any results for 10HB due 
to its non-tubular cross-section. The periodicity seen in these plots for the main bodies 
of the nanotubes reflects the periodicity of the pattern of junctions. The SST tubes
have a 21-slice repeat and only show very small variation in the radius, probably because
their single-crossover junctions do not constrain the helices to come quite so close together 
as the double-crossover junctions in the origamis. The MT origamis have a clear pattern with 
a periodicity of 63 slices for 4HB-MT and 42 slices for 6HB-MT. The twisted tubes, by
contrast, have a much more irregular pattern of junctions and a more noisy variation
in the radius, but there is still a rough repeating pattern on the order of 190 slices
(modulated by the insertions and deletions). 
In the case of the origamis, these periodicities can be related to the small features 
seen in the $m$-dependent persistence lengths and elastic moduli at multiples of the 
repeat.

\subsection{Persistence lengths}

For the SST series of nanotubes, we can compare the persistence lengths to theoretical 
expectations. A common approach is as follows.
For a uniform elastic medium, the persistence length can be related to the Young's modulus through
\begin{equation}
l_b=\frac{Y I}{k_B T}
\end{equation}
where $I$ is the moment of inertia of the cross-section.
Using the parallel-axis theorem to calculate $I$ for a bundle of rods and assuming 
the Young's modulus for DNA is the same in a duplex and in a helix-bundle,
one arrives at the following relationship between the persistence length of a helix bundle
and that for duplex DNA:
\begin{equation}
l_b^\mathrm{tube}=n l_b^\mathrm{duplex}\left(1+2\left(\frac{R}{r}\right)^2\right).
\label{eq:Sparallel}
\end{equation}
where $R$ is the radius of the bundle and $r$ the radius of duplex DNA.\cite{Wang12b}
In the experimental papers where this formula has been 
applied \cite{Rothemund04,Liedl10,Wang12b,Schiffels13} 
one difficulty is that $R$ cannot be measured directly; therefore,
one first approximation has been to assume that the helices are in direct 
contact \cite{Liedl10,Wang12b}. 
By contrast, for oxDNA we can measure the tube radii 
(e.g. Fig.\ \ref{sfig:radius_SST}). Thus, together with the oxDNA duplex values of 
$r=1.15$\,nm and $l_b=41.2$\,nm (Table \ref{table:duplex}), we can use Eq.\ \ref{eq:Sparallel}
to predict the nanotube persistence lengths.
In the main text, we use $r$ as an effective parameter to fit the bending persistence lengths 
of the SST nanotubes.
Other potential causes for discrepancies that have been considered in the experimental studies
are imperfections in the tubes (not relevant here) and that nicks and junctions reduce the duplex 
persistence length for the helices in the nanotubes. 

\begin{figure}
\centering
\includegraphics[width=6.6in]{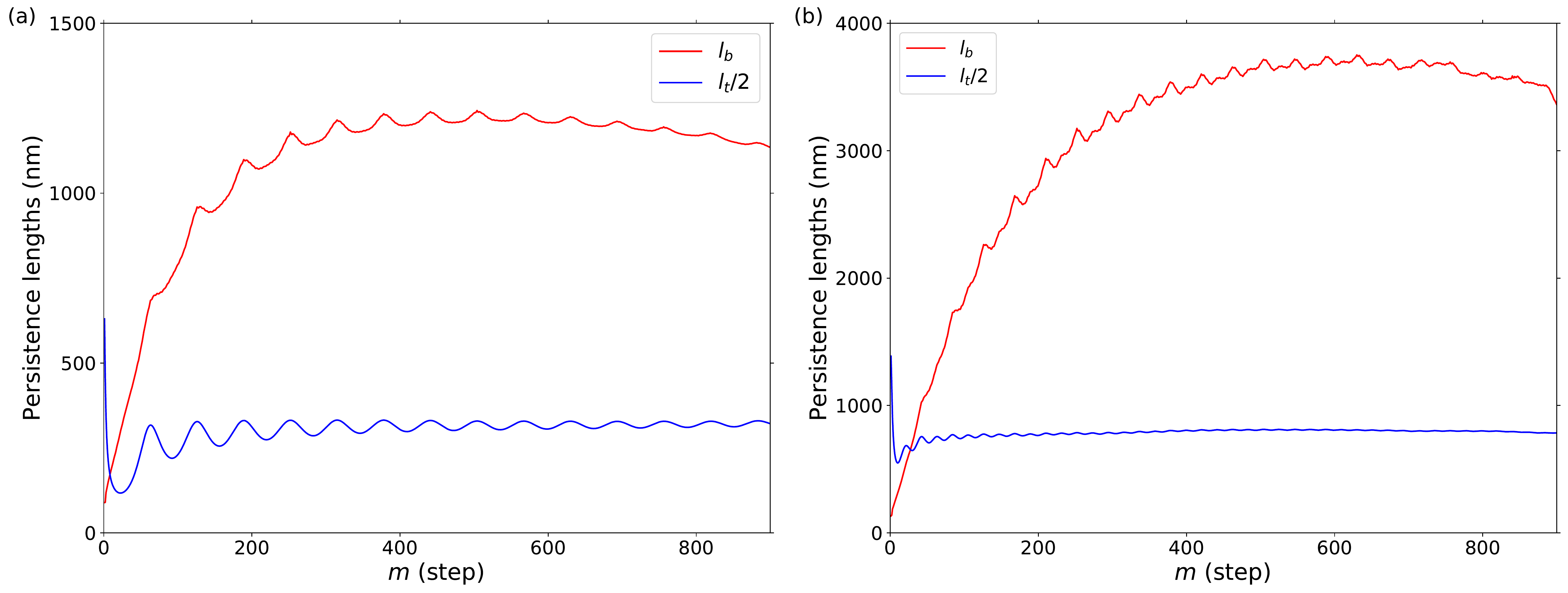}
\caption{$m$-dependent persistence lengths for 
(a) 4HB-MT and (b) 6HB-MT.
}
\label{sfig:lp_MT}
\end{figure}

\begin{figure}
\centering
\includegraphics[width=7.0in]{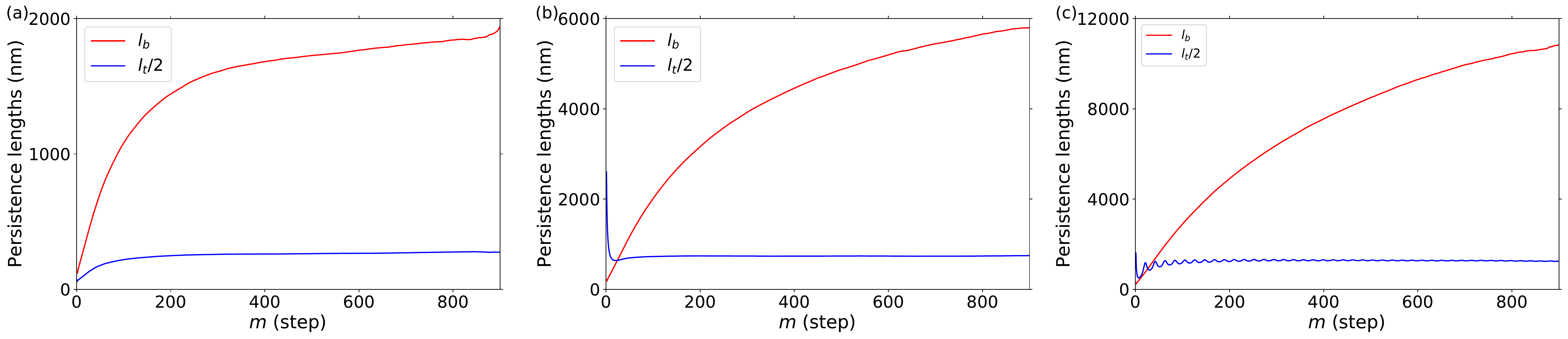}
\caption{$m$-dependent persistence lengths for 
(a) 4HB-SST, (b) 6HB-SST and (c) 8HB-SST.
}
\label{sfig:lp_SST}
\end{figure}

\begin{figure}
\centering
\includegraphics[width=6.6in]{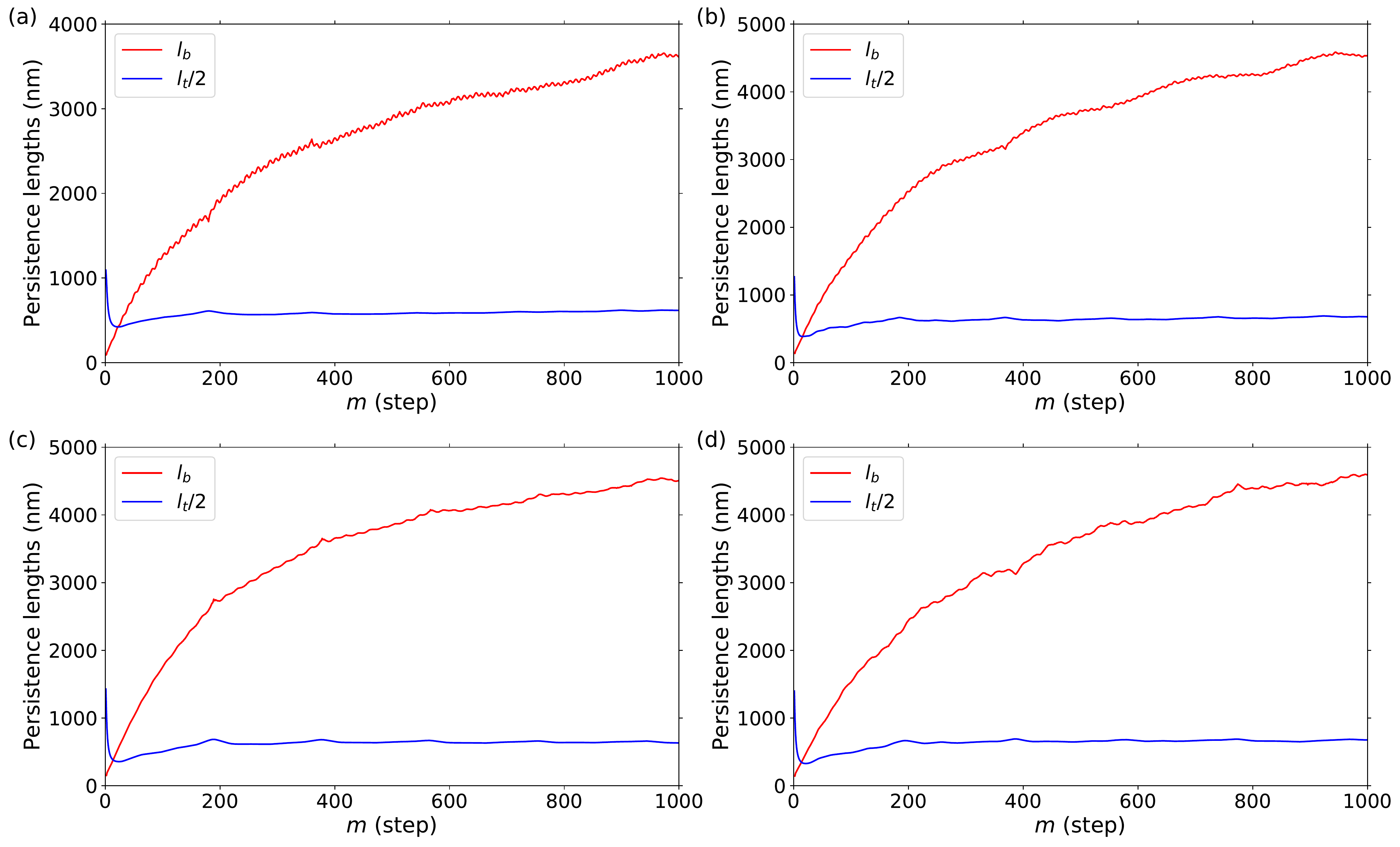}
\caption{$m$-dependent persistence lengths for 
(a) 6HB-2$\times$LH, (b) 6HB-1$\times$LH, (c) 6HB-S and (d) 6HB-1$\times$RH.
}
\label{sfig:lp_twisted}
\end{figure}

\begin{figure}
\centering
\includegraphics[width=3.3in]{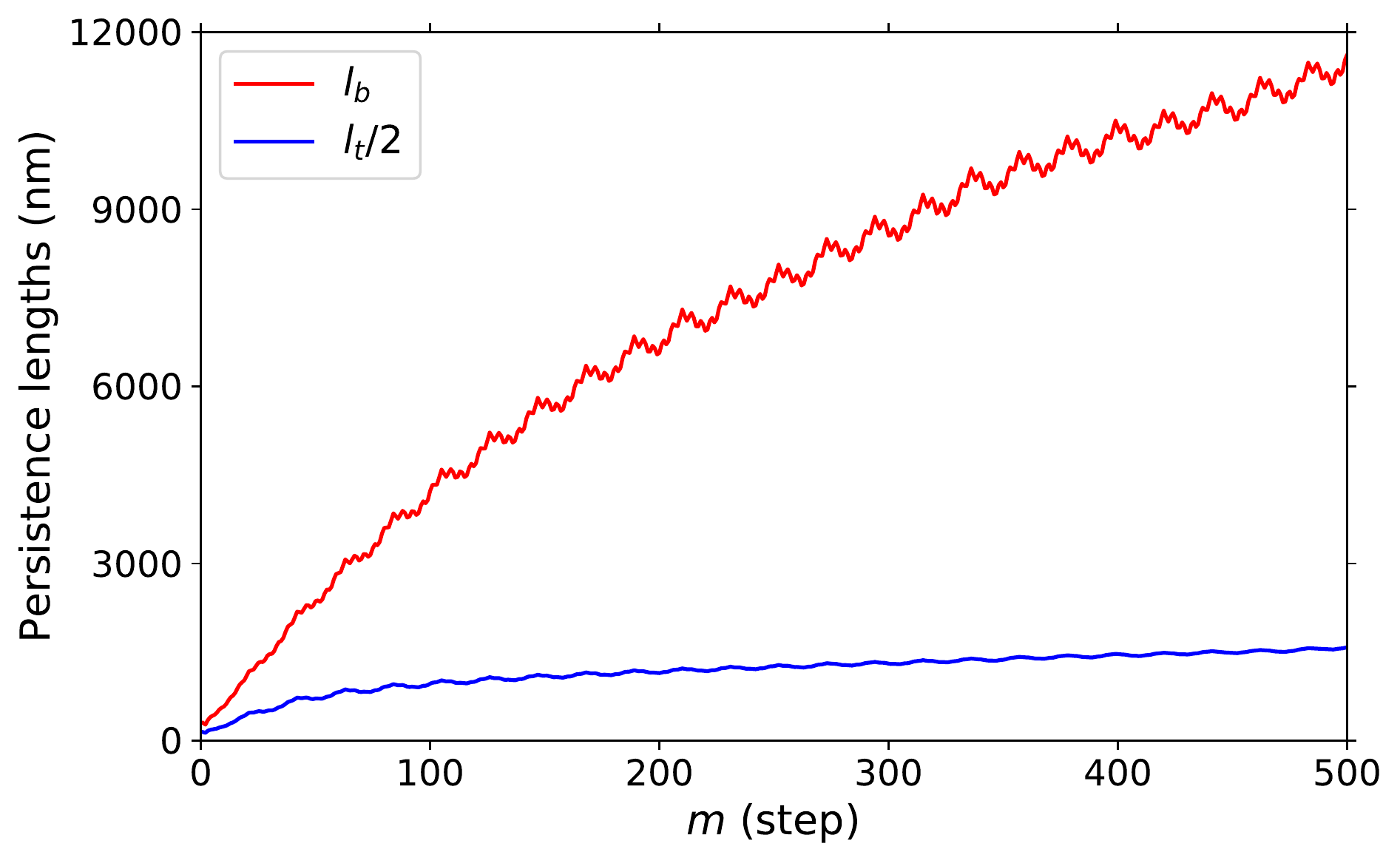}
\caption{$m$-dependent persistence lengths for the 10HB origami. 
}
\label{sfig:lp_10HB}
\end{figure}

The $m$-dependent persistence lengths defined through Eqs.\ 5 and 6 are shown in Figs.\ 
\ref{sfig:lp_MT}-\ref{sfig:lp_10HB}.

\subsection{DNA extensional modulus}

We performed simulations to compute the force-extension curve of double-stranded DNA. 
The strand was 150-bp long and we considered the central 130-bp section.
The simulations were at [Na$^+$]=0.5\,M. Forces up to 40\,pN were considered.

\begin{figure}
\centering
\includegraphics[width=3.3in]{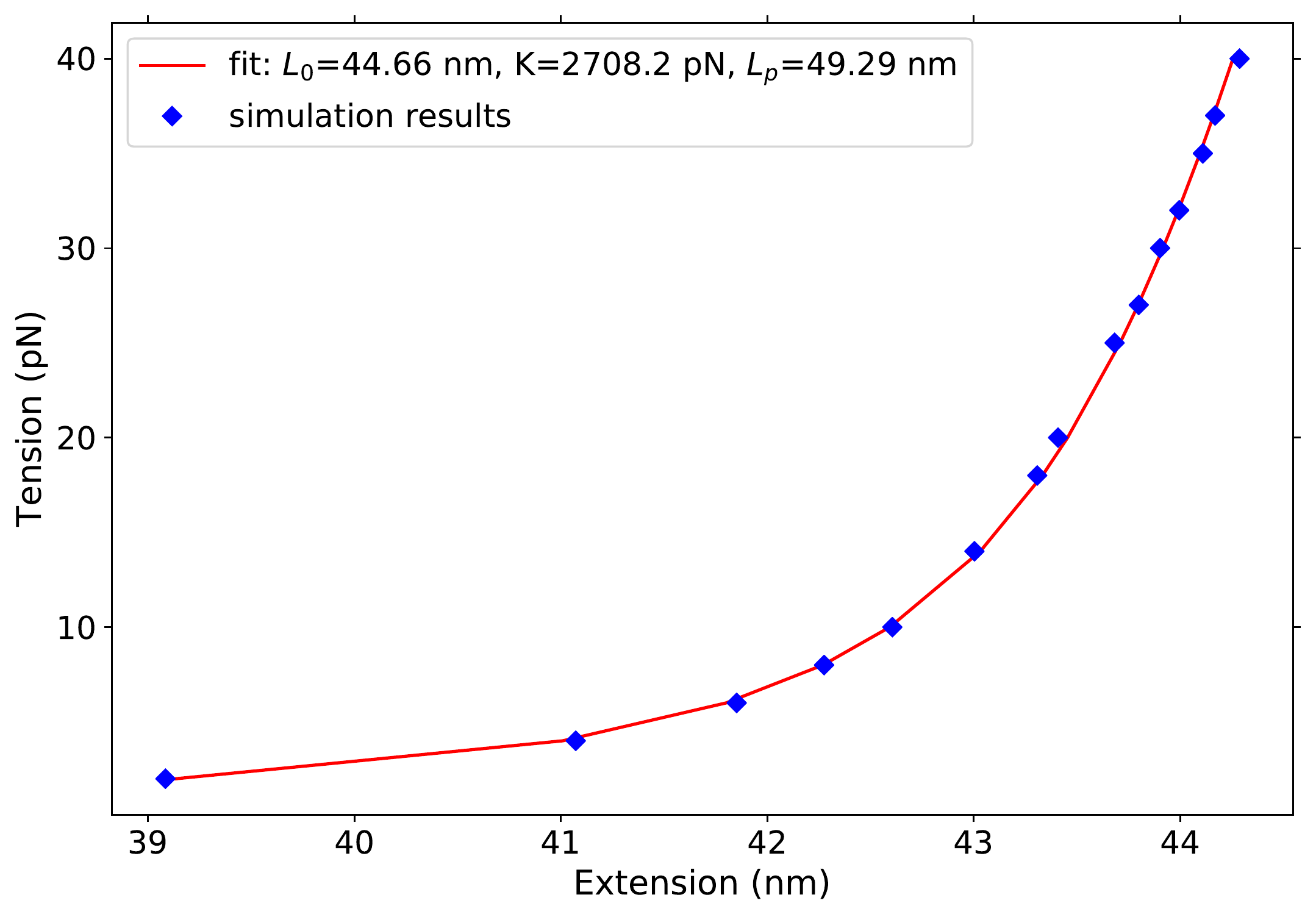}
\caption{Force-extension curve for a 150-bp duplex, neglecting the 10 base pairs at
each end. The data points are from simulations and the solid line a fit to the curve 
using the extensible worm-like chain formula (Eq.\ \ref{eq:EWLC}).}
\label{fig:extension}
\end{figure}

The resulting data (Fig.\ \ref{fig:extension}) was fitted to the extensible worm-like chain 
expression:
\begin{equation}
x=L_0\left(1+\frac{F}{K} -\frac{k_B T}{2 F L_0}\left[1+y\coth y\right]\right),
\label{eq:EWLC}
\end{equation}
where
\begin{equation}
y=\left(\frac{F L_0^2}{l_b k_B T}\right)^{1/2},
\end{equation}
$L_0$ is the contour length, $x$ is the extension, $F$ the force and $K$ the extensional
modulus. An unconstrained three-parameter fit gave $K=2708$\,pN, $L_0=44.66$\,nm and 
$l_b=49.3$\,nm.

\subsection{Elastic moduli}

\begin{figure}
\centering
\includegraphics[width=6.6in]{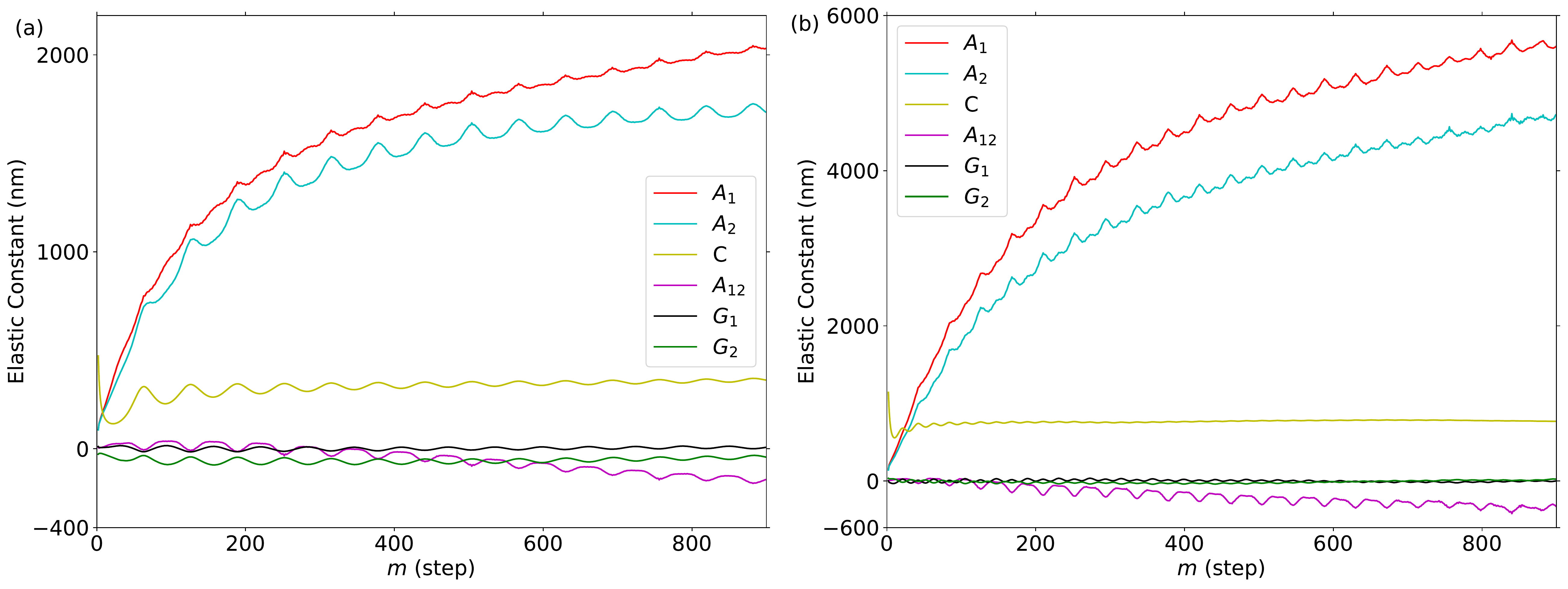}
\caption{Elastic moduli for 
(a) 4HB-MT and (b) 6HB-MT.
}
\label{sfig:elastic_MT}
\end{figure}

\begin{figure}
\centering
\includegraphics[width=7.0in]{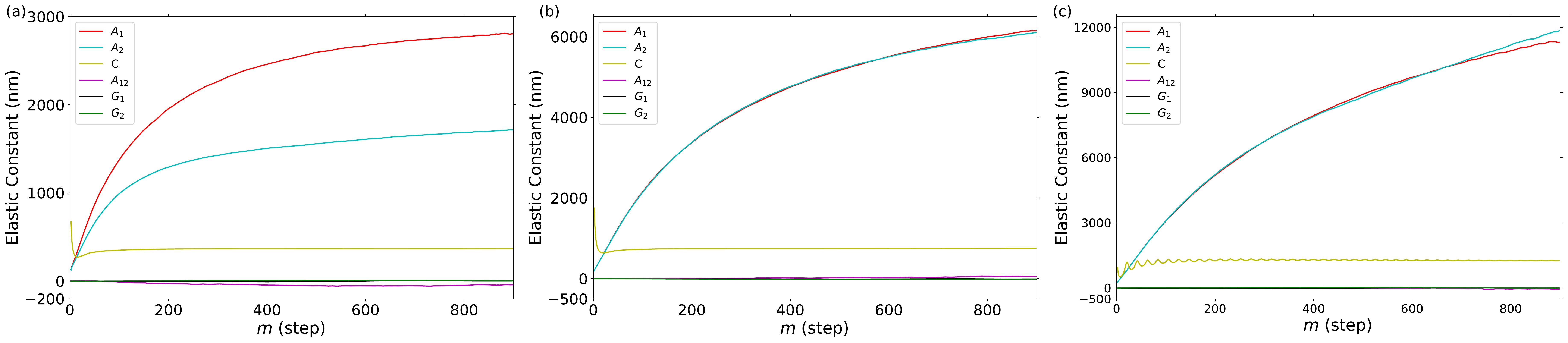}
\caption{Elastic moduli for 
(a) 4HB-SST, (b) 6HB-SST and (c) 8HB-SST.
}
\label{sfig:elastic_SST}
\end{figure}

\begin{figure*}
\centering
\includegraphics[width=6.6in]{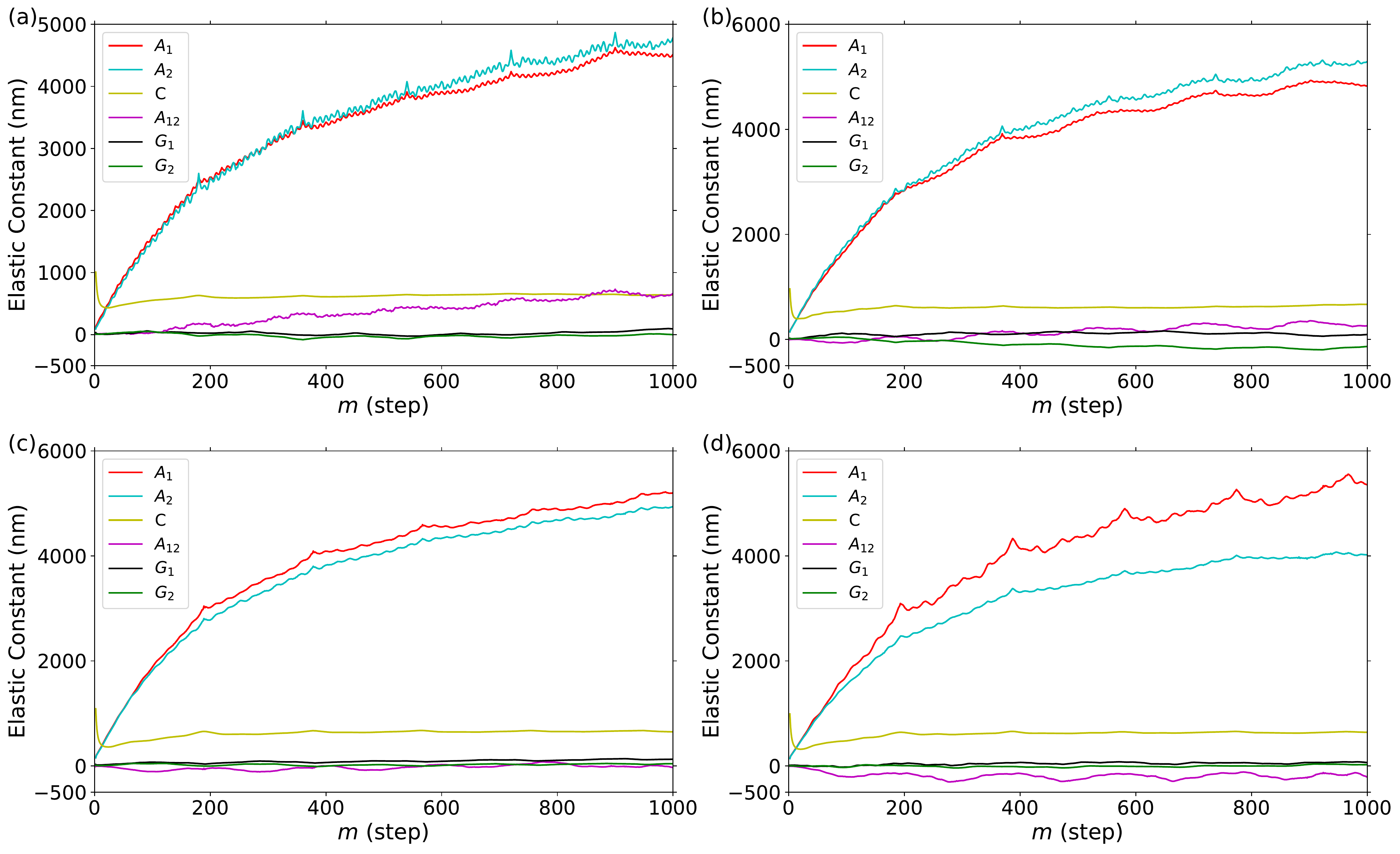}
\caption{Elastic moduli for 
(a) 6HB-2$\times$LH, (b) 6HB-1$\times$LH, (c) 6HB-S and (d) 6HB-1$\times$RH.
}
\label{sfig:elastic_twist}
\end{figure*}

\begin{figure}
\centering
\includegraphics[width=3.3in]{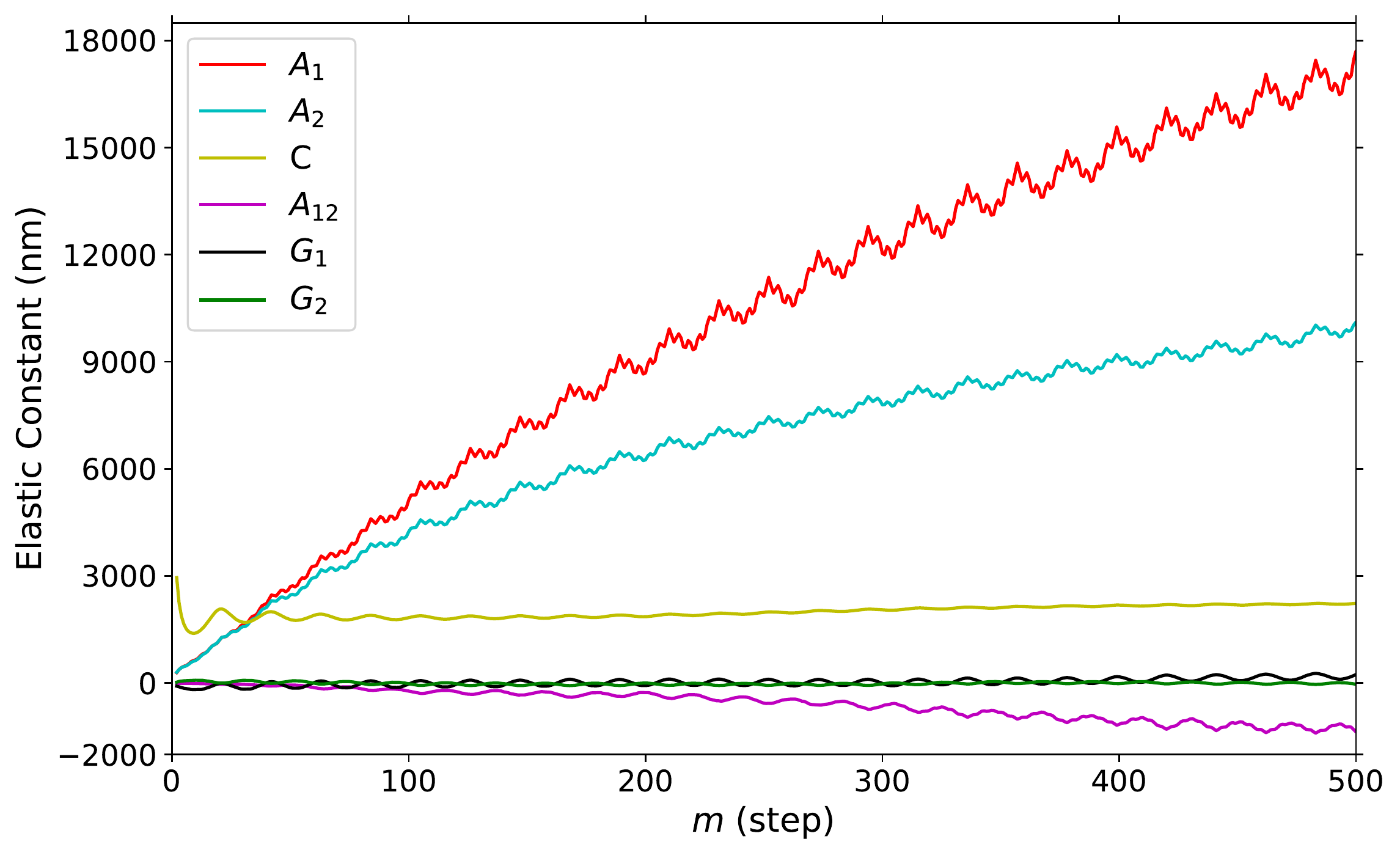}
\caption{Elastic moduli for the 10HB origami at $\alpha=0.06$.
}
\label{sfig:elastic_10HB}
\end{figure}

The $m$-dependent elastic moduli defined through Eqs.\ 7 and 8 are given in Figs.\
\ref{sfig:elastic_MT}--\ref{sfig:elastic_10HB}. Unlike in the main text we also
illustrate the off-diagonal terms. These off-diagonal terms remain closest to
zero for the SST systems, which have both the best statistics and are most
symmetric. 
For the origami systems, although these off-diagonal terms start off close to
zero, there is some tendency to increase with $m$, particularly for $A_{12}$. We
think this behaviour is a result of the worse statistics at longer length
scales. If the off-diagonal elastic constants were actually non-zero, we would
expect that their dependence on $m$ would be much more similar to that for the diagonal
terms, e.g.\ to display a monotonically decreasing slope with increasing $m$,
as is the case for the twist-bend coupling constant for double-stranded DNA.\cite{Skoruppa17}
That $A_{12}$ is typically the largest of the off-diagonal terms is probably
due to the greater difficulty of accurately sampling bending compared to
twisting, because of the much larger value of the bend persistence lengths.
Unfortunately, the non-zero values of the off-diagonal terms at large $m$ are
likely to have a consequent effect on the accuracy of the limiting values of
the other moduli, and to lead to deviations from the expected relationships
between the persistence lengths and elastic moduli that hold when the
off-diagonal terms are zero.  For example, for most systems the value of $C$
quickly converges to its limiting values and is then approximately constant for
larger $m$. However, for the 10HB origami there is an approximately linear rise
in $C$ at large $m$ that seems correlated with the appearance of significant
deviations in $A_{12}$ from zero (Fig.\ \ref{sfig:elastic_10HB}).

\subsection{SST nanotube cross-sections}

\begin{figure}
\centering
\includegraphics[width=6.6in]{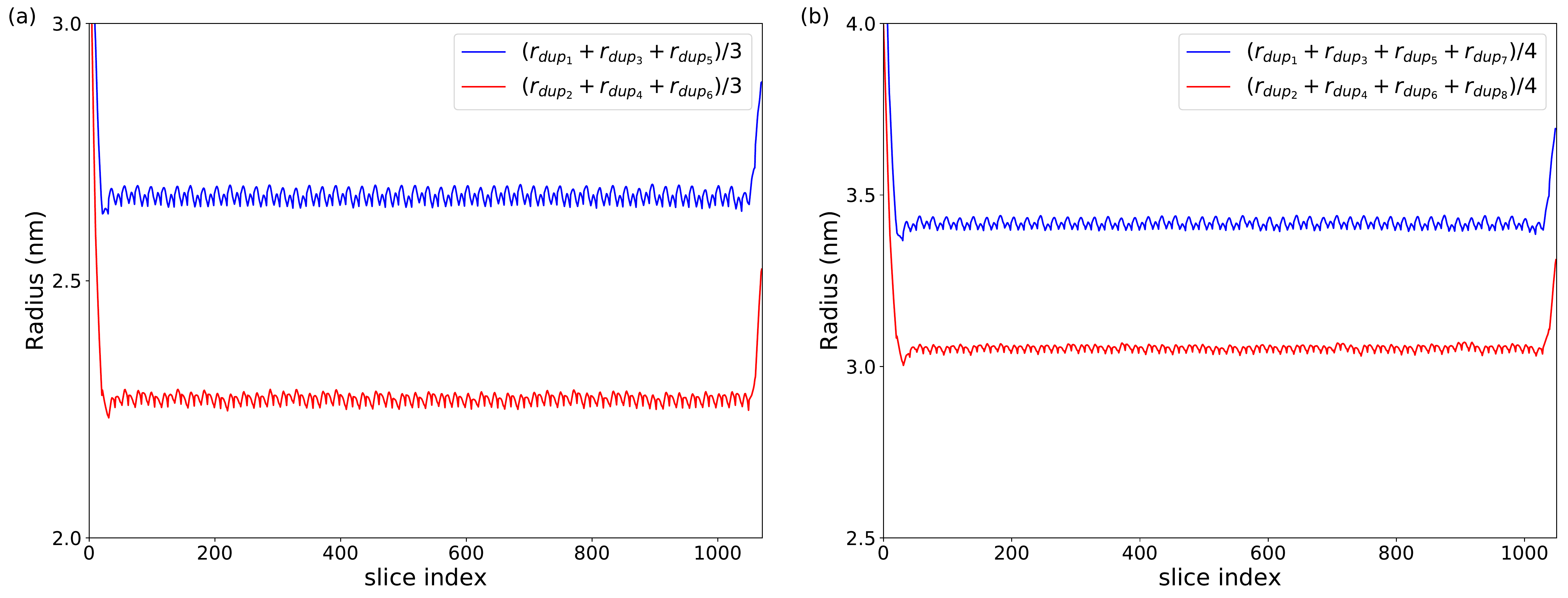}
\caption{Radii for equivalent helices in 6HB-SST and 8HB-SST}
\label{sfig:SST_deform}
\end{figure}

In the main text we showed in Fig.\ 10(b) that equivalent helices in the 4HB-SST nanotube
had different radii,
providing quantitative evidence that the cross-section was diamond-like (Fig. 2(a)) 
rather than square. This is due to the inequivalence of adjacent helices in the SST nanotubes 
and the alternating inter-helix angle as one goes round the tube. Similar plots 
are shown in Fig.\ \ref{sfig:SST_deform} for 6HB-SST and 8HB-SST. The snapshots in Fig.\ 
\ref{sfig:SST_pics} also confirm the deviations of the cross-section away from a regular
polygon.

\end{document}